\documentclass[twocolumn]{aastex631}

\usepackage{epsfig}
\usepackage{graphicx}
\usepackage{tabularx}
\usepackage{float}
\usepackage{placeins}
\usepackage{mathtools}
\usepackage{pgf}
\usepackage{ulem}
\usepackage{lipsum}
\usepackage{placeins}

\usepackage{booktabs}

\usepackage[T1]{fontenc}

\usepackage{amssymb}
\usepackage{hyperref}
\usepackage{tikz}
\usepackage{blindtext}

\shorttitle{LONG-PERIOD VARIABLE STARS IN IC\,10}
\shortauthors{Gholami et al.}

\begin{document}

\title{THE ISAAC NEWTON TELESCOPE MONITORING SURVEY OF LOCAL GROUP DWARF GALAXIES. VII. \\LONG-PERIOD VARIABLE STARS IN THE NEAREST STARBURST DWARF GALAXY, IC\,10}

\author{Mahtab Gholami}
\affiliation{INAF - Osservatorio Astronomico di Capodimonte, Salita Moiariello 16, I-80131 Napoli, Italy}
\affiliation{School of Astronomy, Institute for research in fundamental sciences (IPM), Tehran, P.O.\ Box 19395-5531, Iran}
\email{mahtab.gholami@inaf.it}

\author{Atefeh Javadi}
\affiliation{School of Astronomy, Institute for research in fundamental sciences (IPM), Tehran, P.O.\ Box 19395-5531, Iran}

\author{Hamidreza Mahani}
\affiliation{School of Astronomy, Institute for research in fundamental sciences (IPM), Tehran, P.O.\ Box 19395-5531, Iran}

\author{Jacco Th. van Loon}
\affiliation{Lennard-Jones Laboratories, Keele University, Staffordshire ST5 5BG, UK}

\author{Habib Khosroshahi}
\affiliation{School of Astronomy, Institute for research in fundamental sciences (IPM), Tehran, P.O.\ Box 19395-5531, Iran}
\affiliation{Iranian National Observatory (INO), Institute for Research in Fundamental Sciences (IPM), Tehran, P.O.\ Box 19568-36613, Iran}

\author{Elham Saremi}
\affiliation{School of Astronomy, Institute for research in fundamental sciences (IPM), Tehran, P.O.\ Box 19395-5531, Iran}
\affiliation{Instituto de Astrof\'{\i}sica de Canarias, C. V\'{\i}a L\'actea s/n, 38205 La Laguna, Santa Cruz de Tenerife, Spain}
\affiliation{Departamento de Astrof\'{\i}sica, Universidad de La Laguna, 38205 La Laguna, Santa Cruz de Tenerife, Spain}

\author{Iain McDonald}
\affiliation{Jodrell Bank Centre for Astrophysics, Alan Turing Building, University of Manchester, M13 9PL, UK}

\author{Samaneh Eftekhari}
\affiliation{School of Astronomy, Institute for research in fundamental sciences (IPM), Tehran, P.O.\ Box 19395-5531, Iran}

\author{Yi Ren}
\affiliation{Department of Astronomy, Beijing Normal University, Beijing 100875, People’s Republic of China}
\affiliation{College of Physics and Electronic Engineering, Qilu Normal University, Jinan 250200, People’s Republic of China}

\author{Hamed Altafi}
\affiliation{Iranian National Observatory (INO), Institute for Research in Fundamental Sciences (IPM), Tehran, P.O.\ Box 19568-36613, Iran}

\correspondingauthor{Atefeh Javadi}
\email{atefeh@ipm.ir}

\begin{abstract}

To identify long-period variable (LPV) stars in IC\,10 -- the nearest starburst galaxy of the Local Group (LG) -- we conducted an optical monitoring survey using the 2.5-m {\it Isaac Newton} Telescope (INT) with the wide-field camera (WFC) in the i--band and V--band from 2015 to 2017. We created a photometric catalog for 53,579 stars within the area of CCD4 of WFC ($\sim$ 0.07 deg$^2$ corresponding to 13.5 kpc$^2$ at the distance of IC\,10), of which we classified 536 and 380 stars as long-period variable candidates (LPVs), mostly asymptotic giant branch stars (AGBs) and red supergiants (RSGs), within CCD4 and two half-light radii of IC\,10, respectively. By comparing our output catalog to the catalogs from Pan-STARRS, {\it Spitzer} Space Telescope, {\it Hubble} Space Telescope (HST), and carbon stars from the Canada-France-Hawai'i Telescope (CFHT) survey, we determined the success of our detection method. We recovered $\sim$ 73\,\% of {\it Spitzer}'s sources in our catalog and demonstrated that our survey successfully identified 43\,\% of the variable stars found with {\it Spitzer}, and also retrieved 40\,\% of the extremely dusty AGB stars among the {\it Spitzer} variables. In addition, we successfully identified $\sim$ 70\,\% of HST variables in our catalog. Furthermore, we found all the confirmed LPVs that {\it Gaia} DR3 detected in IC\,10 among our identified LPVs. This paper is the first in a series on IC\,10, presenting the variable star survey methodology and the photometric catalog, available to the public through the Centre de Donn\'ees Astronomiques de Strasbourg.
\end{abstract}


\section{Introduction} \label{sec:intro}

The Local Group (LG), with a diameter of approximately 3 Mpc \citep{mcconnachi12}, is mainly populated by dwarf galaxies. Dwarfs with various luminosities, 
metallicities, and stellar populations help explain galaxies' evolution and structure, and how the LG components came together \citep{Weisz14}. The dwarf irregular 
galaxy IC\,10 (UGC\,192) is the nearest starburst dwarf, with a star formation rate ranging from 0.02 M$_\odot$ yr$^{-1}$ to 0.04 M$_\odot$ yr$^{-1}$ \citep{Yin10} 
and a total stellar mass of (4--6) $\times$ 10$^8$ M$_\odot$ \citep{Jarrett03, Sakai99, Vaduvescu07}, located within the LG, and has an oxygen abundance similar to 
that of the Large Magellanic Clouds (LMC) \citep{Tehrani17, Gerbrandt15}. In IC\,10, the young, intermediate, and old-age stellar populations have all been studied \citep{Massey95, Sakai99, Borissova00, Gholami19}. The high concentration of Wolf--Rayet stars (WRs) in this galaxy indicates recent star formation activity associated with H\,{\sc ii} regions \citep{Mateo98, Richer01, Crowther03, Cosens22, Tramper15, Cosens22}.

IC\,10 lies at a low Galactic latitude ($\ell$ = 118$^{\circ}$.9, $b$ = -3$^\circ$.3), where significant foreground reddening is expected, making it challenging to determine the precise distance and reddening for this galaxy \citep{Richer01}. Table \ref{table:dis} lists distance estimations for IC\,10 using various methods, resulting in substantial variation in the values for the distance modulus ($\mu$), ranging from $\sim$ 22 mag to $\sim$ 25 mag, partly attributed to the adoption of different reddening estimations, E(B--V), ranging from 0.4 mag \citep{deVaucouleurs78} to $\sim$ 2 mag \citep{Yang93}. In our study, we adopted a distance modulus of 24.51 $\pm$ 0.08 mag from \cite{Sanna08} (see Sec. \ref{sec:ex}).

\begin{table*}
\begin{center}
\caption{A brief history of distance estimation for IC\,10.
\label{table:dis}
}
\begin{tabular}{cccccc}
\hline
References    &  Method  &   Filter  &  E(B--V)  &  $\mu$  &  d  \\
         &          &          &          &    (mag)      &  (kpc) \\

\hline
\cite{deVaucouleurs65} & Integrated B--V color & UBV & 0.87 & 24.06 & 649 \\
\cite{deVaucouleurs78} & H\,{\sc ii} rings & BV  & 0.4 & 24.9 & 955 \\
\cite{Yang93} & H\,{\sc ii} regions &   & 1.7--2.0 & 22.5--21.9 & 316--240 \\
\cite{Massey95} & WRs & sp/BV & 0.75--0.8 & 24.9 & 950 \\
\cite{Wilson96} & Cepheids & JHK & 0.6--1.1 & 24.8--23.6 & 912--525 \\
\cite{Saha96} & Cepheids & gri & 0.97 & 23.87 & 594 \\
\cite{Sakai99} & Cepheids & VI & 1.16 $\pm$ 0.08 & 24.10 $\pm$ 0.20 & 661 \\
\cite{Sakai99} & TRGB & VI & 1.16 $\pm$ 0.08 & 23.51 $\pm$ 0.19 & 504 \\
\cite{Borissova00} & RSGs & JHK & 1.05 $\pm$ 0.10 & 23.86 $\pm$ 0.12 & 592 \\
\cite{Hunter01} & TRGB & F555W, F814W & 0.77 & 24.95 $\pm$ 0.20 & 700--800 \\
\cite{Demers04} & Carbon stars & R, I, CN, TiO & 0.79 & 24.35 $\pm$ 0.11 & 741 \\
\cite{Leroy06} & TRGB & CO luminosity & 0.77 & 24.9 & 950 \\
\cite{Vacca07} & TRGB &  F814W, K' & 0.95 & 24.48 $\pm$ 0.08 & 787 \\
\cite{Sanna08} & TRGB & F555W, F814W & 0.78 $\pm$ 0.06 & 24.51 $\pm$ 0.08 & 798 \\
\cite{Kim09} & TRGB & JHK$_{\rm s}$ & 0.98 $\pm$ 0.06 & 24.27 $\pm$ 0.03 $\pm$ 0.18 & 715 \\
\cite{Gon12} & PNLF & m(5007)$\AA$ & 0.77 & 24.1 $\pm$ 0.25 & 660 \\
\cite{Lim15} & TRGB & UBVRI & 0.52 $\pm$ 0.04 & 24.27 $\pm$ 0.03 & 715 \\
\cite{Kristen17} & TRGB & 3.6 $\mu$m, F814W & 0.78 & 24.43 $\pm$ 0.03 & 769 \\
\cite{Boyer15a, Boyer15b} & TRGB, SFD98 & F814W, UBVRI & 0.81 & 24.09 & 660 \\
\cite{Dell'Agli18} & Carbon stars & NIR, mid-IR & 1.14 & 24.43 & 770 \\\hline
\end{tabular}
\end{center}
\end{table*}

Since asymptotic giant branch stars (AGBs) and red supergiants (RSGs) in the long-period variable (LPV) phase exhibit larger amplitudes in optical bands, they 
are easier to find in the optical than at longer wavelengths; especially RSGs, which tend to have smaller amplitudes anyway. Additionally, optical photometry better constrains the circumstellar dust envelope through optical depth, as circumstellar dust is more opaque in the optical than in the infrared. Consequently, the reddening of the star's light by circumstellar dust is more readily noticed in the optical. To determine the quantity of mass loss and dust generated by LPVs, it is necessary to analyze their spectral energy distribution (SED) \citep{Javadi13, Abdollahi23}. The SED of these stars is derived from their observed magnitudes across various wavelengths. While the DUSTiNGs survey (Sec.~\ref{sec:dustings}) covers the mid-infrared range, complementing these observations with optical band data allows for a more precise determination of the SED. Therefore, optical studies provide valuable insights into dust production and mass loss.

The coolest (T $\sim$ 3000--4000 K) and most luminous ($\sim$ 10\,000--60\,000 L$_\odot$) AGBs generate a substantial amount of dust in their atmosphere, which they then eject into the interstellar medium (ISM) at rates of up to 10$^{-4}$ M$_\odot$ yr$^{-1}$ \citep{Herwig05}. These stars can significantly contribute to the interstellar dust budget \citep{Goldman17}, playing a crucial role in altering the chemical composition of galaxies and enhancing the rate of star formation \citep{Javadi15, Tikhonov10}. Nevertheless, it is still unclear how metallicity affects the mass loss of AGBs.

AGBs and RSGs are in their final stages of post-main-sequence stellar evolution. Low to intermediate-mass stars (0.8--8\,M$_\odot$) reach the AGB phase 
\citep{Marigo08}, whereas more massive stars reach the RSG phase \citep{Levesque05, Levesque10}. These two phases trace the evolution of stellar populations spanning 10 Myr to 10 Gyr of cosmic time. During the final evolutionary stage of AGBs, the degenerate carbon--oxygen core is surrounded by hydrogen and helium-burning shells \citep{Habing03}. Over time, AGBs experience thermal pulses due to the instability of the helium shell \citep{Marigo13}, during which the nuclear products of helium burning are transported to the stellar surface via the third dredge-up process, alerting the abundance of elements in their atmosphere \citep{Busso99, Fishlock14, Karakas14, Cristallo15}. 

Towards the end of the thermal pulse AGB phase, these stars exhibit long-period variability on the order of months to over 1000 days due to the radial pulsation of 
their cool atmospheres \citep{Wood92, Wood98, Pierce00, Whitelock03}. 

RSGs are stars with masses up to $\sim$ 30 M$_\odot$ that terminate in supernovae \citep{Levesque05, Levesque10, Yang12}; they are valuable probes for calculating recent star formation in a galaxy. RSGs also exhibit long-period variability on timescales of years, though this may not be solely due to radial pulsation.

IC\,10's LPV candidates are reported in the {\it Gaia} DR3 catalog \citep{Lebzelter23, Rimoldini23}, which contains the brightest sources in this galaxy. Our catalog covers a wide range of magnitudes and is the most comprehensive LPV candidate catalog of IC\,10.

We surveyed most dwarf galaxies in the LG between June 2015 and October 2017, using the 2.5-m {\it Isaac Newton} Telescope (INT) over nine epochs \citep{Saremi17, Saremi20} with cadences for an expected period to adequately identify LPVs. The primary goal of our survey is to compile the most comprehensive catalog to date of LPV candidates in LG dwarf galaxies observable from the Northern Hemisphere. In this paper, we identified LPV candidates in the IC\,10 dwarf irregular galaxy. In subsequent work, we will determine the star formation history (SFH) of the galaxy based on the luminosity distribution of the LPVs in our next paper, employing a technique we have successfully applied to other LG members \citep{Javadi11b, Javadi15, Javadi17, Rezaeikh14, Golshan17, Hashemi18, Saremi21, Navabi21, Parto23, Abdollahi23, Gholami23}. In the process, we obtain precise time-averaged photometry for the LPV candidates and pulsation amplitudes. In our next work, we will use the variability information to determine SEDs, photospheric radius variations, and the relationship between mass loss and stellar characteristics.

This is the first paper in a series on IC\,10 and paper VII in the series of our monitoring survey of LG dwarf galaxies, organized as follows: In Sec.~\ref{sec:obs}, we explain the observational data. A description of the photometry methods and calibration is presented in Sec.~\ref{sec:Phot}. In Sec.~\ref{sec:var}, we discuss the method of identification of variable candidates in detail. In Sec.~\ref{sec:DIS}, we describe the LPV candidates and cross-correlate our detected sources with other photometric catalogs. A summary is given in Sec.~\ref{sec:CON}.

\section{OBSERVATIONS} \label{sec:obs}

We conducted imaging with the 2.5-m INT at the Observatorio del Roque de Los Muchachos (La Palma), covering nine epochs spaced a few months apart between June 2015 and October 2017, with cadences for an expected period to adequately identify LPVs. The log of the wide-field camera (WFC) observations of IC\,10 is presented in Table~\ref{table:log}. We used the WFC, an optical mosaic camera containing four 2048 $\times$ 4096 pixels CCDs with a pixel size of 0$\rlap{.}^{\prime\prime}$33. The observations were performed in the Sloan i and Harris V filters, except for the night of 21 June 2015, which used the Landolt I filter. Therefore, we used the transformation equation (Eq.~\ref{eq:jordi}) determined by \cite{Jordi06} to obtain the magnitudes for all stars in the Sloan i--band. For the estimation of color (R--I), we used the image observed in the R-band on 21 June 2015. The I--band magnitudes were then transformed into Sloan i--band magnitudes using the equation below.
\begin{equation}\label{eq:jordi}
i - I = (0.251 \pm 0.003) \times (R - I) + (0.325 \pm 0.002).
\end{equation}

Further investigation of the images taken on June 21, 2015, revealed that due to the high seeing values on that night, a significant fraction of stars were affected by blending. As a result, their magnitudes cannot be reliably used for detecting variable stars. Consequently, the data from this night has been excluded from further analysis.

\begin{table}
\begin{center}
\caption{Log of WFC observations of IC\,10 dwarf galaxy.
\label{table:log}
}
\begin{tabular}{cccccc}
\hline
    Date   &  Filter  &   Epoch  &  t$_{\rm exp}$  &  Airmass  &  Seeing  \\
(y m d)    &          &          &      (sec)      &           &  (arcsec) \\

\hline
2015 06 21 &   I &  1 & 430 & 1.748 & 1.87 \\
2015 06 21 &   R &  1 & 181 & 1.582 & 1.80 \\
2015 06 21 &   V &  1 & 430 & 1.634 & 1.90 \\
2016 06 13 &   i &  1 & 555 & 1.450 & 1.10 \\
2016 06 15 &   i &  2 & 555 & 2.271 & 1.82 \\
2016 08 10 &   i &  3 & 555 & 1.168 & 1.11 \\
2016 08 13 &   V &  2 & 735 & 1.331 & 1.69 \\
2016 10 19 &   i &  4& 555 & 1.303 & 1.55 \\
2016 10 19 &   V &  3 & 735 & 1.361 & 1.62 \\
2017 08 01 &   i &  5 & 555 & 1.496 & 1.35 \\
2017 08 01 &   V &  4 & 735 & 1.411 & 1.40 \\
2017 09 01 &   i &  6 & 555 & 1.263 & 1.47 \\
2017 09 01 &   V &  5 & 735 & 1.225 & 1.62 \\
2017 09 02 &   i &  7 & 555 & 1.634 & 1.18 \\
2017 09 02 &   V &  6 & 735 & 1.525 & 1.43 \\
2017 10 06 &   i &  8 & 555 & 1.263 & 1.24 \\
2017 10 08 &   V &  7 & 735 & 1.184 & 1.11 \\\hline
\end{tabular}
\end{center}
\end{table}


\section{DATA PROCESSING, PHOTOMETRY AND CALIBRATION} \label{sec:Phot}

Before commencing the photometry process, we utilized Transforming HEavenly Light into Images (THELI) for the automated pre-reduction of astronomical images \citep{Erben05}. Photometry for all sources in IC\,10's crowded stellar field was conducted using the {\sc daophot/allstar} package \citep{Stetson87, Stetson94, Stetson90}, fitting the Point-Spread Function (PSF) on each image separately in both the i--band and V--band. A constant PSF model was determined based on a 
selection of approximately 50 bright and isolated stars.

\begin{figure*}[]
\centering
\includegraphics[width=1\linewidth,clip]{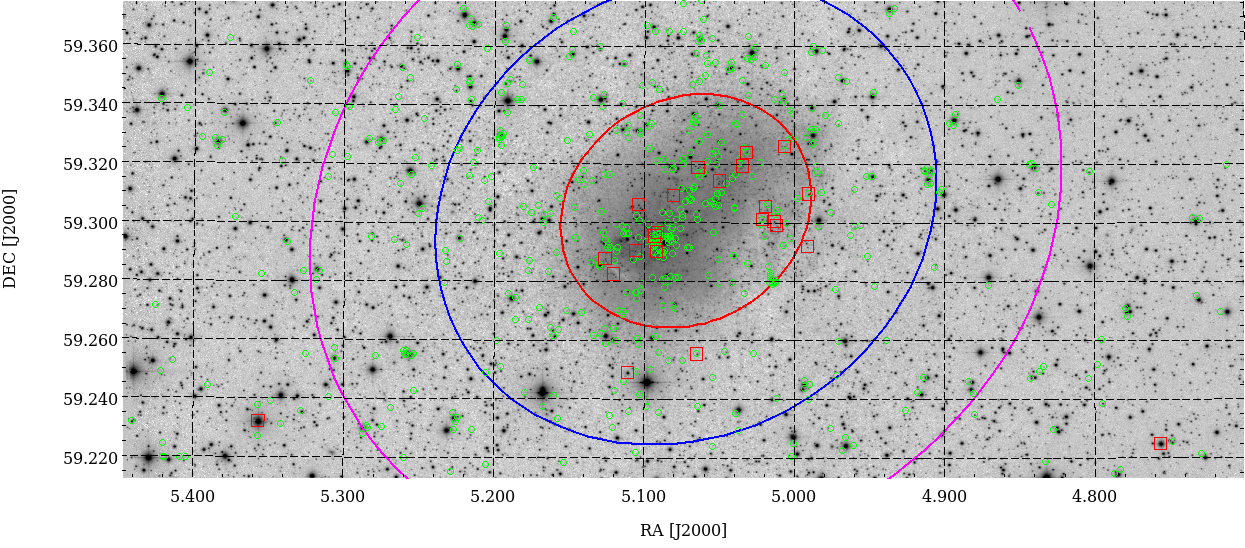}
\caption{Median image of IC\,10. The red, blue, and magenta ellipses in the middle of the image indicate the half-light radius (r$_{\rm h}$), two half-light radii (2r$_{\rm h}$), and three half-light radii (3r$_{\rm h}$) from the center of IC\,10, respectively. The LPV candidates are displayed in green. The {\it Gaia} DR3 LPVs are represented by red open squares (Sec.~\ref{sec:cont}).}
\label{fig:lpv-median}
\end{figure*}

The {\sc daomatch} routine was employed to obtain rough coordinate transformations between the images, followed by the {\sc daomaster} routine to refine these 
transformations. The {\sc montage} routine was then executed to combine the individual images in both i and V bands, creating a median image (Fig.~\ref{fig:lpv-median}). A master list of stars was produced using the {\sc allstar} routine, which implements PSF photometry.

Residual magnitudes between the PSF-fitted values and magnitudes in large aperture photometry were obtained using the {\sc daogrow} routine \citep{Stetson90}. Aperture corrections were calculated by measuring the difference between the magnitude of the PSF-fitting and the largest aperture's magnitude for PSF stars via the {\sc collect} routine \citep{Stetson93}. The {\sc ccdave} routine was then applied to calibrate the magnitudes of stars on the master list. Finally, the {\sc 
newtrial} routine was executed to calculate and apply the aperture corrections, convert instrumental magnitudes to the standard system, and estimate the weighted mean magnitude for all stars. A zero-point for each frame was derived from the standard star images (ranging from a few to several tens each night) to 
determine photometric calibration. For nights lacking observations of standard stars, the average of all zero points over all frames was used. Using 
Eq.~\ref{eq:jordi}, we determined the precise zero points for the Sloan i filter.

Airmass-dependent atmospheric extinction correction was applied, adopting extinction coefficients of 0.0197 mag and 0.1036 mag in the i--band and V--band, 
respectively, as determined for La Palma \citep{Garcia-Gil10}. In addition to atmospheric extinction, corrections for external extinction caused by dust clouds in the Milky Way (MW) and internal extinction generated by dust within IC\,10 were also applied (see \cite{Tikhonov10}; Sec.~\ref{sec:ex} \& Sec.~\ref{sec:cont}).

We calibrated the images relative to one another to improve the accuracy of variability detection \citep{Saremi20}. To perform this relative calibration, we selected $\sim$ 1000 stars that were in common between all images in each band. Then, the magnitude deviation at each epoch was calculated concerning the weighted mean magnitude for that star over all epochs; these deviations were then averaged for each epoch and filter. Even though the relative values were minor (between $-$0.005 and $+$0.006 mag), we applied them to the images in each band to obtain the best internally calibrated photometry. Consequently, the images were properly calibrated, with magnitude corrections of less than 1\,\%.

\begin{figure}[]
\centering
\includegraphics[width=1\linewidth,clip]{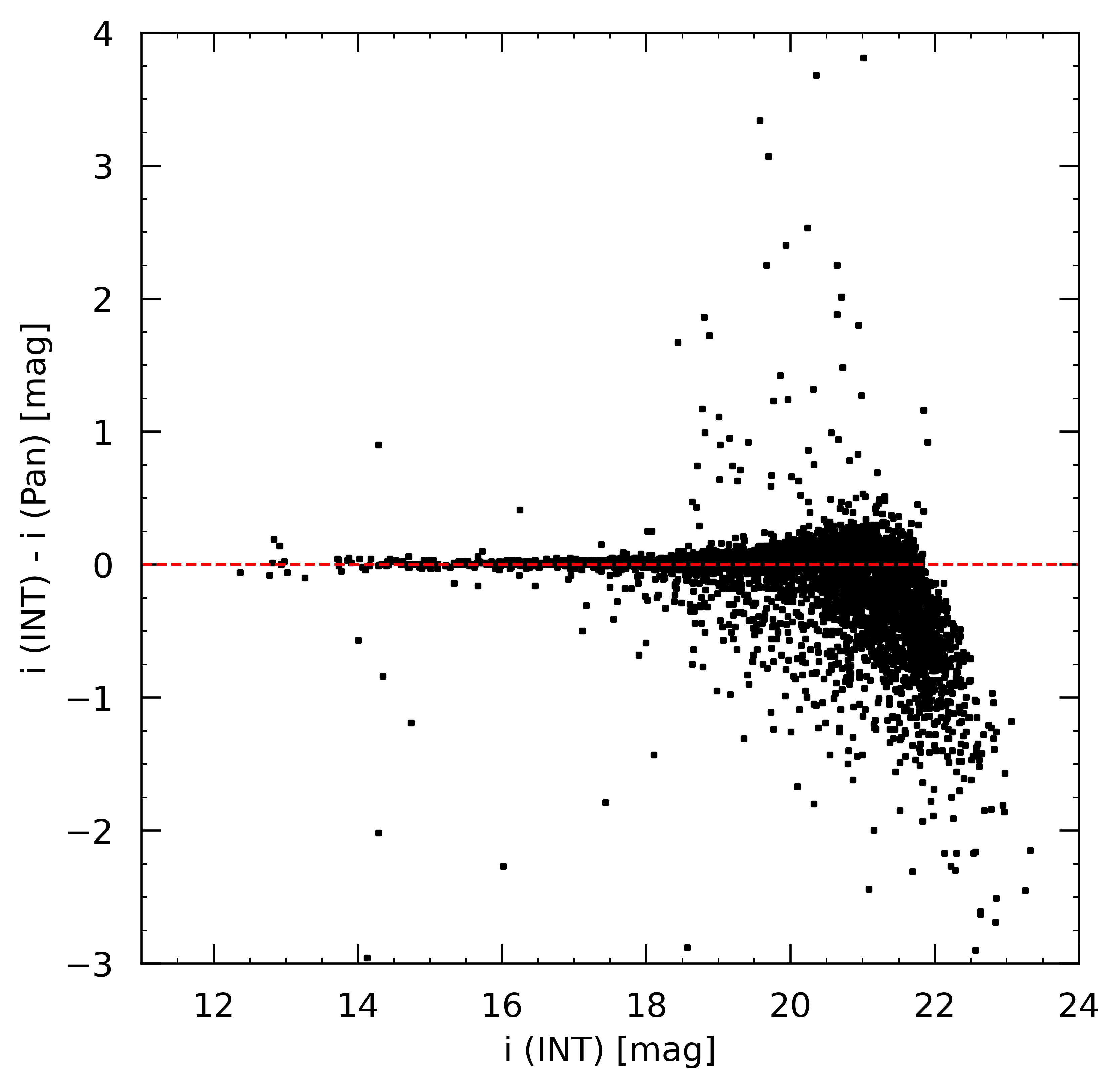}
\caption{Magnitude differences in the i--band between INT and Pan-STARRS catalogs for IC\,10 sources within 2r$_{\rm h}$ from the galaxy's center.}
\label{fig:cross}
\end{figure}

In order to determine the precision of our calibration approach, we conducted a cross-match identification between our sources within two half-light radii (2r$_{\rm h}$) from the center of IC\,10 and the Pan-STARRS data release 1 (PS1) catalog \citep{Chambers16}. By arranging the principal photometry in decreasing brightness order, radii were searched iteratively in steps of 0$\rlap{.}^{\prime\prime}$1 to 1$^{\prime\prime}$ until a match was found. Fig.~\ref{fig:cross} demonstrates a good match in the i--band between our photometric catalog and the PS1 survey up to $\sim$ 19 mag. Due to the systematic errors for fainter stars, there is more dispersion in their magnitude disparities \citep{Chambers16}. However, our survey is deep enough to detect AGBs in the relevant magnitude range.

\subsection{Quality assurance}\label{sec:Q}

To estimate our survey's completeness limit, we used the {\sc addstar} routine of the {\sc daophot/allstar} software package \citep{Stetson87}. The {\sc addstar} 
routine was employed to add 3000 artificial stars across ten trials for each of the two individual frames of the i--band and V--band, starting at i = 17 mag and ending at i = 26 mag. All photometry steps (Sec.~\ref{sec:Phot}) were then applied to the reconstructed frame to estimate star-finding efficiency by comparing the results to the input catalog.

\begin{figure}[]
\centering
\includegraphics[width=1\linewidth,clip]{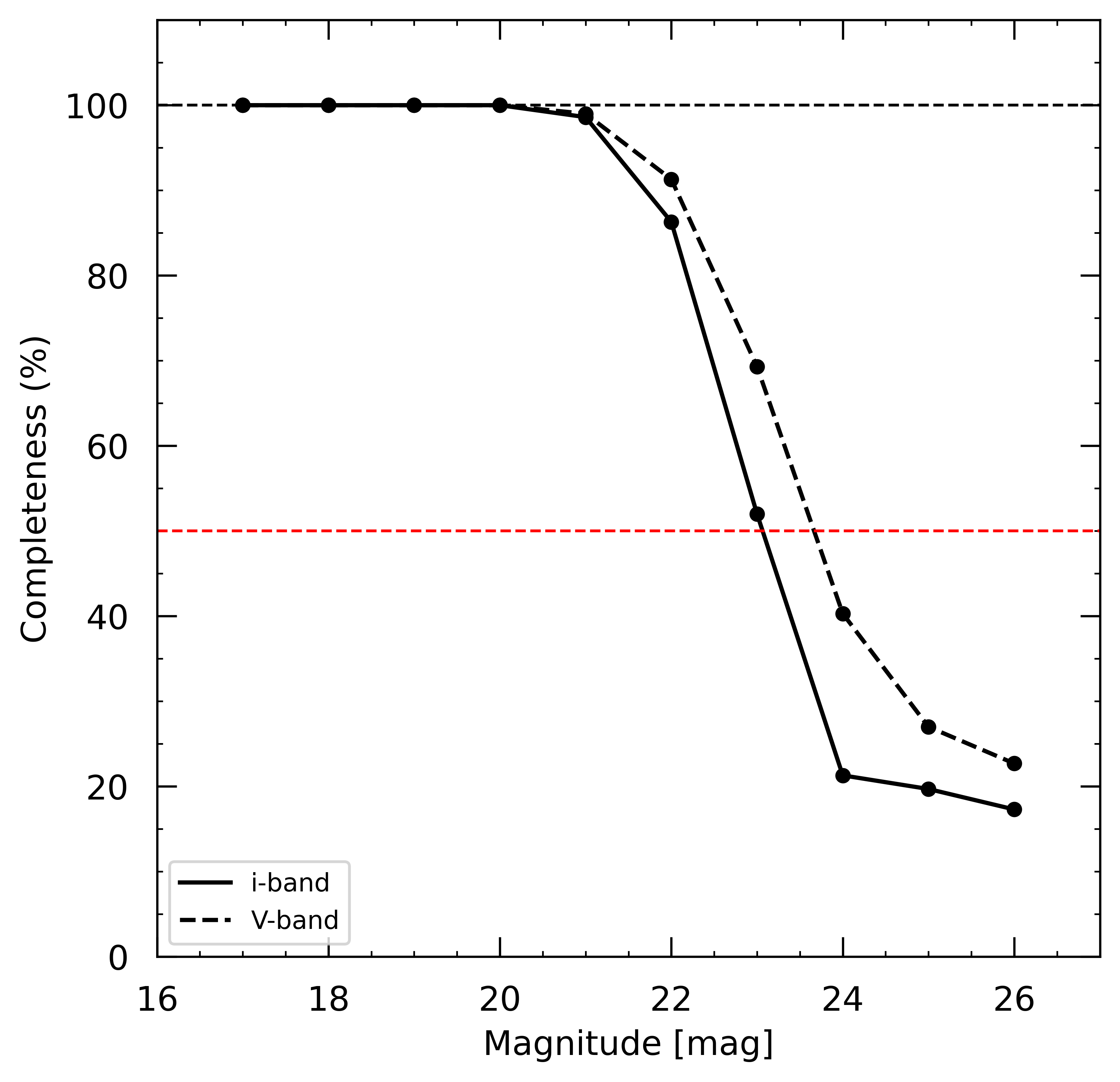}
\caption{Completeness as a function of the i--band (solid line) and V--band (dashed line) magnitudes. The red horizontal dashed line represents the completeness 
level of 50\,\% for stars with magnitudes of 23.07 and 23.66 mag in the i--band and V--band, respectively.}
\label{fig:comp}
\end{figure}

Our catalog is 100\,\% complete for stars brighter than i = 21 mag, drops to 80\,\% at i = 22.5 mag, and falls to less than 50\,\% at i $\sim$ 23.07 mag 
(Fig.~\ref{fig:comp}). As shown in Fig.~\ref{fig:comp}, the V--band 50\,\% completeness is at V $\sim$ 23.66 mag.

\begin{figure}[]
\centering
\includegraphics[width=1\linewidth,clip]{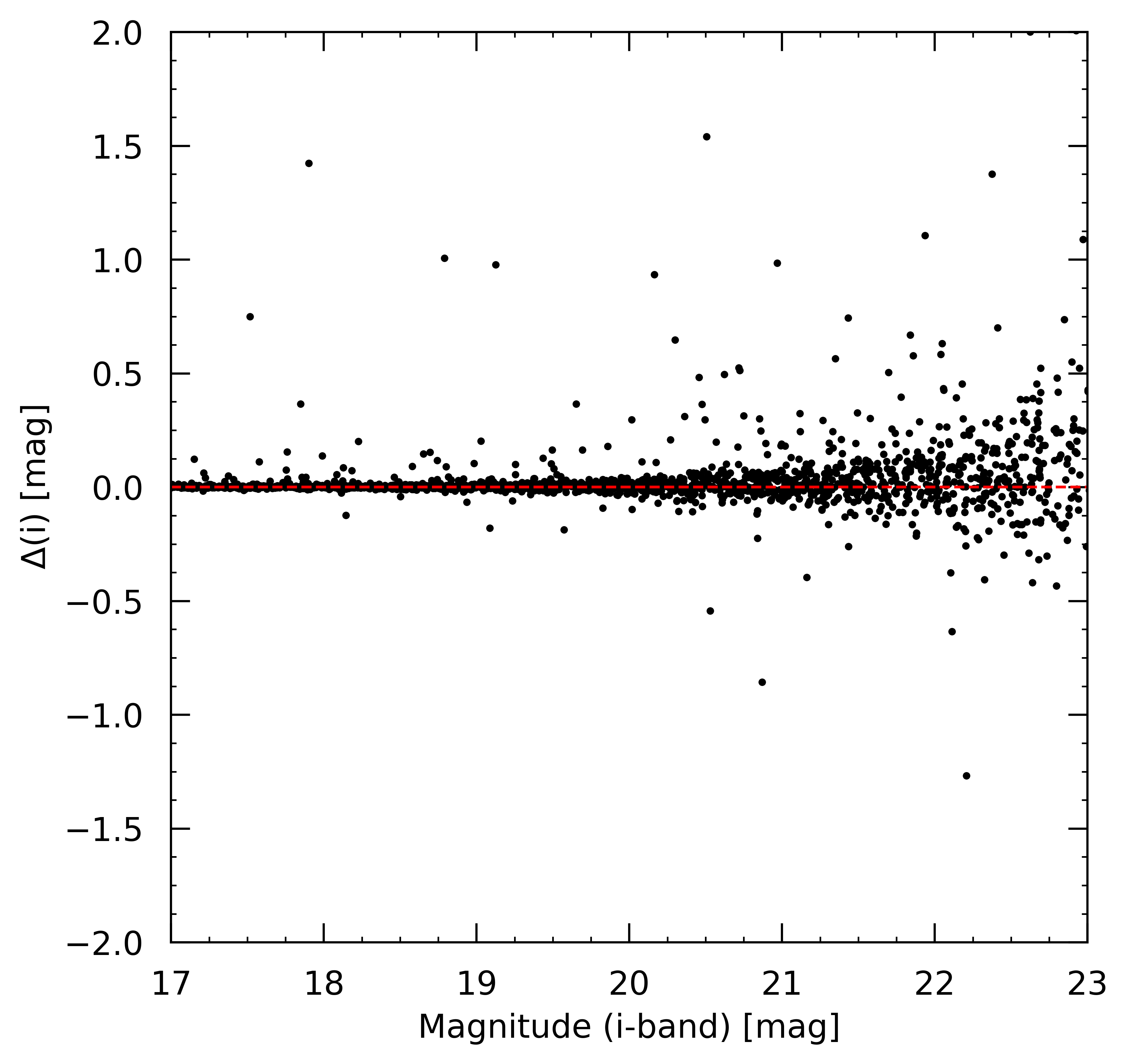}
\caption{Magnitude differences between artificial input stars and their recovered counterparts in the i--band.}
\label{fig:delt}
\end{figure}

Fig.~\ref{fig:delt} shows that the magnitude difference between artificial stars and recovered ones ($\Delta$i) is very small (|$\Delta$(i)| $<$ 0.1 mag) until i $
\sim$ 22 mag, but increases for fainter magnitudes. The photometry is deep and accurate enough to discover LPV candidates in the i--band, which in our case generally have i $<$ 22 mag.

Additionally, we employed the {\sc addstar} routine to add 1500 artificial stars to one of the individual images in each of the six trials, with magnitudes of i $\in$ (18, 19, 20, 21, 22, 23) mag. This was done to evaluate the accuracy of our photometry concerning the distance from the galaxy's center. In the top two panels of Fig.~\ref{fig:r}, the difference between the artificial input magnitude and the recovered magnitude is very small, |$\Delta$(i)| $<$ 0.1 mag. The closest stars to the center of the galaxy, at r $<$ 50$^{\prime\prime}$, also exhibit magnitude differences up to this value. As demonstrated, the recovered stars are brighter than the artificial stars, which is related to blending due to the crowdedness in the center of IC\,10. For stars fainter than 21 mag, the magnitude difference 
exceeds 0.2 mag in the galaxy's center.

\begin{figure}[]
\centering
\includegraphics[width=0.49\textwidth]{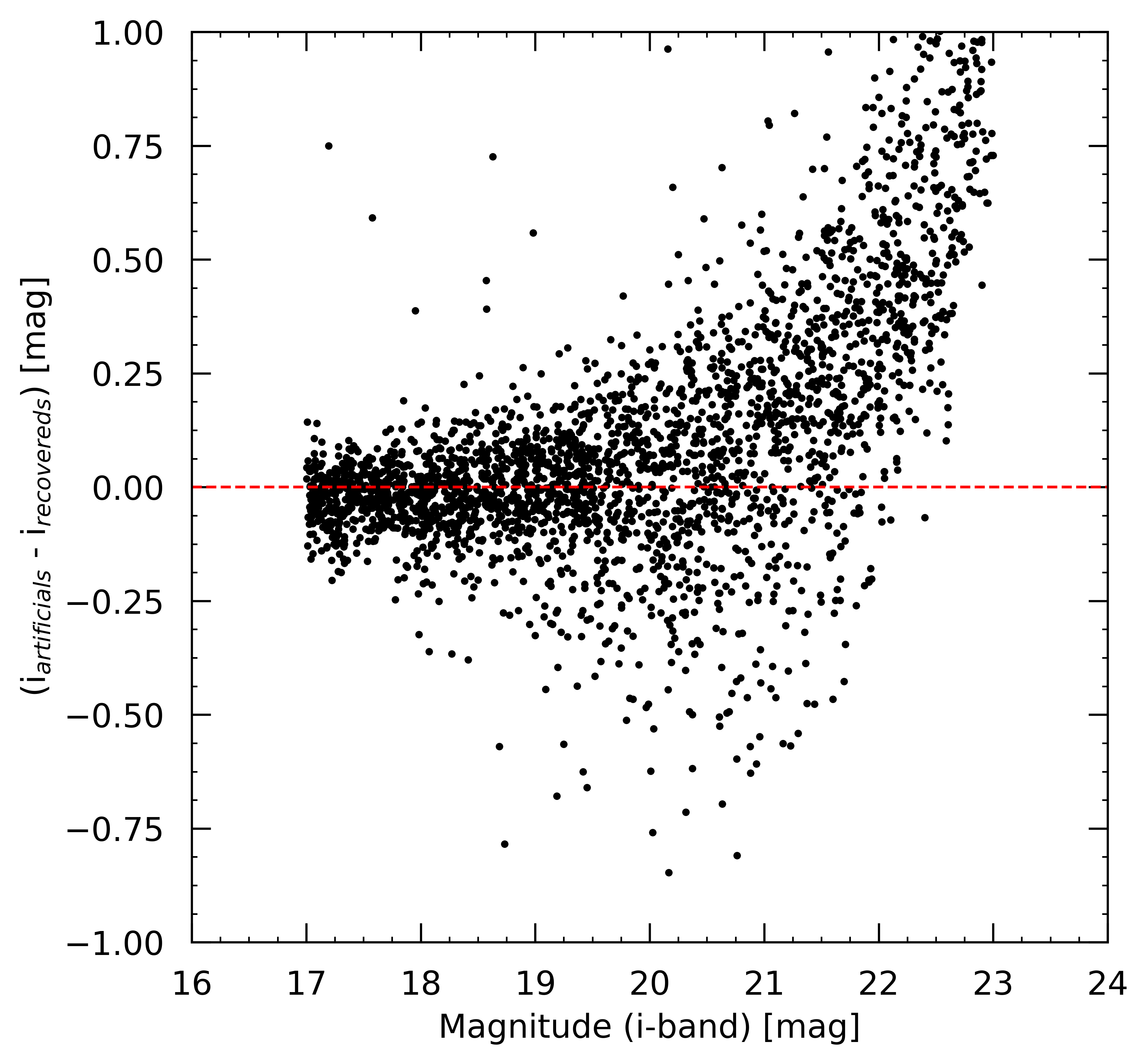}
\caption{The magnitude difference between artificial input stars and their recovered counterparts in the i--band for the night of June 15, 2016.}
\label{fig:June}
\end{figure}

\begin{figure}[]
  \centering
  \makebox[\textwidth][l]{\includegraphics[width=90mm, height=90mm]{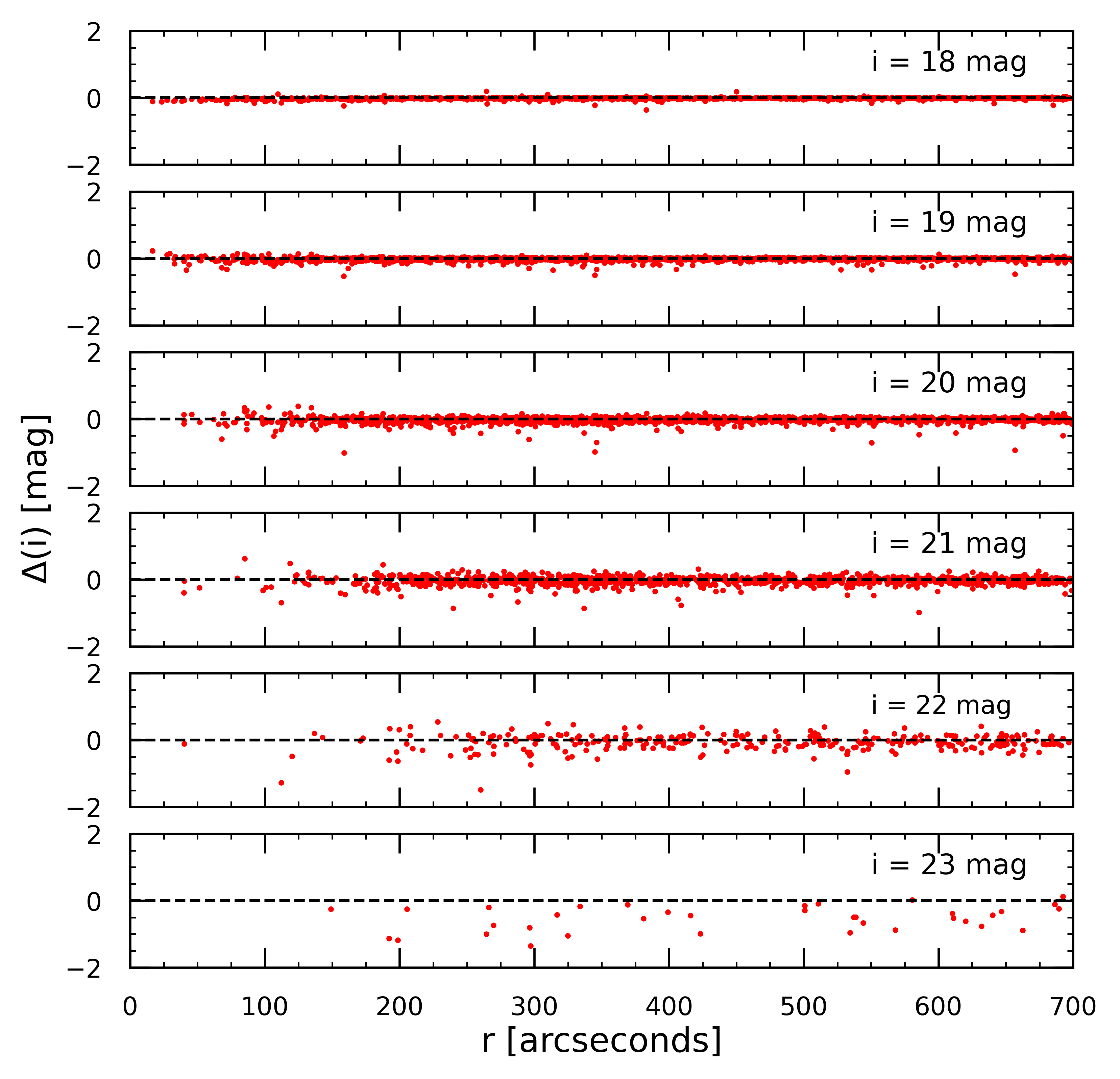}}
  \caption{Relationship between the artificial magnitude and recovered magnitude versus the distance from the center of IC\,10 for six input magnitudes.}
  \label{fig:r}
\end{figure}

Furthermore, images taken on 15 June 2016 exhibited high seeing values (1.82 arcseconds) (see Table~\ref{table:log}), which were notably higher than the rest of the observations. This increased seeing could have caused blending, making stars appear brighter and potentially leading to false detections of variable stars. To assess the extent of blending on this night, we used the {\sc addstar}  routine. As shown in Fig.~\ref{fig:June}, the magnitude of stars fainter than approximately 21 mag was notably affected by blending.

To further investigate the impact of blending on our variable star detection, we identified the variables (see Sec.~\ref{sec:var} for the method of variability detection) by including and excluding the 15 June night observation. The number of common variable stars detected between the two datasets dropped from 80\,\% for stars brighter than $\sim 21$ mag to less than 50\,\% for fainter stars. Since most of our LPVs are concentrated in the magnitude range of 20--22 mag (see Fig.~\ref{fig:3cmd}, left panel), and given the significant effect of blending at these magnitudes, we decided to exclude the 15 June night observation from further analysis. Additionally, we applied a similar approach to the rest of the images with approximately higher seeing values and found that excluding none of these images significantly affected the variability detections.

It is also worth mentioning that, on the night of 13 June, we obtained another observation in the i--band with good seeing (1.10 arcseconds). Since the goal of this study is the detection of LPVs, which typically have periods longer than 60 days \citep{McDonald2016}, excluding the 15 June observation does not result in the loss of any critical epochs.

The photometric catalog including all variable and non-variable stars is made publicly available at the Centre de Donn\'ees Astronomiques de Strasbourg (CDS). 
Table~\ref{table:catalog1} and Table~\ref{table:catalog2} describe the catalog, including the mean properties of sources and light curve information, 
respectively. The variability status listed in Table~\ref{table:catalog1} indicates whether a star is classified as an LPV--candidate, a non--LPV 
star (a variable star with an i--band amplitude less than 0.2 mag, or one with an inadequate measurement to reliably estimate the amplitude, or a color $(V-i)_0<0$; see Sec.~\ref{sec:var}), or a non-variable star.  
 Foreground sources have also been labeled in the catalog. The magnitudes in these tables are given without applying any extinction correction.

\begin{table}
\centering
\small
\caption{Description of the photometric catalog, mean properties of sources.
\label{table:catalog1}
}
\begin{tabular}{ll}\hline
    Column no. & Descriptor  \\
\hline
\multicolumn{2}{l}{Stellar mean properties (53,579 lines)}\\
1   &  \tablenotemark{a}Star number \\
2   &  Mean V--band magnitude \\
3   &  Error in $\langle$V$\rangle$ \\
4   &  Mean i--band magnitude \\
5   &  Error in $\langle$i$\rangle$ \\
6   & Mean $\chi$ value from {\sc daophot}  \\
7   & Mean sharpness value from {\sc daophot}  \\
8   & Variability index J \\
9   & Kurtosis index K \\
10 & Variability index L \\
11 &  Amplitude in i--band for LPVs \\
12 &  Variability status \\
13 &  Right ascension (J2000) \\
14 &  Declination (J2000) \\[1ex]\hline
\end{tabular}
\tablenotetext{a}{Stars that cannot be classified as either foreground stars or IC\,10 sources due to the quality of the astrometric solution are marked with an asterisk in the catalog.} 
\end{table}

\begin{table}
\centering
\small
\caption{Description of the photometric catalog, light curve information.
\label{table:catalog2}
}
\begin{tabular}{ll}\hline
    Column no. & Descriptor  \\
\hline
\multicolumn{2}{l}{Light curve data}\\
1  & Star number \\
2  & Epoch (HJD $-$ 245\,0000)\\
3  & Filter (i \& V)\\
4  & Magnitude\\
5  & Error in magnitude\\
6  & $\chi$ value from {\sc daophot}\\
7  & Sharpness value from {\sc daophot}\\[1ex]\hline
\end{tabular}
\end{table}

\section{LPV CANDIDATES IDENTIFICATION METHOD} \label{sec:var}

To identify variable stars, we employed a method similar to the {\sc newtrial} routine developed by \cite{Stetson96}, which was first introduced by \cite{Welch93}. In this method, a J index is calculated as follows \citep{Stetson96}:

\begin{equation}\label{eq:jj}
J = \frac{ \sum_{k=1}^nw_{k}\, \text{sign}(P_{k})\sqrt{|P_{k}}|}{ \sum_{k=1}^nw_k}.
\end{equation}

To calculate the variability indices, observations are paired based on the timespan between them, ensuring that the timespan for each pair is shorter than 
the expected shortest period for the type of variable star of interest. Based on \cite{Stetson96}'s method, $w$$_{k}$ in Eq. \ref{eq:jj} represents the weight of each pair of frames. It is set to 1 for each pair (with i and V measurements) and to 0.5 for a single frame (only one measurement) compared to epochs in both the i and V bands. For a single frame, a reduced weight is desirable. This is because a single frame would otherwise contribute as much as a pair of frames, leading to an inaccurate estimation of a star's variability index. Two independent but contemporary frames offer stronger constraints on the light curve of a star than a single one. As a result, we set the weight for a single frame to 0.5. Also, observations i and j are supposed to be a pair k with a given weight $w$$_{k}$. The product of normalized residuals, P$_{k}$, is defined as follows:

\begin{equation} \label{eq:ttt}
P_{k} = \delta_{i}(k) \delta_{j}(k).
\end{equation}

The $\delta_{i}$ = $\sqrt{\frac{N}{N-1}}$ $\frac{m_{i} - \bar{m}}{\sigma_{i}}$ represents the deviation of the magnitude in a filter (m$_{i}$) from the mean ($\bar{m}$), normalized by the measurement's standard error ($\sigma_{i}$, photometric errors), where N is the total number of observations.
In this method, $\delta_{i}$ and $\delta_{j}$ may correspond to measurements taken with different filters. If i = j, then the $P_{k}$ = $\delta^{2} - 1$.
For the most probable variable stars, the J index has a large positive value, while it approaches zero for data with random noise. A Kurtosis index 
(Eq.~\ref{eq:kkk}) is also defined as a value that depends on the shape of the light curve. For a sinusoidal light curve, K = 0.9 and for a Gaussian distribution, K = 0.798.

\begin{equation} \label{eq:kkk}
K = \frac{\frac{1}{N} \sum_{i=1}^{N}|\delta_{i}|}{\sqrt{\frac{1}{N} \sum_{i=1}^{N} \delta_{i}^{2}}}.
\end{equation}

In this equation, the index i refers to all observations (N) regardless of pairing.
In the next step, the variability index L is calculated \citep{Stetson96} and used in the {\sc newtrial} routine:

\begin{equation}
L = \frac{J \times K}{0.798} \frac{\sum_{i=1}^{N}w_{i}}{w_{\text{all}}}.
\end{equation}

The $\sum$ $w$$_{i}$ represents the total weight assigned to a star based on the number of detections, and $w$$_{all}$ is the total weight for a star if the star had been detected at all epochs.

\begin{figure}[]
\centering
\includegraphics[width=90mm, height=100mm]{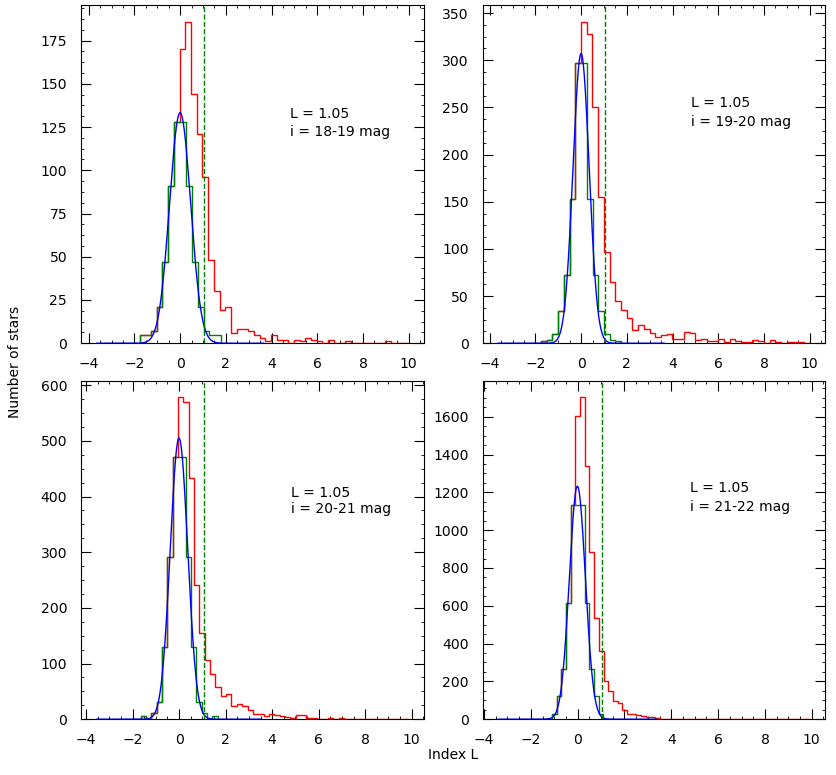}
\caption{Histograms of the variability index L for several magnitude bins in the i--band. The negative part of each histogram is mirrored (green bins). The fitted Gaussian function for each histogram is shown with a blue solid line, representing the expected distribution of non-variable sources. The most appropriate variability index thresholds are indicated with vertical green dashed lines, representing a constant value of L = 1.05 in each histogram.}
\label{fig:l-hist}
\end{figure}

To determine the index L threshold for identifying variable candidates, we plotted the histograms of index L for i--band magnitude bins in the range of 18 to 22 mag 
for all stars in our catalog (Fig.~\ref{fig:l-hist}). Since variables are not expected to have an L $<$ 0, we mirrored the negative part of each histogram to 
(statistically) separate the distribution of non-variables from variable candidates by fitting a Gaussian function to them. The Gaussian fit represents the non-variable distribution, while deviations in the histogram for larger values of index L represent the variable star distribution. We chose an index L threshold for 
each magnitude interval at which at least 90\,\% of the stellar distribution is above the Gaussian fit. In other words, $\sim$ 10\,\% of these might be non-variable. As can be seen in Fig.~\ref{fig:l-hist}, the index L is magnitude independent, and a threshold of L = 1.05 is derived. After applying this threshold, we identified 1,052 variable stars in CCD4.  It is worth noting that, although some stars have an L-index greater than 1.05, an inspection of their locations in the image reveals they are situated near very bright or saturated stars. As a result, their photometry is significantly affected by blending, leading to their removal from the final list of variables. Additionally, some stars were excluded from the  list because they are classified as foreground sources (see Sec.~\ref{sec:cont}).

\begin{figure}[]
\centering
\includegraphics[width=1\linewidth,clip]{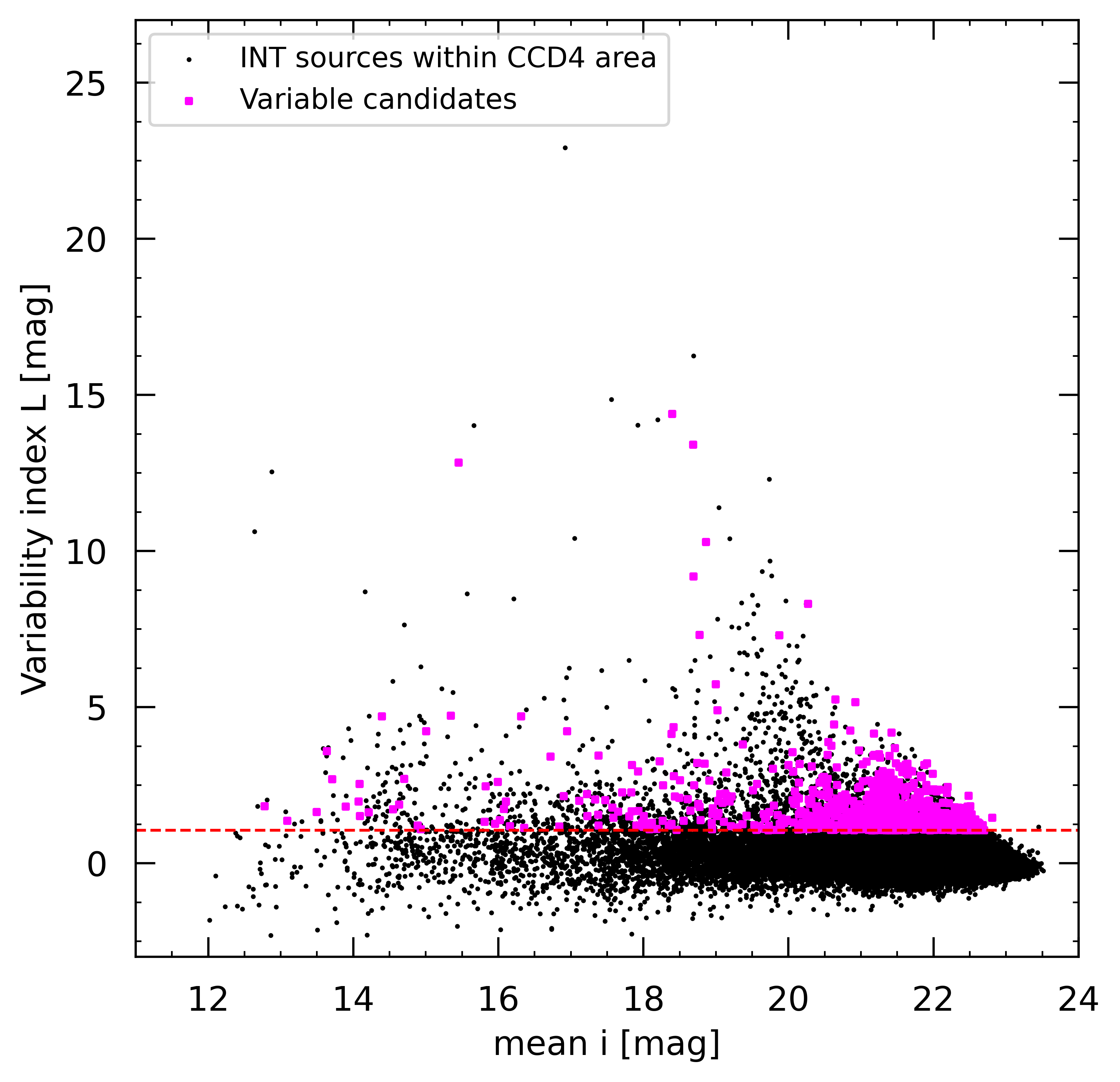}
\caption{For all sources in IC\,10, the variability index L is plotted against the mean i--band magnitude. The red horizontal dashed line represents the index L 
threshold (L = 1.05) for identifying variable candidates (see Fig.~\ref{fig:l-hist}). Variable candidates that have a higher index L are shown in magenta. Additionally, stars that exceed the threshold are shown with black colors are classified as saturated stars, foreground stars (Sec.~\ref{sec:cont}), or unresolved stars located in the crowded center of the galaxy.}
\label{fig:lmag}
\end{figure}

\begin{figure}[]
\centering
\includegraphics[width=0.4\textwidth]{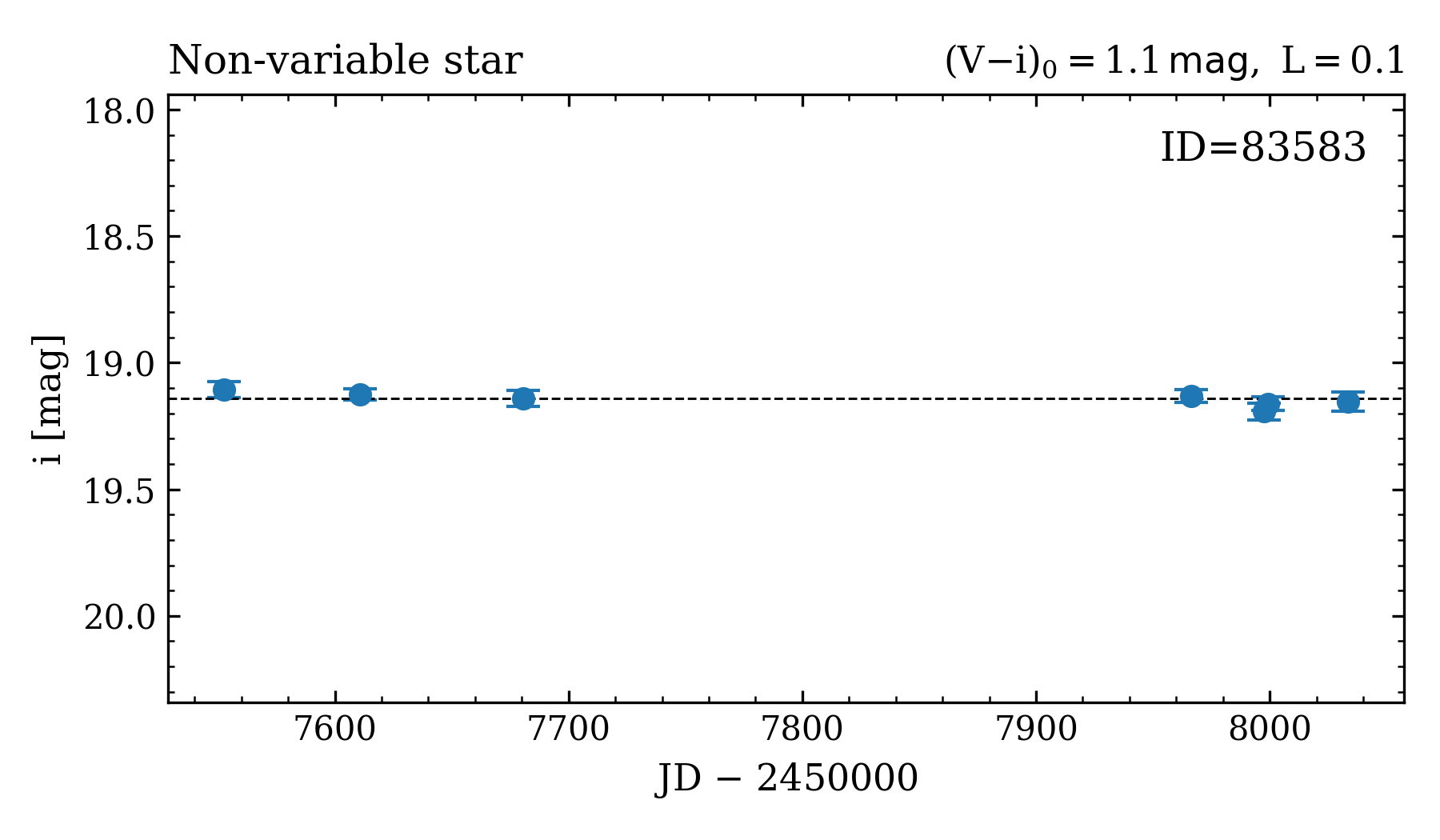}
\vfill
\includegraphics[width=0.4\textwidth]{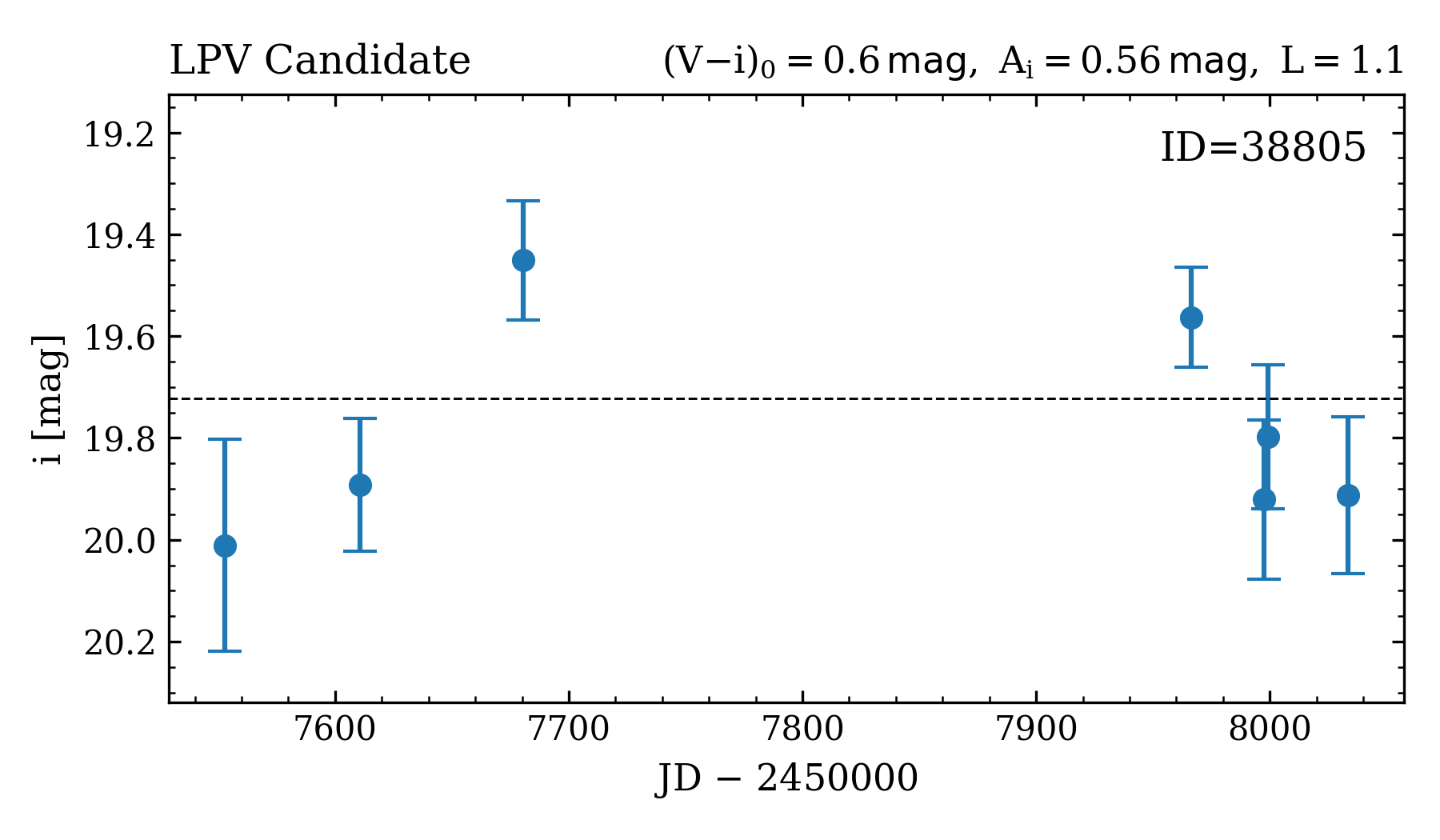}
\vfill
\includegraphics[width=0.4\textwidth]{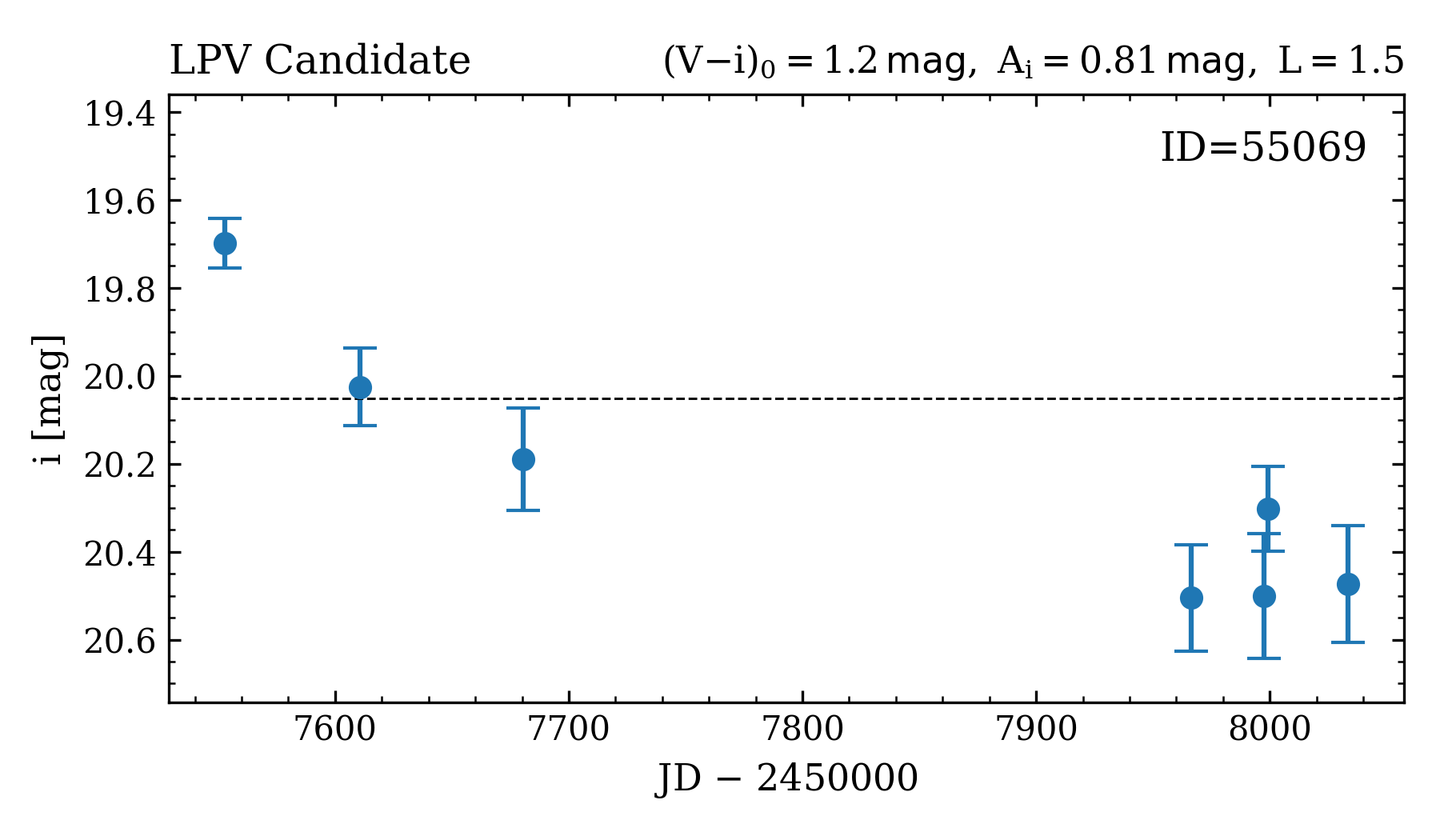}
\vfill
\includegraphics[width=0.4\textwidth]{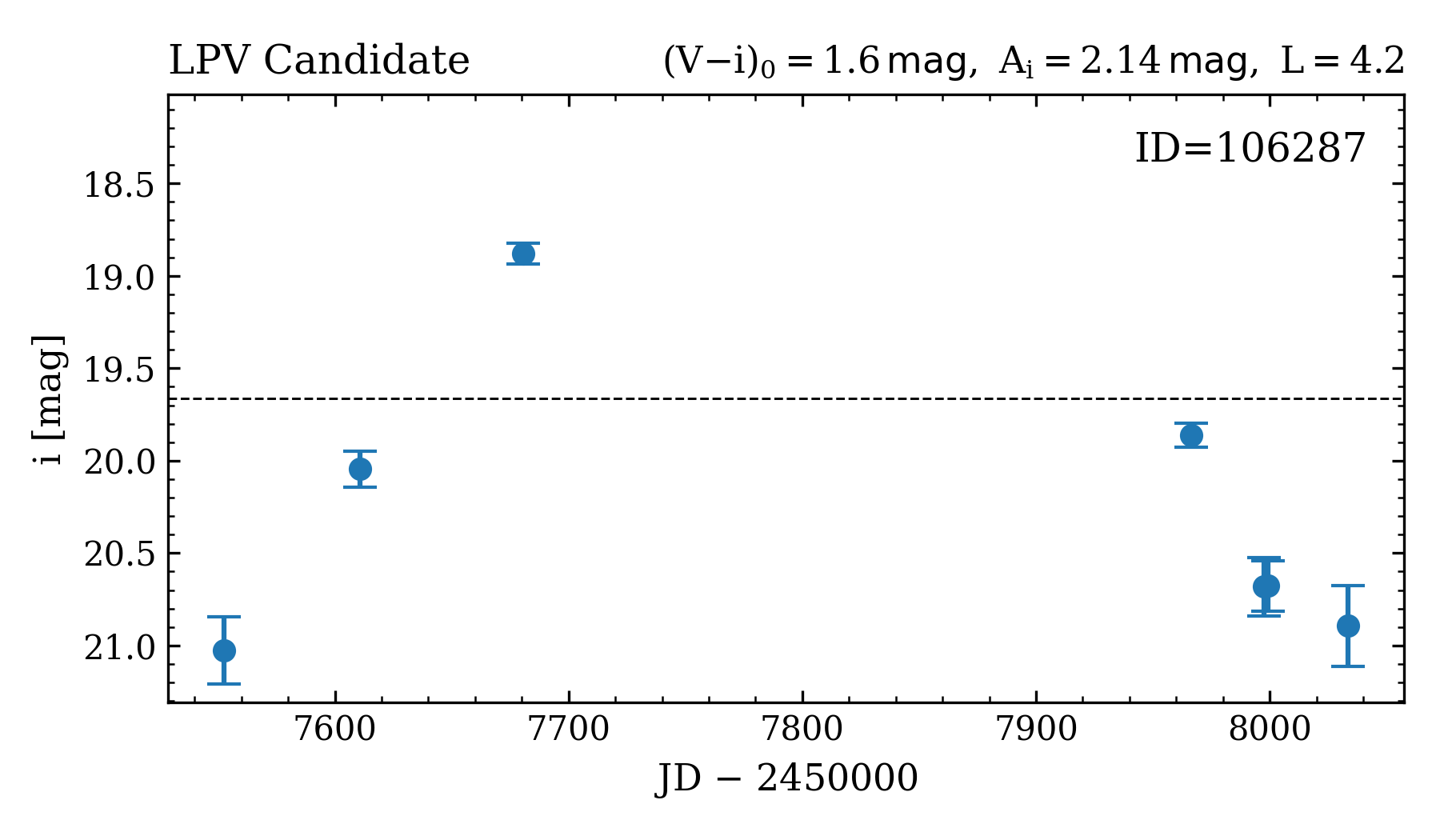}
\vfill
\caption{Examples of light--curves within the CCD4 area: a non--variable star,  and three LPV--candidates.  The horizontal dashed line represents the weighted mean magnitude of each star across all epochs. The light--curves of all LPV candidates are provided in the Appendix A1.}
\label{fig:lcv}
\end{figure}

Fig.~\ref{fig:lmag} illustrates the relationship between the index L and i-band magnitude for all stars in the output catalog. Among the stars with $L > 1.05$, there is a population of candidate variables with $A_i < 0.2$ mag. However, such small amplitudes must be regarded with a great deal of caution. Therefore, we limited our LPV candidate catalog to those stars with variability amplitudes larger than 0.2 mag.  Additionally, some variable candidates have fewer than six measurements, making their amplitudes unreliable \citep{Javadi11a}; these stars are also removed from the LPV list. Variable sources with $(V-i)_0<0$ are also excluded from the list of LPVs. These variable sources, which satisfy the variability criteria ($L > 1.05$), but due to their amplitude or color cannot be classified as LPV candidates, are labeled as non--LPV source in the catalog (Table~\ref{table:catalog1}). 
After applying these conditions, a total of 536 LPV candidates were identified in CCD4, with 380 located within 2r$_{h}$ of IC\,10.  Fig.~\ref{fig:lcv} shows the light--curve of a non--variable star, along with three LPV candidates with different amplitudes: $A_i = 0.56$ mag, $A_i = 0.81$ mag, and $A_i = 2.14$ mag.  The light--curves of all LPV candidates are provided in the Appendix A1.

Inspection of some of the light curves reveals that certain stars exhibit significant differences in magnitude (greater than 0.2 mag, which is assumed to be the minimum amplitude for LPVs) between two consecutive night observations, taken on September 1, 2017, and September 2, 2017. One of these nights had better seeing conditions (the seeing for the September 2 observation was 1.18 arcseconds in the i-band, compared to 1.47 arcseconds for the September 1 observation). Examination of these stars in the image shows that they are located in very crowded regions, which may affect their photometry due to blending.
Similar to the method outlined  in Sec.~\ref{sec:Q}, to verify the reliability of these detected LPVs, we excluded the September 1 observation and reanalyzed our variability detection. We found that all of these stars were again detected as LPV candidates, confirming they are not false detections. In the final analysis, we retained the September 1 observation; however, for the variables showing significant differences in magnitude between the two nights, we calculated the amplitude of variability using only the September 2 observation.

\subsection{Amplitudes of variability}\label{sec:amp}

We obtained the amplitude of variability for our detected variable candidates by considering a sinusoidal light--curve shape. The amplitude of variability is then given by:

\begin{equation}\label{eq:amplitude}
A = \frac{2 \times \sigma}{0.707},
\end{equation}

Where $\sigma$ represents the standard deviation in our measurements, and 0.707 corresponds to the standard deviation of the sinusoidal function. \cite{Javadi11a} demonstrated that the standard deviation depends on the number of measurements for a small data set and will drop below the asymptotic value (0.707 for the unit sine function) that is reached for well-sampled light--curves, thus becoming less reliable. To estimate a more accurate amplitude using Eq.~\ref{eq:amplitude}, they set the minimum number of measurements to six for each variable candidate (see Fig. 10 in \citep{Javadi11a}). These epochs are spaced at least 30 days apart.

\begin{figure}[]
\centering
\includegraphics[width=1\linewidth,clip]{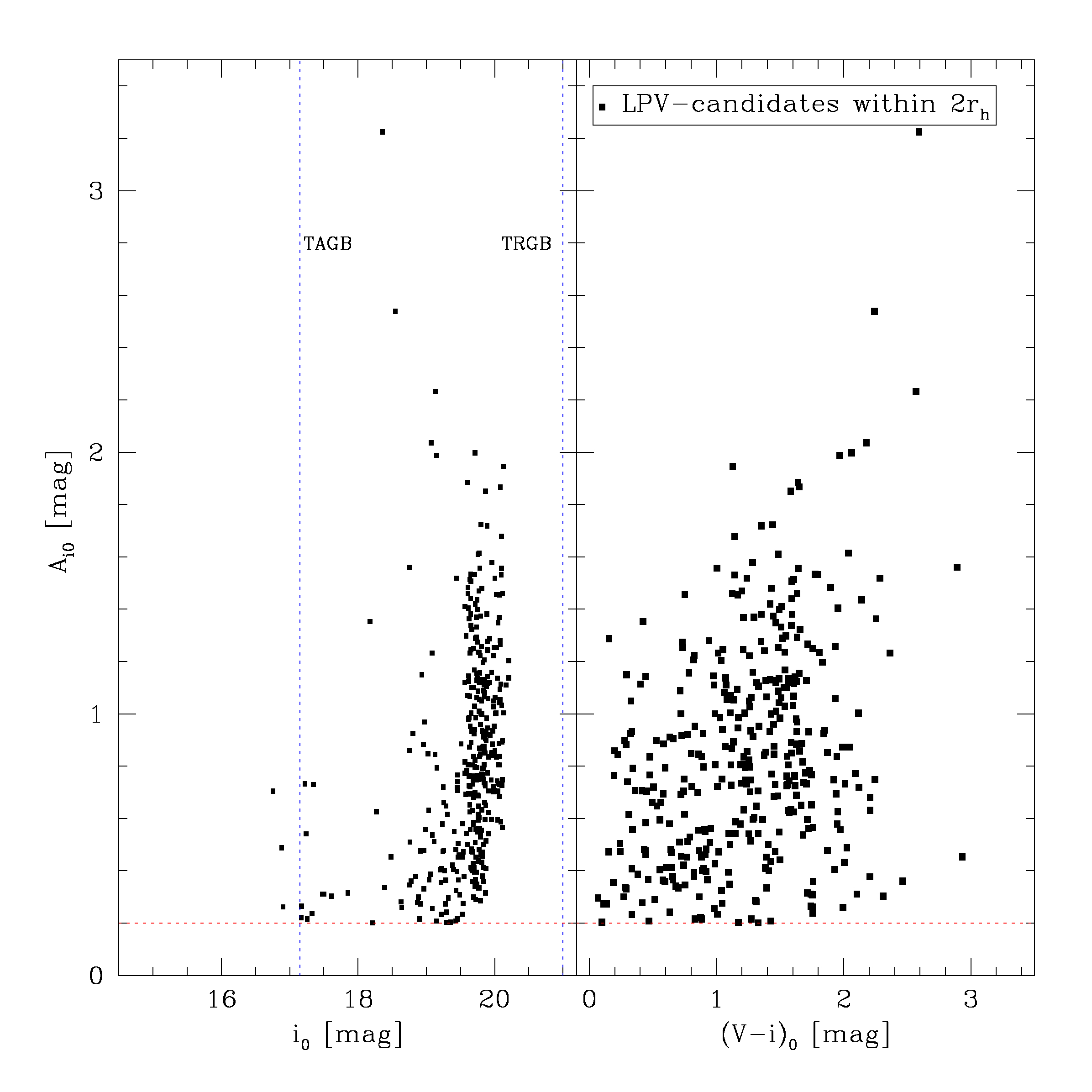}
\caption{Estimated amplitude, A$_i$, of variability versus i$_0$--band magnitude (left panel) and color (V--i)$_0$ within the 2r$_{\rm h}$ area, centered on IC\,10 for LPV candidates, corrected for a constant extinction. The blue vertical dashed lines denote the TAGB and TRGB. The red horizontal dashed line indicates the amplitude of 0.20 mag.}
\label{fig:amp}
\end{figure}

\begin{figure}[]
\centering
\includegraphics[width=1\linewidth,clip]{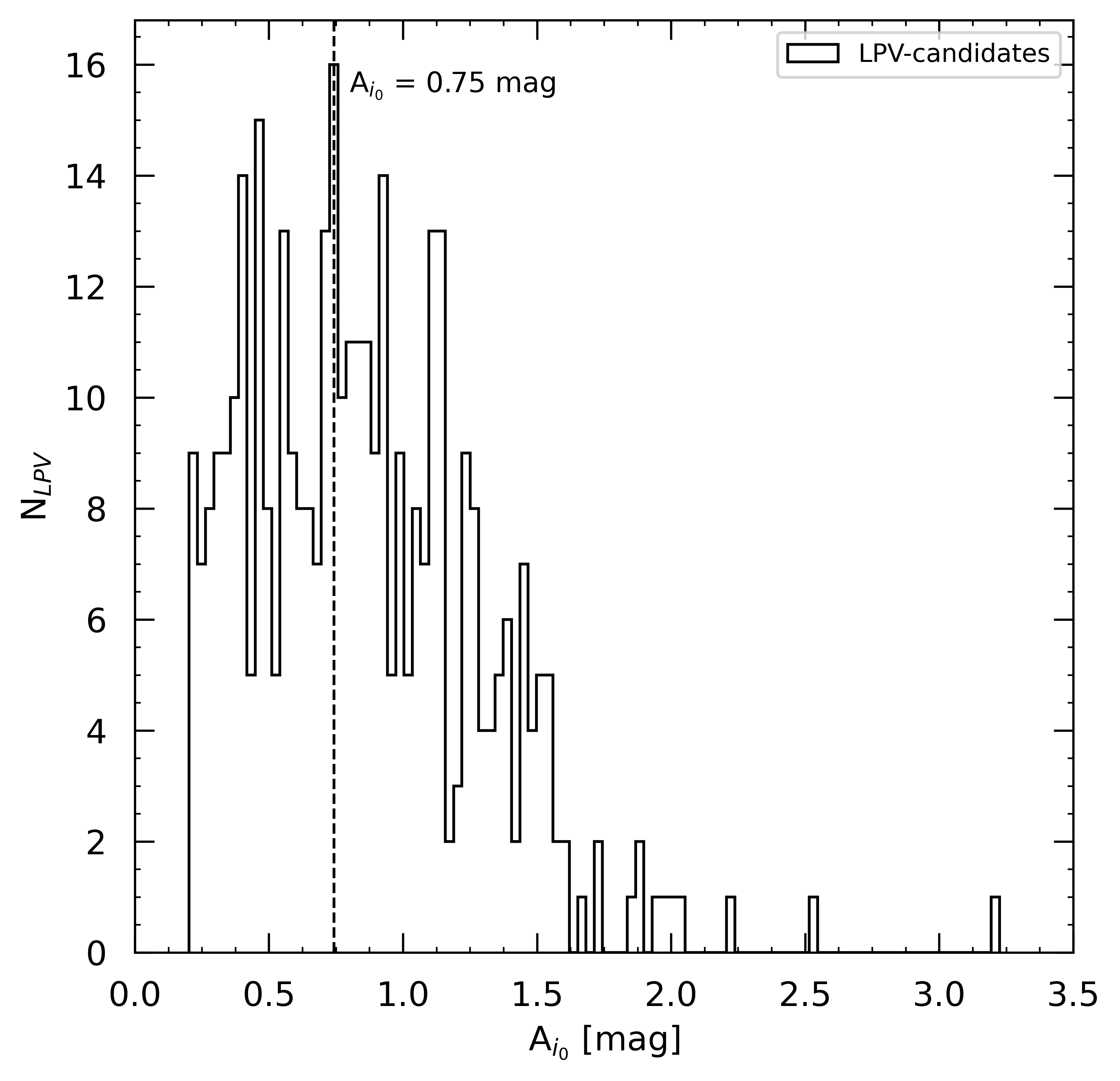}
\caption{Histogram of the amplitude of variability for LPV candidates detected in the i--band within the 2r$_{\rm h}$. The vertical green dashed line marks an 
amplitude of 0.75 mag.}
\label{fig:amp-hist}
\end{figure}

Fig.~\ref{fig:amp} (left panel) illustrates the estimated amplitude for LPV candidates as a function of i$_0$--band magnitude. The i$_0$ and V$_0$ magnitudes are corrected for extinction (see Sec. \ref{sec:cred}). The amplitudes are generally less than 3.3 mag and are predominantly concentrated around A$_i$ = 0.75 mag (Fig.~\ref{fig:amp-hist}). Variable candidates with amplitudes less than 0.20 mag are not considered LPV candidates in this paper. Among the variable candidates are three LPV candidates brighter than the tip of the AGB (TAGB) (i$_{\rm TAGB}$ = 17.15 mag, Sec.~\ref{sec:spatial}) with amplitudes ranging from 0.3 mag to 0.7 mag. 

Fig.~\ref{fig:amp} (right panel) also shows the estimated amplitude of variability in the i$_0$--band against the color (V--i)$_0$. The right panel indicates an approximate upward trend, suggesting that the amplitude increases as the stars' colors become redder.  
This is not unexpected, as large-amplitude variability has been linked to significant dust formation \citep{Whitelock03}. However, some stars exhibit a modest amplitude despite having a red color. These stars are located in the central regions of the IC\,10 galaxy, and this discrepancy may be due to an underestimated extinction correction, suggesting that these stars are situated behind more dust.


\subsection{Extinction correction} \label{sec:ex}

IC\,10 is experiencing significant extinction, resulting in red colors and fainter magnitudes. This is because the galaxy is located near the Galactic plane 
($\ell$ = 118$\rlap{.}^\circ$9, b = $-$3$\rlap{.}^\circ$3) and contains dust. Therefore, there are various reddening estimations for this galaxy, which are listed in Table~\ref{table:dis}. In our study, we utilized the SFD98 reddening map (as described in Sec.~\ref{sec:sfd}) to calculate reddening. Our calculations revealed a reddening range of 0.8 mag to 1.5 mag, whereas studies based on the TRGB \citep{Sakai99, Sanna08} or Cepheids \citep{Wilson96} provide a range of 0.6 mag to 1.1 mag for reddening.

In this paper, we corrected the extinction for each star in each band using two distinct methods: (1) the reddening values obtained from the dust map (Sec.~\ref{sec:sfd}), and (2) by adopting a constant reddening derived by \citep{Sanna08} for all stars (Sec.~\ref{sec:cred})

When adopting a constant reddening,  we used the reddening (E(B--V) = 0.78 $\pm$ 0.06 mag) and distance modulus 24.51 $\pm$ 0.08 mag (corresponding to a distance of 0.81 Mpc) from \cite{Sanna08}, as their results align with most previous studies that used TRGB stars as distance indicators \citep{Richer01, Vacca07, Dell'Agli18}. Another advantage of using this distance modulus is its compatibility with theoretical isochrones, which align well with the magnitude at which variable stars become more abundant.

\cite{Sanna08} determined the distance modulus of IC\,10 using TRGB stars, comparing it to the Small Magellanic Cloud (SMC) and two globular clusters, 47\,Tuc and $\omega$\,Cen.  Their photometric catalog, generated using Hubble Space Telescope (HST) archival data, covers a wide range of magnitudes from 21 mag (for bright main sequence stars) to 25.5 mag (for faint red giant stars) in the F814W band. To calculate the TRGB magnitude in the F814W band, they selected stellar samples around the TRGB magnitude, ensuring there were enough stars for precise detection \citep{Madore95}. In this study, we adopted the TRGB magnitude of 21.90 $\pm$ 0.03 mag  in the F814W band ($\sim$ 21 mag in the i$_{0}$--band magnitude).

\subsubsection{2D dust map (SFD98)} \label{sec:sfd}

The 2D dust map constructed by \cite{Schlegel98} (hereafter referred to as the SFD98 map) provides the E(B--V) value for each star. We converted these values to the extinction in the V--band, adopting R$_V$ $\equiv$ A$_V$ / E(B--V) = 3.1 \citep{Cardelli89, Odon94}. The extinction in the i--band is calculated based on 
equations 9 \& 10 in \cite{Wang19}.

\begin{figure}[]
\centering
\includegraphics[width=1\linewidth,clip]{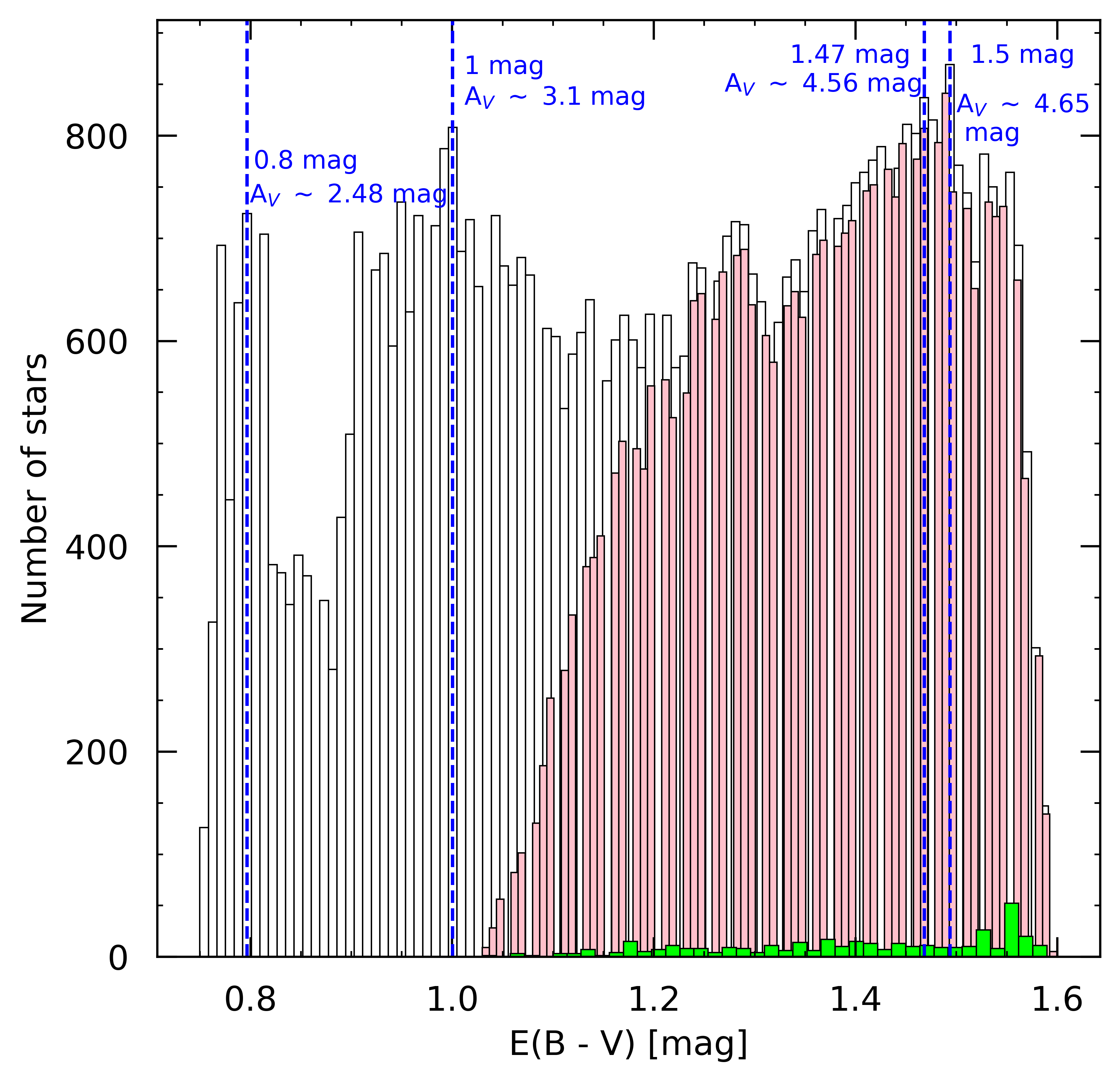}
\caption{Histogram of E(B--V) values for all sources (white bins) and those within 2r$_{\rm h}$ from the center of IC\,10 (pink bins). LPV candidates within 2r$_{h}$ are shown in green. The blue vertical dashed lines indicate the main concentrations of E(B--V) values.}
\label{fig:hist-ex}
\end{figure}

We plotted the histogram of E(B--V) values for all sources in our catalog, and for the LPV candidates (green) within the 2r$_{\rm h}$ from the galaxy's center in Fig.~\ref{fig:hist-ex}. The E(B--V) values of the sources are mainly concentrated at 0.8 mag, 1 mag, 1.47 mag, and 1.5 mag, corresponding to A$_V$ values of 2.48 mag, 3.1 mag, 4.56 mag, and 4.65 mag, respectively. Additionally, the E(B--V) values of the LPV candidates are primarily concentrated at 1.56 mag.

\begin{figure}[]
\centering
\includegraphics[width=1\linewidth,clip]{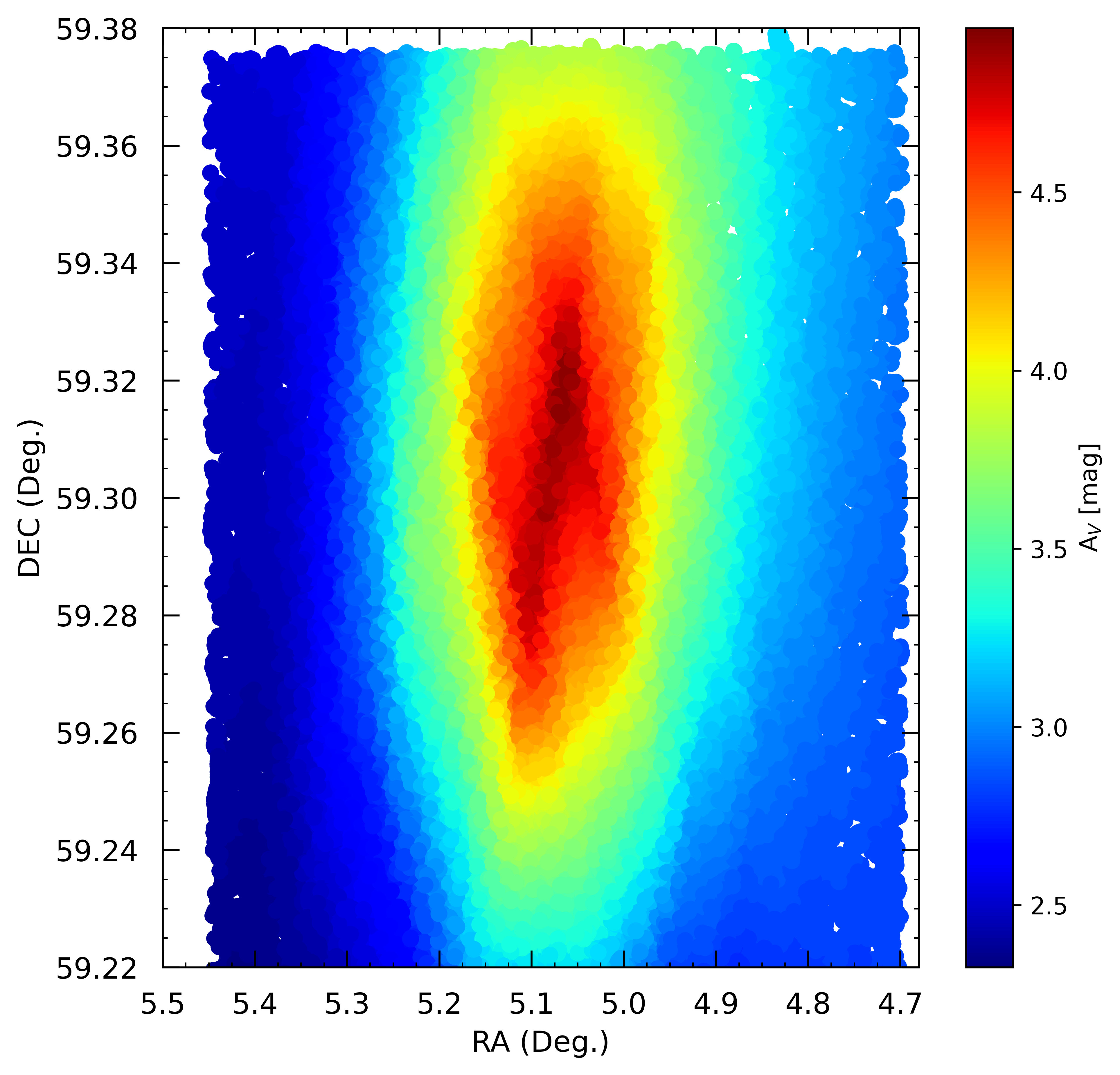}
\caption{Spatial distribution of the visual extinction (A$_V$) of IC\,10 sources, color-coded according to the reddening values from the SFD98 map.}
\label{fig:3dplot}
\end{figure}

Fig.~\ref{fig:3dplot} illustrates the variation in visual extinction (A$_V$) over the field of view towards IC\,10. Reddening is seen to increase significantly 
toward the center of the galaxy (star-forming regions, \citet{Sakai99}).

\subsubsection{A constant reddening} \label{sec:cred}

\begin{figure*}[]
\centering
\includegraphics[width=0.33\textwidth]{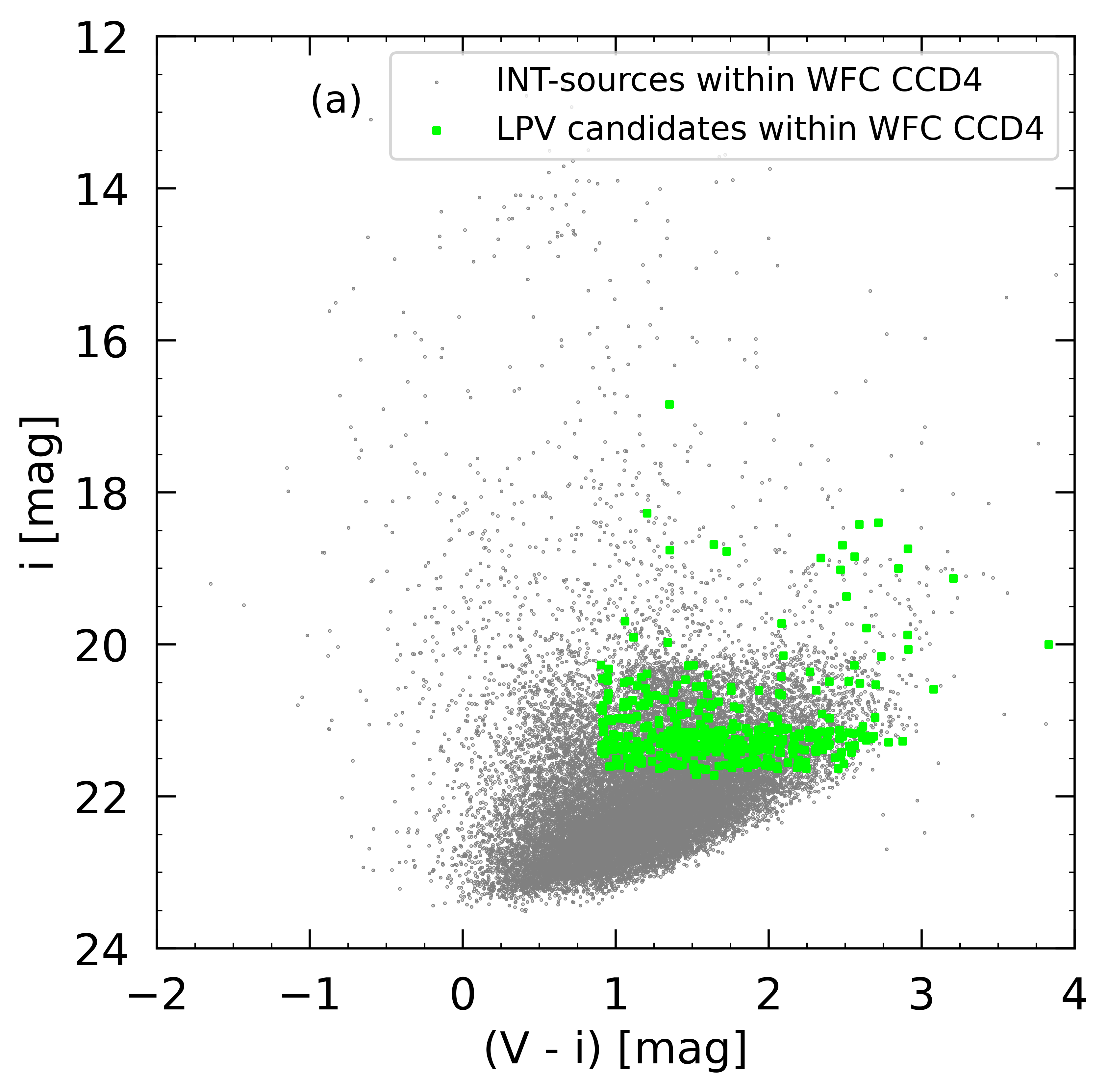}\hfill
\includegraphics[width=0.33\textwidth]{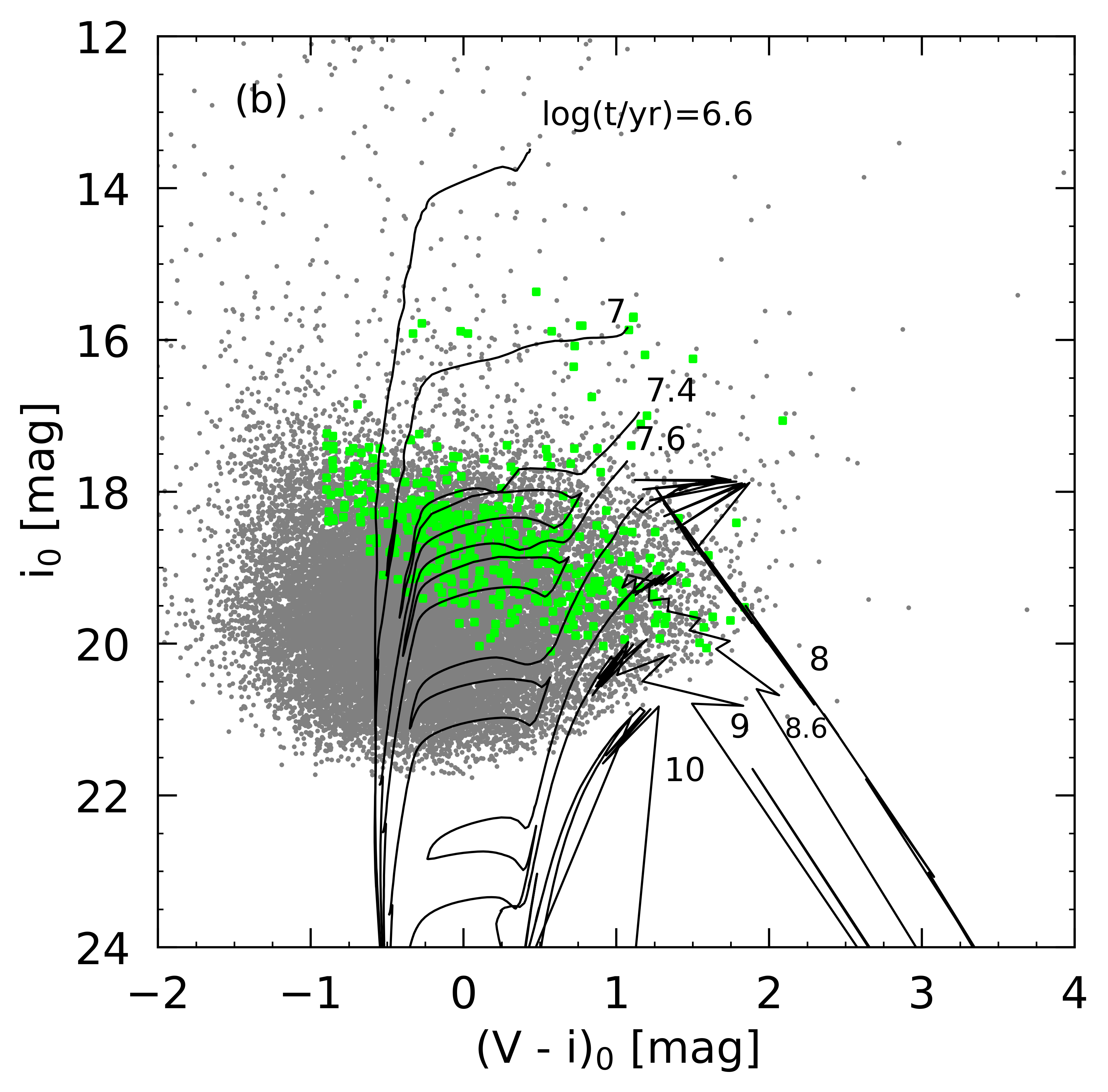}\hfill
\includegraphics[width=0.33\textwidth]{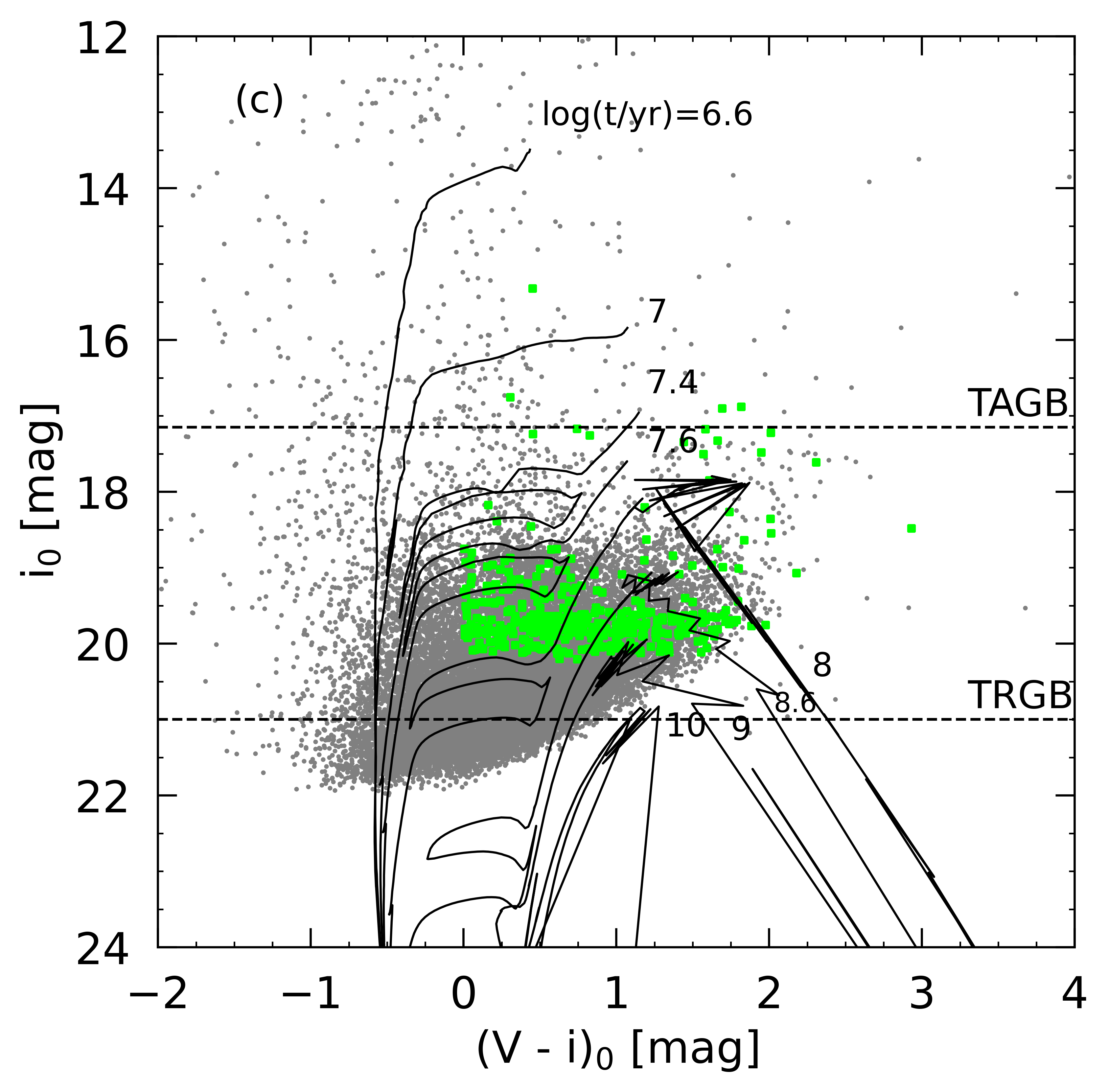}
\caption{(a) Original CMD; (b) CMD corrected for extinction using the SFD98 dust map for each star; (c) CMD corrected for a constant extinction from \cite{Sanna08}. 
The LPV candidates are shown in green. Overplotted are isochrones from \cite{Marigo17} for a distance modulus of 24.51 $\pm$ 0.08 mag \citep{Sanna08} and a constant metallicity of [Fe/H] = $-$1.28 dex. In the right panel, the horizontal black dashed lines indicate TAGB and TRGB.}
\label{fig:3cmd}
\end{figure*}

We also corrected the extinction using a constant reddening of E(B--V) = 0.78 $\pm$ 0.06 mag \citep{Sanna08} applied to all stars in our catalog. In Fig.~\ref{fig:3cmd}, the extinction is corrected for both constant and non-constant reddening for all sources in the i--band and V--band. The i$_0$ and V$_0$ indicate the magnitudes after extinction correction. We displayed all sources and LPV candidates within the entire WFC CCD4 area, corrected using the SFD98 dust map and for constant extinction in panels b and c, respectively. The overplotted are the Padova isochrones \citep{Marigo17} for a constant metallicity of [Fe/H] = $-$1.28 dex (Z = 0.0008) \citep{Tikhonov10, mcconnachi12}.

In panel c of Fig.~\ref{fig:3cmd}, the horizontal black dashed lines represent the positions of the TAGB and TRGB. Stars at TAGB undergo mass loss, resulting in a partial or complete obscuration at optical wavelengths, making it challenging to conduct a comprehensive count using optical imaging alone. Previous studies \citep{Jackson07a, Jackson07b, Boyer09} have suggested that up to 30\,\% to 40\,\% of TAGB stars can be obscured at optical wavelengths. To determine the TAGB for IC\,10, we utilized the relation of the classical core-luminosity for a Chandrasekhar core mass, which corresponds to an approximate magnitude of M$_{\rm bol}$ = -7.1 mag \citep{Zij96}. By adopting a constant metallicity of Z = 0.0008 and a distance modulus of $\mu$ = 24.51 mag \citep{Sanna08}, the peak of the log(t/yr) $\sim$ 6.68 isochrone aligns with an equivalent magnitude of i$_0$ = 17.15 mag.

As shown in Fig.~\ref{fig:3cmd}, the LPV candidates are concentrated within the magnitude range of 15 mag $<$ i$_0$ $<$ 20 mag, with colors between approximately $-$1 mag $<$ V$_0$--i$_0$ $<$ 2 mag based on the SFD98 extinction correction (panel b), and 0 mag $<$ V$_0$--i$_0$ $<$ 3 mag for constant extinction correction (panel c). Furthermore, the isochrones align perfectly with our CMD in panel c but not in panel b. It is clear that the SFD98 dust map does not work for IC\,10.  This is because, firstly, in the original CMD (Panel a), between $V-i\sim2-3$ mag,  at least one branch (the giant branch) is visible, which is also seen on the constant dust-corrected CMD  between  $V-i\sim1-2$ mag. However, it disappears when applying individual reddening corrections derived from the SFD98 map, causing the original CMD to blur. Secondly, the reddening map suggests
significant extinction within IC\,10. While there likely is some internal extinction, it is challenging to correct individual stars for it as we lack precise information about the relative positions of stars and dust along the
 line of sight. Thus, the SFD98 model might overestimate the reddening of the LPV candidates, causing their color to become negative, while the constant--extinction model does not show such negative colors. Additionally, the SFD98 values are determined from emission data, assuming diffuse Galactic ISM conditions. However, IC\,10 differs from these conditions. The emission is likely more intense due to a stronger radiation field, leading to warmer dust. Therefore, we chose to apply the constant extinction of A$_V$ $\sim$ 2.5 mag from \cite{Sanna08}.

This is because firstly, in the original CMD (panel a), at least one branch is visible, but it
 disappears when applying individual reddening corrections derived from the SFD98 map, causing the original CMD to blur. Secondly, the reddening map suggests
significant extinction within IC\,10. While there likely is some internal extinction, it is challenging to correct individual stars for it as we lack precise information about the relative positions of stars and dust along the
 line of sight.

\subsection{Contamination}\label{sec:cont}

The IC\,10 dwarf irregular galaxy is located near the Galactic plane of the MW, resulting in significant foreground contamination. We simulated the spatial distribution of foreground stars in the direction of the galaxy ($\ell$ = 118$\rlap{.}^\circ$9, b = $-$3$\rlap{.}^\circ$3) and Galactic extinction A$_V$ $
\sim$ 2.5 mag \citep{Sanna08} using the {\sc trilegal} (TRIdimensional modeL of thE GALaxy) code \citep{Girardi05}. The simulation was conducted in three different 
areas: the full area of CCD4 (0.070 deg$^2$), within the 2r$_{\rm h}$ radius (0.020 deg$^2$), and the r$_{\rm h}$ (0.005 deg$^2$) from the center of IC\,10.

\begin{figure}[]
\centering
\includegraphics[width=1\linewidth,clip]{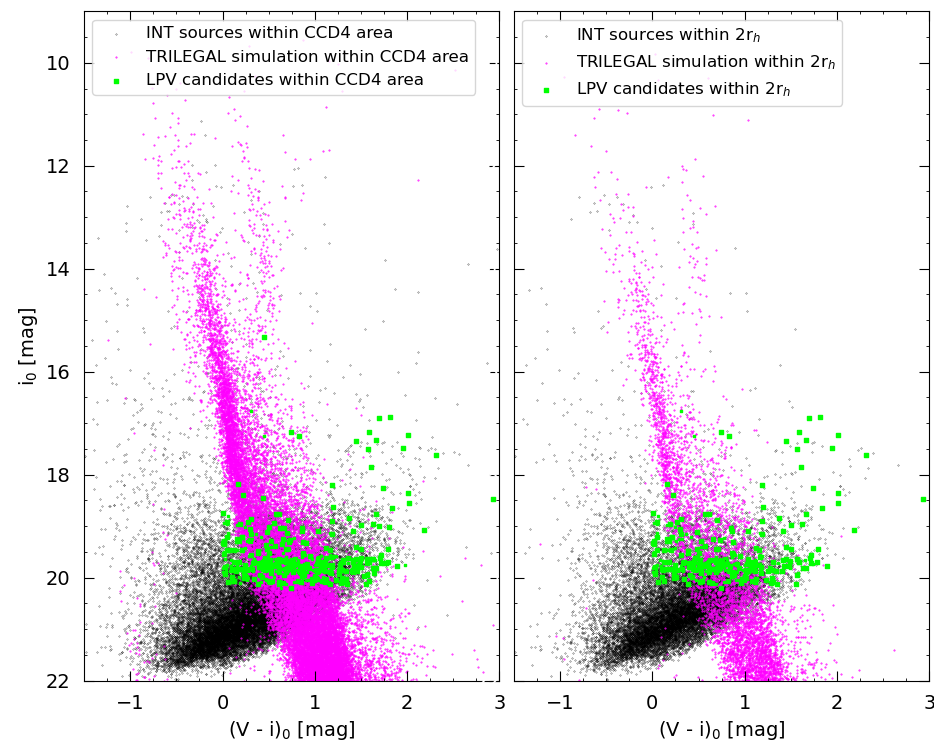}
\caption{The {\sc trilegal} simulation \citep{Girardi05} of the foreground stars within CCD4 area (left panel) and 2r$_{\rm h}$ (right panel) centered on IC\,10. 
The foreground stars and LPV candidates are shown in magenta and green, respectively.}
\label{fig:tri}
\end{figure}

The number of contaminated stars using the {\sc trilegal} simulation in CCD4, 2r$_{\rm h}$, and r$_{\rm h}$ areas was 31,815, 10,035, and 2,257, respectively. As shown in Fig.~\ref{fig:tri}, the foreground contamination is substantial.

\begin{figure}[]
\centering
\includegraphics[width=9cm,height=10cm]{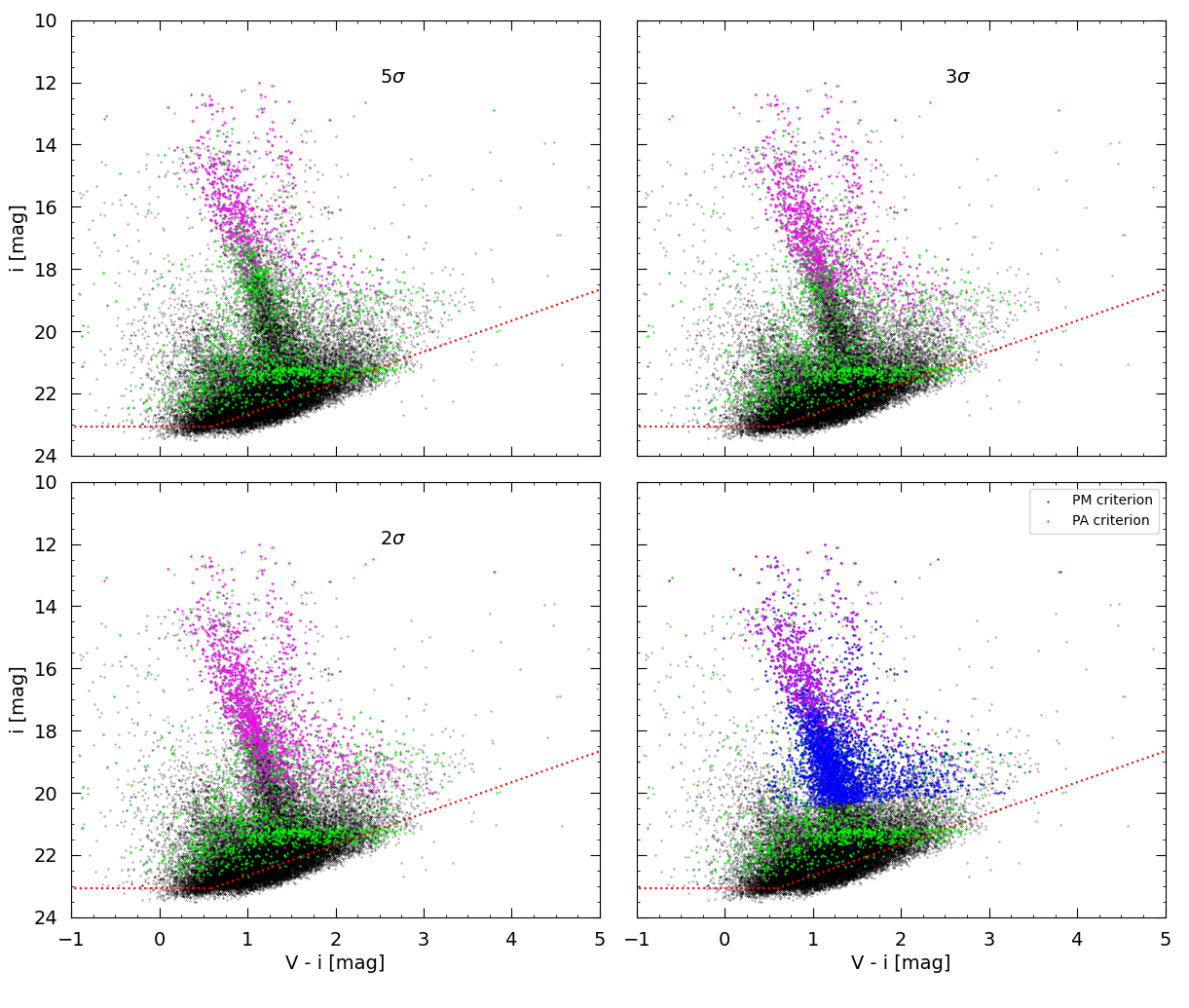}
\caption{Identification of foreground stars based on DR3 data in our catalog within the CCD4 area. Besides the variables shown in green, the foreground stars 
characterized based on their PA measurements are highlighted in magenta. Blue indicates the foreground stars as determined by a PM criterion. According to the different criteria for PA and PM, the number of foregrounds and variables changes.}
\label{fig:pm-pa}
\end{figure}

To obtain a more realistic determination of foreground stars, we identified stars common to both our catalog and the third {\it Gaia} Data Release (DR3) catalog \citep{Lindegren21}. {\it Gaia} conducted multi-epoch observations over the 34 months for 1.812 billion sources in the magnitude range G = 3--21 mag. {\it Gaia} DR3 provides both astrometry and photometry for these sources, including their positions at the epoch J2016.0, parallaxes (PA), and proper motions (PM). Additionally,  {\it Gaia} DR3 includes the second {\it Gaia} catalog of LPVs under the Specific Object Study (SOS), which contains 1,720,558 sources  with G-band amplitude variations greater than 0.1 mag \citep{Lebzelter23}. The variability selection criteria were based on a combination of color and brightness, the number of epochs, and the signal-to-noise ratio in the G-band. We found 4,586 stars in common, including 348 of our identified variable candidates. In the next stage, we selected foreground stars based on their parallax measurements. Fig.~\ref{fig:pm-pa} shows the CMD for INT sources with the selected foreground stars based on their parallax measurements at three distinct levels of 2$\sigma$, 3$\sigma$, and 5$\sigma$.

As can be seen, the number of variable candidates changes with applying different criteria to the parallax. We consider the threshold at 5$\sigma$. The PM criterion 
is calculated according to the method of \cite{Roeland19} based on the PM in both $\mu_\alpha$ and $\mu_\delta$ using $\sqrt{\mu_\alpha^2 + \mu_\delta^2}$ $>$ 0.28 
mas yr$^{-1}$ + 2.0 error. Consequently, a star is classified as foreground if it satisfies either the 5$\sigma$ parallax or proper motion criteria. Based on the 
comparison of the number of foregrounds identified using {\sc trilegal} and DR3, it can be seen that the selection made for parallax does not recognize all the 
foreground sources. Therefore, we also considered proper motion criteria, which is a more decisive criterion \citep{Roeland19}. Using this criterion, most foreground sources are eliminated, as shown in Fig.~\ref{fig:pm-pa}. There is, however, no significant change in the number of recognized foreground stars when only the PA criterion is changed.

\begin{figure}[]
\centering
\includegraphics[width=1\linewidth,clip]{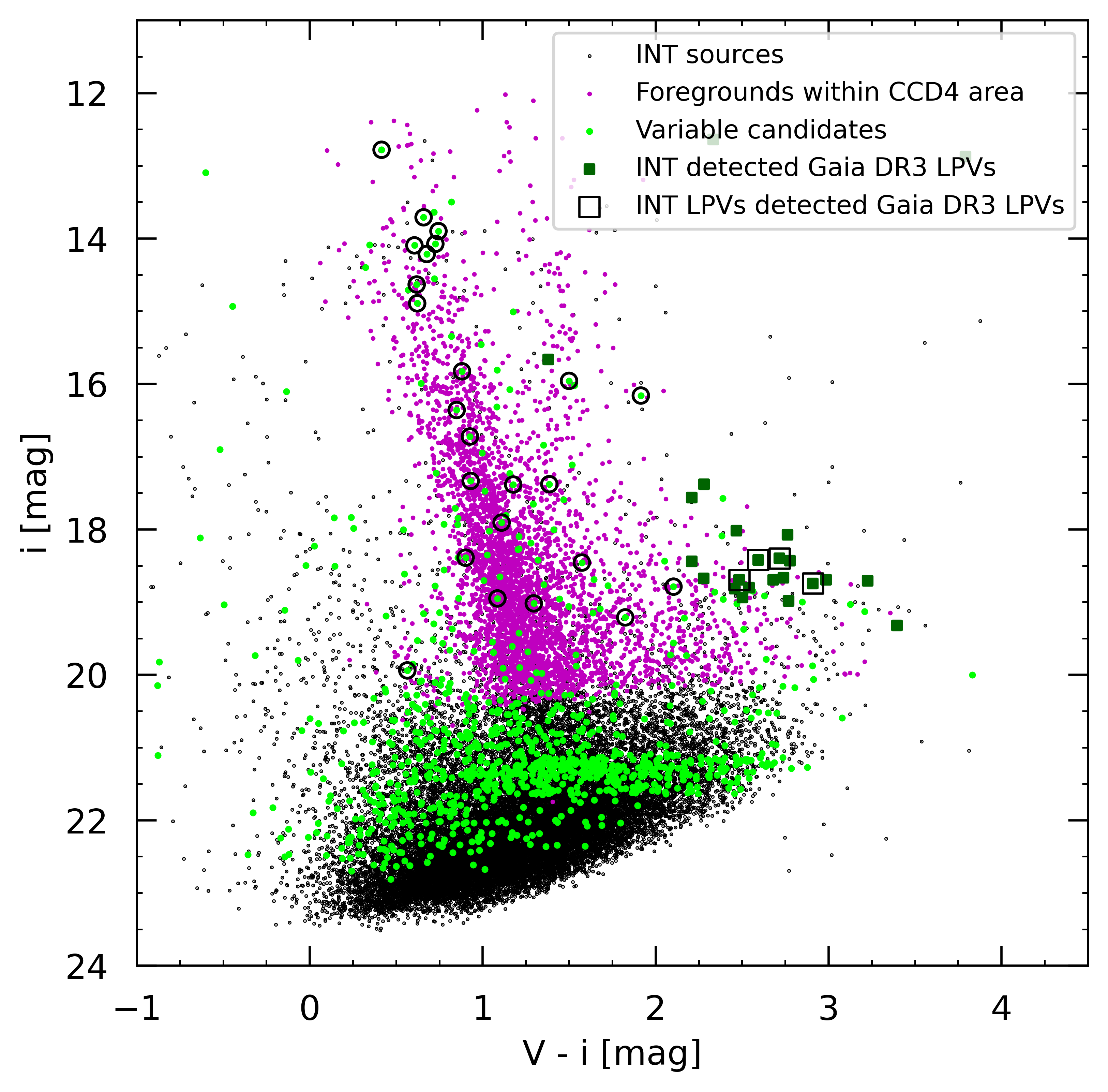}
\caption{CMD of i--band versus color (V--i) for sources in IC\,10. {\it Gaia}'s foreground stars are shown in magenta within the entire CCD4 area. The INT variable 
candidates are represented by green points. Variable candidates having RUWE $>$ 1.4 are represented by black open circles. Additionally, the {\it Gaia} DR3 LPV candidates are shown with dark green squares, and those detected in INT LPVs are represented by black open squares.}
\label{fig:ruwe}
\end{figure}

By applying these criteria, 3,955 stars were recognized as foreground stars, including 309 variable stars (Fig.~\ref{fig:pm-pa}). The change in the number of 
foreground stars is shown in Table~\ref{table:tril} by applying the PM and $\sigma$ thresholds. The identified foregrounds are depicted in Fig.~\ref{fig:ruwe} 
within the entire CCD4 area.

In comparison with the {\sc trilegal} simulation, the DR3 catalog does not cover faint stars with i $>$ 20.50 mag in our catalog. Therefore, we limited our {\sc trilegal} calculation to i $\leq$ 20.50 mag to enable a better comparison in Table~\ref{table:tril}. As can be seen, along with the change in the number of variable candidates, the number of foreground stars identified using the {\sc trilegal} simulation and {\it Gaia} catalog (5$\sigma$ PA + PM) in the area of 2r$_{\rm h}$ is very close to each other, with an estimated number of 1,570 and 1,443, respectively. Therefore, we consider the criterion of 5$\sigma$ PA + PM for the detection of foreground stars. We estimated the percentage of catalog contamination by foreground stars to be approximately 7\,\%.

\begin{table}
\begin{center}
\caption{Number of foreground stars for IC\,10 within CCD4, r$_{\rm h}$, and 2r$_{\rm h}$, comparing the results of cross-correlation with the {\it Gaia} DR3 data and {\sc trilegal} simulation. The results are limited to i $\leq$ 20.50 mag.
\label{table:tril}
}
\begin{tabular}{cccc}
\hline
Criterion  &  N$_{\rm CCD4}$  &  N$_{r_{\rm h}}$  &  N$_{2r_{\rm h}}$  \\
\hline
{\sc trilegal} sim.  &  5,688  &  393  &  1,570  \\
2$\sigma$   &  1,590  &  144  &  535  \\
3$\sigma$   &  1,115  &  100  &  374  \\
5$\sigma$   &  703  &  61  &  235  \\
2$\sigma$ + PM  &  3,964  &  494  &  1,453  \\
3$\sigma$ + PM  &  3,955  &  490  &  1,447  \\
5$\sigma$ + PM  &  3,947  &  487  &  1,443  \\\hline
\end{tabular}
\end{center}
\end{table}

Hereafter, the data are corrected for extinction in both the i--band and V--band, and the foreground sources are removed from our catalog. A newly added parameter calculated by DR3 is the re-normalized unit weight error (RUWE) for each star, which identifies unreliable sources. Based on the information provided by {\it Gaia} DR3 \citep{Lindegren21}, the threshold for retaining sources with high-quality astrometry and the full astrometric solution is RUWE $<$ 1.4. Among the variable candidates, 24 stars had RUWE $>$ 1.4, indicating that we cannot rely on high-quality photometry, astrometry, and distance measurements of these stars in the DR3 data. 

Additionally, we cross-matched our INT LPVs with those of {\it Gaia} DR3. \cite{Rimoldini23} identified 20 LPV candidates in IC\,10, and we found all these DR3 LPVs in our photometric catalog except for one, which was located outside the CCD4 area. These recovered DR3 LPVs are depicted by dark green squares in Fig.~\ref{fig:ruwe} and red open squares in Fig.~\ref{fig:lpv-median}. Furthermore, \cite{Lebzelter23} applied different selection criteria to minimize contamination from variable types other than LPVs. They confirmed only four LPVs in the IC\,10 galaxy, all of which we detected as INT LPV candidates in our optical catalog. We showed the common sources between our INT LPVs and those of {\it Gaia} DR3 by black open squares in Fig.~\ref{fig:ruwe}.

\section{DISCUSSION}\label{sec:DIS}
\subsection{Distribution of LPV candidates} \label{sec:spatial}

In the final list of variable candidates, we found 1,052 stars in the entire CCD4 area (0.07 deg$^{2}$), of which 752 are located within 2r$_{\rm h}$. Of these, 536 and 380 candidates are identified as LPV candidates within the CCD4 and 2$r_{\rm h}$ areas, respectively. Fig.~\ref{fig:lpv-median} shows the LPV candidates on the median image, marked with green open circles. LPV candidates are mainly concentrated towards the center of IC\,10; a total of 63 LPV candidates were identified outside the 3r$_{\rm h}$ area.

\begin{figure}[]
\centering
\includegraphics[width=90mm, height=90mm]{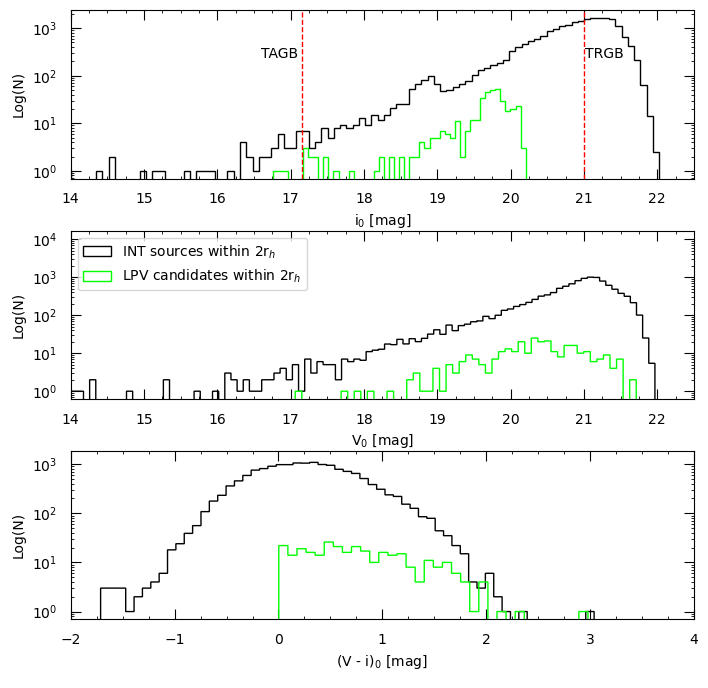}
\caption{Logarithmic distribution of brightness (top and middle panels) and color (bottom panel) for IC\,10 sources (black) and LPV candidates within 2r$_{\rm h}$. The vertical dashed lines in the top panel indicate the TAGB and TRGB.}
\label{fig:hist}
\end{figure}

Fig.~\ref{fig:hist} shows the magnitude distribution of the IC\,10 sources and LPV candidates within 2r$_{\rm h}$ of the galaxy's center in the i$_0$--band, V$_0$--band, and color (V--i)$_0$. The {\it Gaia} DR3 catalog, used to distinguish foreground contamination in our catalog, operates for bright foreground sources (i $<$ 20.5 mag) but not for faint stars. Therefore, to avoid a statistical discontinuity between bright populations corrected for foregrounds and faint populations that include foreground stars, we corrected the luminosity distribution for faint stellar populations statistically. For this purpose, we compare the density of the stars in our catalog with those of the foreground artificial stars (from the {\sc trilegal} simulation) in each magnitude bin. 

The i$_0$--band histogram shows that approximately 73\,\% of our detected sources are brighter than i$_0$ $\sim$ 21.25 mag, as are the LPV candidates, which are brighter than i$_0$ $\sim$ 19.90 mag. Around 80\,\% of LPV candidates are observed in the range of 18 mag to $\sim$ 20 mag in the i$_0$--band. From 17 mag to 20 mag, the fraction of LPV candidates drops to about 7\,\% of all sources. However, this magnitude interval consists of the highest number of variables. The bottom panel shows that about 78\,\% of LPV candidates have a color redder than 1 mag, indicating they are mostly dusty AGB stars.

\subsection{Cross-identifications with other catalogs}

The following list presents cross-matching identifications between our detected sources and those from other catalogs. In the CMDs, the colors of stars not detected in the V--band are calculated based on the completeness limit of 50\,\% for the V--band (23.66 mag). Additionally, all sources displayed in the CMDs have been corrected for constant extinction.

1) \textbf{DUSTiNGS survey}: Using 3.6 and 4.5 $\mu$m imaging data from the {\it Spitzer} Space Telescope \citep{Boyer15a, Boyer15b}.

2) \textbf{CFHT survey}: Images obtained with the Canada-France-Hawai'i Telescope (CFHT) using R, I, CN, and TiO filters \citep{Demers04}.

3) \textbf{HST survey}: Images obtained with WFPC2, ACS, and WFC3 onboard the {\it Hubble} Space Telescope (HST), providing the first and deepest catalog of variables \citep{Bonanos19}.

\subsubsection{DUSTiNGS survey}\label{sec:dustings}

\begin{figure*}[]
\centering
\includegraphics[width=0.5\textwidth]{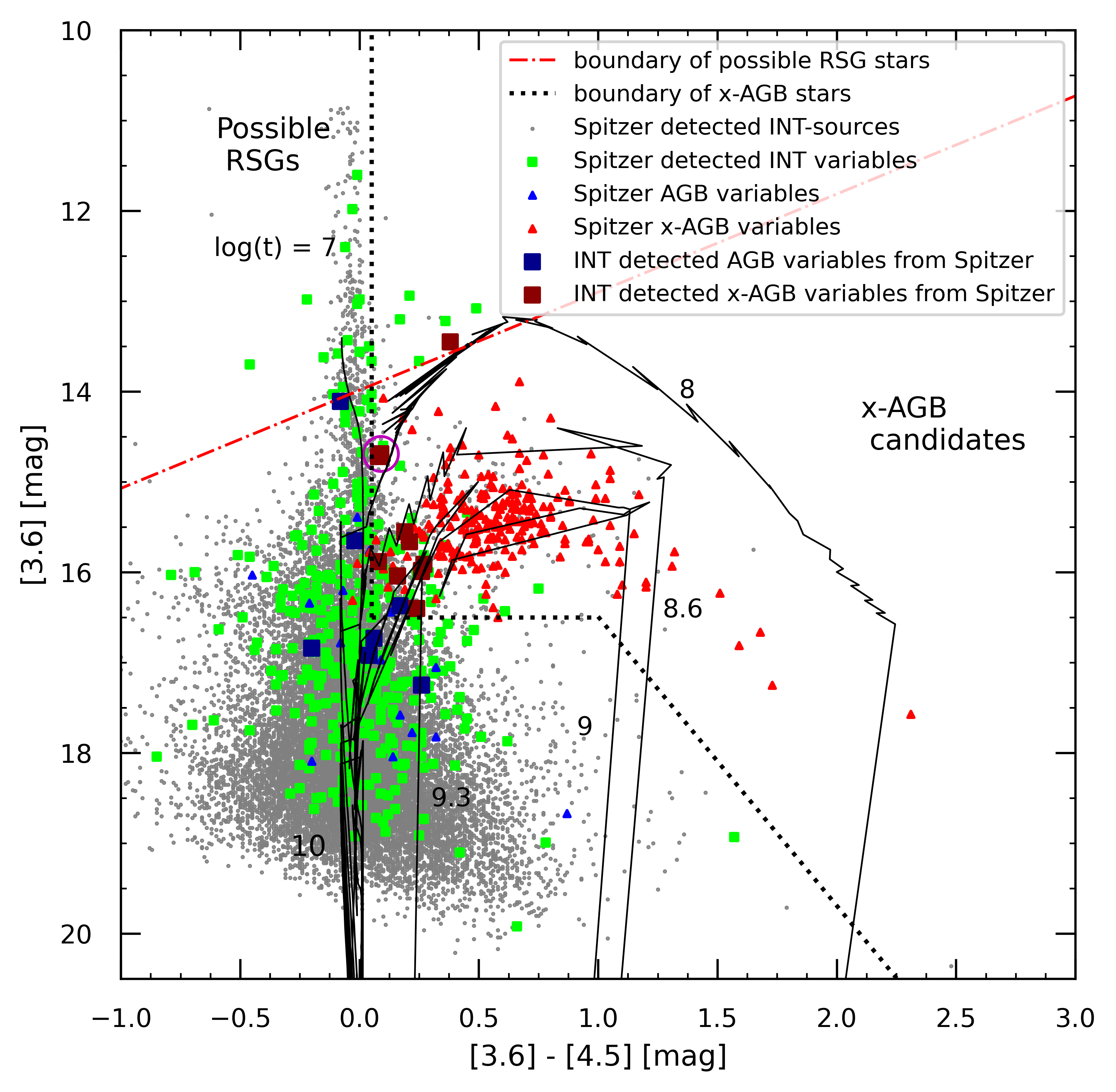}\hfill
\includegraphics[width=0.5\textwidth]{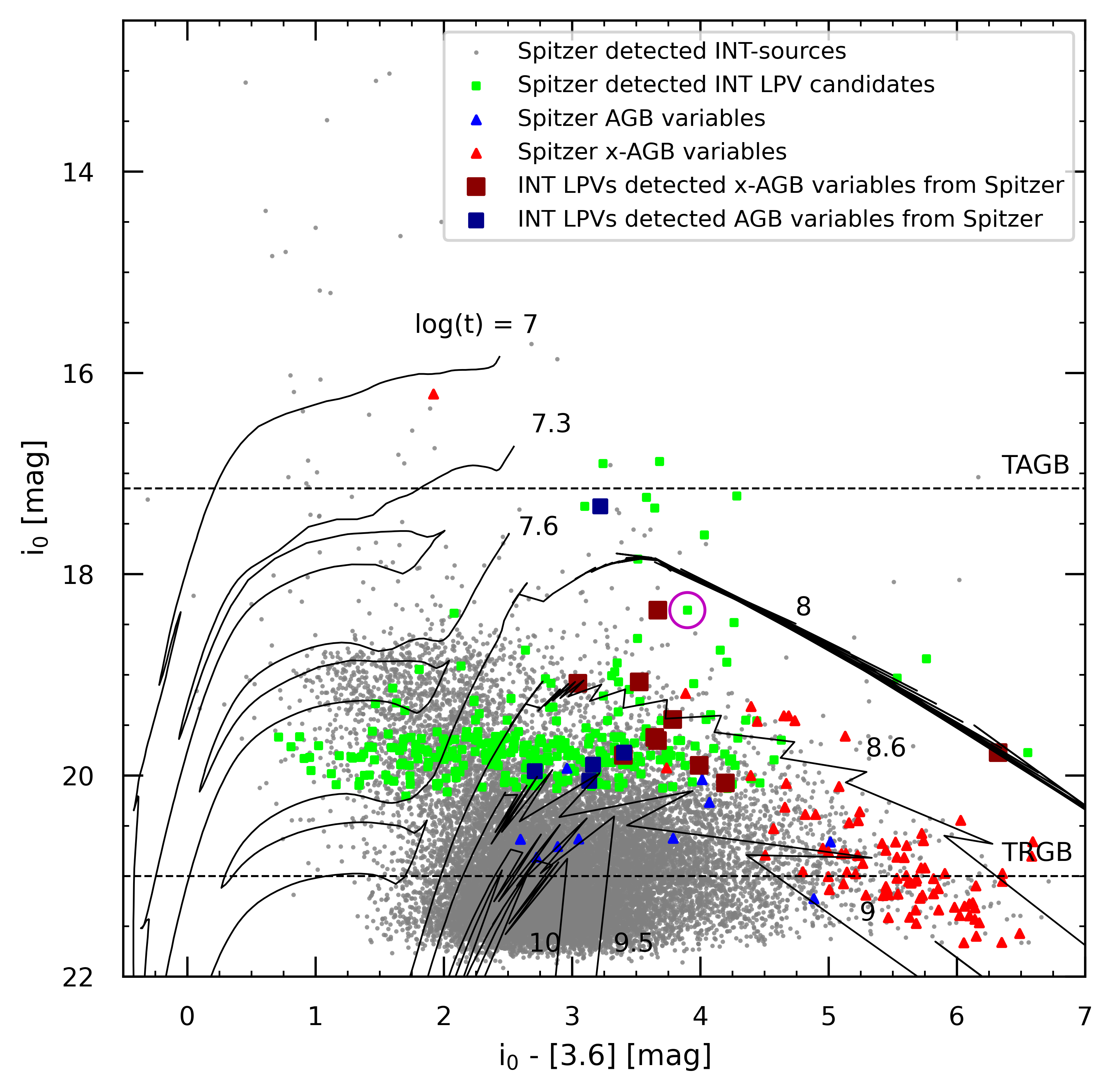}
\caption{CMD for the sources common to our photometric catalog and the DUSTiNGS survey within CCD4 area, shown in [3.6] vs.\ [3.6]--[4.5] (left panel) and i$_{0}$--band vs.\ i$_0$--[3.6] (right panel). In both panels, INT variables recovered by {\it Spitzer} are depicted in green, while the less dusty AGB and x-AGB variables from DUSTiNGS \citep{Boyer15b} are represented by blue and red triangles, respectively. Additionally, the {\it Spitzer} variables recovered among INT variable candidates are displayed as dark red and blue squares for x-AGB and less dusty AGB stars, respectively. Isochrones from \cite{Marigo17} are overplotted for ages between 10 Myr and 10 Gyr. The black dotted lines in the left panel indicate the approximate boundaries for x-AGB stars, and the red dashed-dotted line marks the approximate boundary for potential RSGs. In both panels, a magenta open circle highlights the INT LPV candidate with the greatest amplitude of 3.23 mag in the i$_0$--band.}
\label{fig:spitzer}
\end{figure*}

The mid--infrared imaging survey of DUSTiNGS (DUST in Nearby Galaxies with {\it Spitzer}) was used to identify dust--producing AGB stars and massive stars in 50 nearby dwarf galaxies within 1.5\,Mpc in two epochs, mapped about six months apart. In this survey, \cite{Boyer15a,Boyer15b} classified the variable stars into extremely dusty AGB (x-AGB) stars, less dusty AGB stars, and unknowns based on their position on the CMD. They also classified the variables as AGB or x-AGB stars, or potential RSGs if they met a brightness threshold depending on their color ([3.6]--[4.5]; Fig.~\ref{fig:spitzer}).

The x-AGB stars are the dustiest AGBs in the superwind phase and are near the end of their evolution \citep{Goldman19}. These stars are defined as those brighter than M$_{[3.6]}$ = $-$8 mag with colors [3.6]--[4.5] $>$ 0.1 mag. Despite producing more than 75\,\% of the dust generated by cool evolved stars, they 
account for less than 6\,\% of the entire AGB population (e.g., \cite{Riebel12, Boyer12}). The stars classified as "unknowns" are variables with magnitudes 
fainter than M$_{3.6}$ = $-$6 mag, which is the assumed TRGB in their analysis.

Based on the method of \cite{Boyer15a}, \cite{Boyer15b} detected 235 x-AGB stars and 22 less dusty AGB stars in IC\,10. We found 28,485 stars in common within the area covered by CCD4, recovering approximately 73\,\% of DUSTiNGS sources within the 2r$_{\rm h}$ area in our catalog. Among the 257 variables identified with {\it Spitzer}, we recovered 110 stars in our catalog, comprising approximately 40\,\% x-AGB and $\sim$ 68\,\% less dusty AGB stars. They found 22 less dusty AGB and 235 x-AGB variable candidates in IC\,10, of which 7 AGB and 11 x-AGB variable candidates were recovered among our identified variables within the CCD4 area. Our results are consistent with the study by \cite{Boyer15a, Boyer15b}.

The unrecovered variables in our catalog mostly consist of faint stars with unreliable photometry, characterized by high $\chi$ values, so we removed them from our 
catalog. The sources from DUSTiNGS that are not found in our catalog are generally located closer to the center of IC\,10. A concentration of dust in the central 
region of the galaxy makes it difficult to identify AGB stars, so our survey was not able to detect all AGB stars in the central region. In contrast, {\it Spitzer} 
is more complete for dusty or reddened sources than optical surveys. As a result, DUSTiNGS was able to identify more x-AGBs and optically obscured AGBs.

Fig.~\ref{fig:spitzer} (left panel) presents a CMD of the {\it Spitzer} survey's mid-IR sources, which were recovered in our catalog along with the INT variable candidates. Overplotted on the CMD are isochrones for ages between 10 Myr and 10 Gyr from \cite{Marigo17} at a constant metallicity of [Fe/H] = $-$1.28 dex. We also indicated the approximate boundaries between x-AGB stars and others (black dotted lines in the left panel), and for potential RSGs (red dashed-dotted line), based on the study by \cite{Boyer15b}. In addition, less dusty AGB and x-AGB stars from \cite{Boyer15b} are shown, and those sources recovered from our variable candidates are highlighted.

Among the {\it Spitzer} variables, a five less dusty AGB variable candidates that overlap with our identified variable candidates (totaling seven) are classified as LPV candidates in our catalog. Additionally, 10 out of 11 x-AGB variable candidates detected in our variable list are classified as LPV candidates, due to their amplitude exceeding 0.20 mag in the i--band. We also identified our dustiest (reddened) LPV candidate, with an amplitude of 3.23 mag, among the x-AGB variable stars in the DUSTiNGS survey, indicated by magenta open circles. The cross-matching results between our photometric catalog and the {\it Spitzer} survey are confined to the CCD4 area in our catalog, as illustrated in Fig.~\ref{fig:spitzer}.

Fig.~\ref{fig:spitzer} (right panel) further presents a CMD for the sources common to both the {\it Spitzer} and INT surveys in the i$_{0}$--band vs.\ color (i$_0$--[3.6]). It is evident that the recovered x-AGB stars among our variable candidates are dusty and faint, with i$_0$--[3.6] $\ge$ 3 mag, most of which exhibit amplitudes greater than 1.0 mag in the i--band.

\subsubsection{CFHT survey: Carbon stars}

\cite{Demers04} studied IC\,10 using the wide-field CFH12K camera at the 3.58-m Canada-France-Hawai'i Telescope (CFHT) with R, I, CN, and TiO filters in a field of 
42$^\prime$ $\times$ 28$^\prime$. They detected 676 carbon stars based on their location in the color--color diagram of (CN--TiO) versus (R--I) using PSF-fitting 
photometry; these were then used to calculate the distance to IC\,10. By cross-correlating \cite{Demers04}'s carbon star catalog with our catalog within the area of 
CCD4, we recovered 510 out of 676 carbon stars, of which 25 stars are in our variable candidate list. Among these, 24 carbon stars are identified within the 536 LPV candidates found by us.

\begin{figure}[]
\centering
\includegraphics[width=1\linewidth,clip]{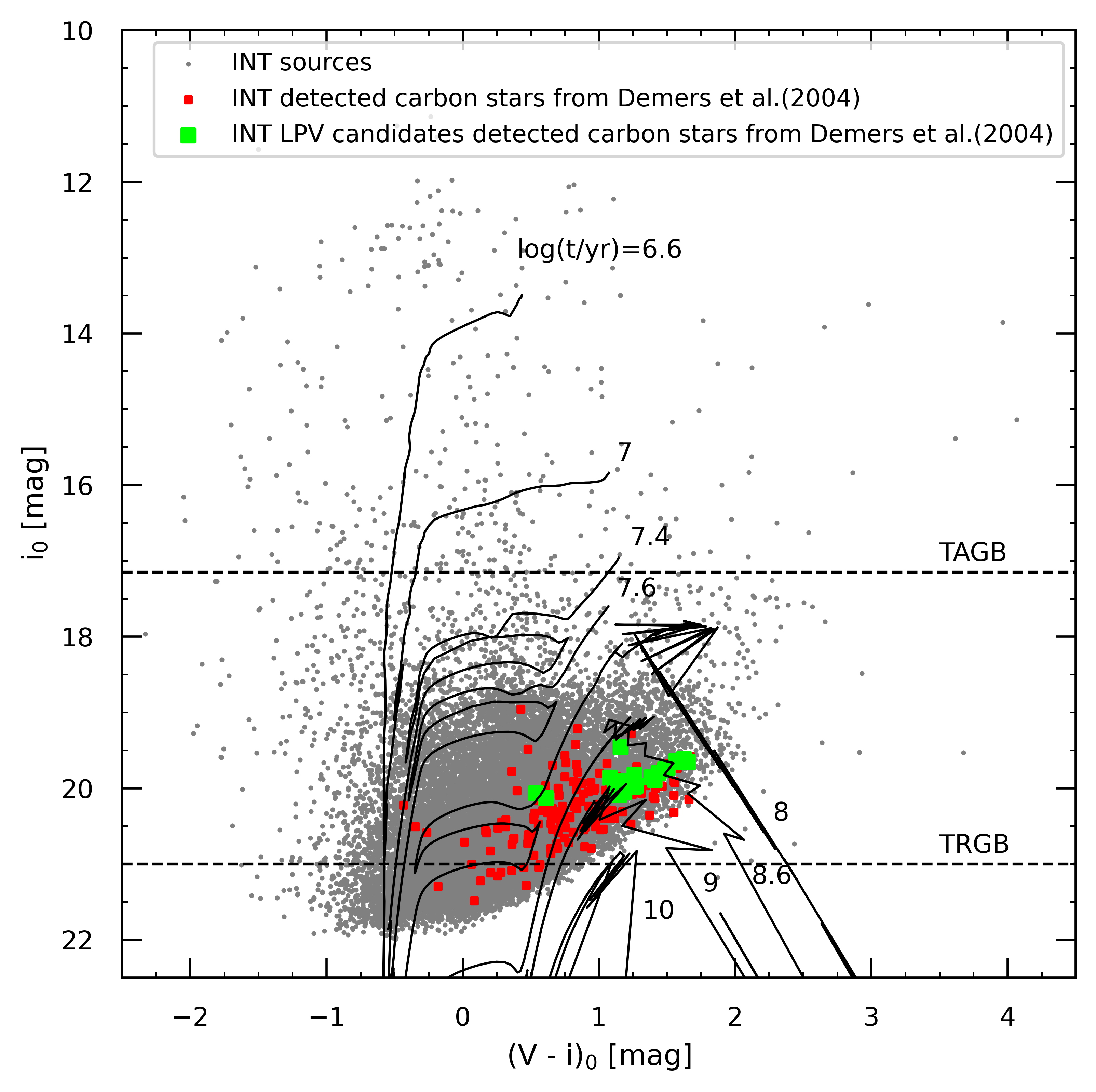}
\caption{CMD of the i$_0$-band versus color (V--i)$_0$ for IC\,10 sources. The carbon stars identified by \cite{Demers04} and recovered in our catalog are shown as red squares. Green squares represent the LPV candidates within CCD4 area classified as carbon stars. Isochrones from \cite{Marigo17} are also overplotted. All sources have been corrected for extinction. The horizontal black dashed lines indicate the TAGB and TRGB magnitudes.}
\label{fig:demers}
\end{figure}

In cross-correlating the {\it Spitzer} catalog with \cite{Demers04}'s carbon stars, \cite{Boyer15a, Boyer15b} recovered 356 carbon stars, whereas we detected the 
vast majority of carbon stars in our survey (approximately 75\,\%). As shown in Fig.~\ref{fig:demers}, most of the recovered carbon stars fall within the magnitude range of 19 mag $<$ i$_0$ $<$ 21.5 mag with a color of -0.5 mag $<$ V$_0$--i$_0$ $<$ 2.0 mag (four recovered carbon stars in our catalog have negative colors due to large photometric errors), which is generally consistent with the location of our detected LPV candidates on the CMD (see Fig.~\ref{fig:3cmd}, middle and right panels). 

\subsubsection{HST survey}

Based on archival images acquired with WFPC2, ACS, and WFC3 onboard the HST, \cite{Bonanos19} compiled a comprehensive catalog of variable sources, known as the Hubble Catalog of Variables (HCV). The HST survey, conducted between 1994 and 2017, utilized 108 different filters to cover an area of 40.6 deg$^2$, approximately 0.1\,\% of the sky. To identify reliable variable sources, they applied a magnitude-dependent threshold of 5$\sigma$ to data obtained with the same 
instrument and filter combination.

The catalog contains 84,428 variables with a magnitude of V $\leq$ 27 mag in the Milky Way and nearby galaxies. These variables were categorized into two groups 
based on the number of filters through which a source exhibited variability. A source was classified as a multi-filter variable candidate (MFVC) if it displayed 
variability through more than one filter; Otherwise, it was labeled a single-filter variable candidate (SFVC).

\begin{figure}[]
\centering
\includegraphics[width=1\linewidth,clip]{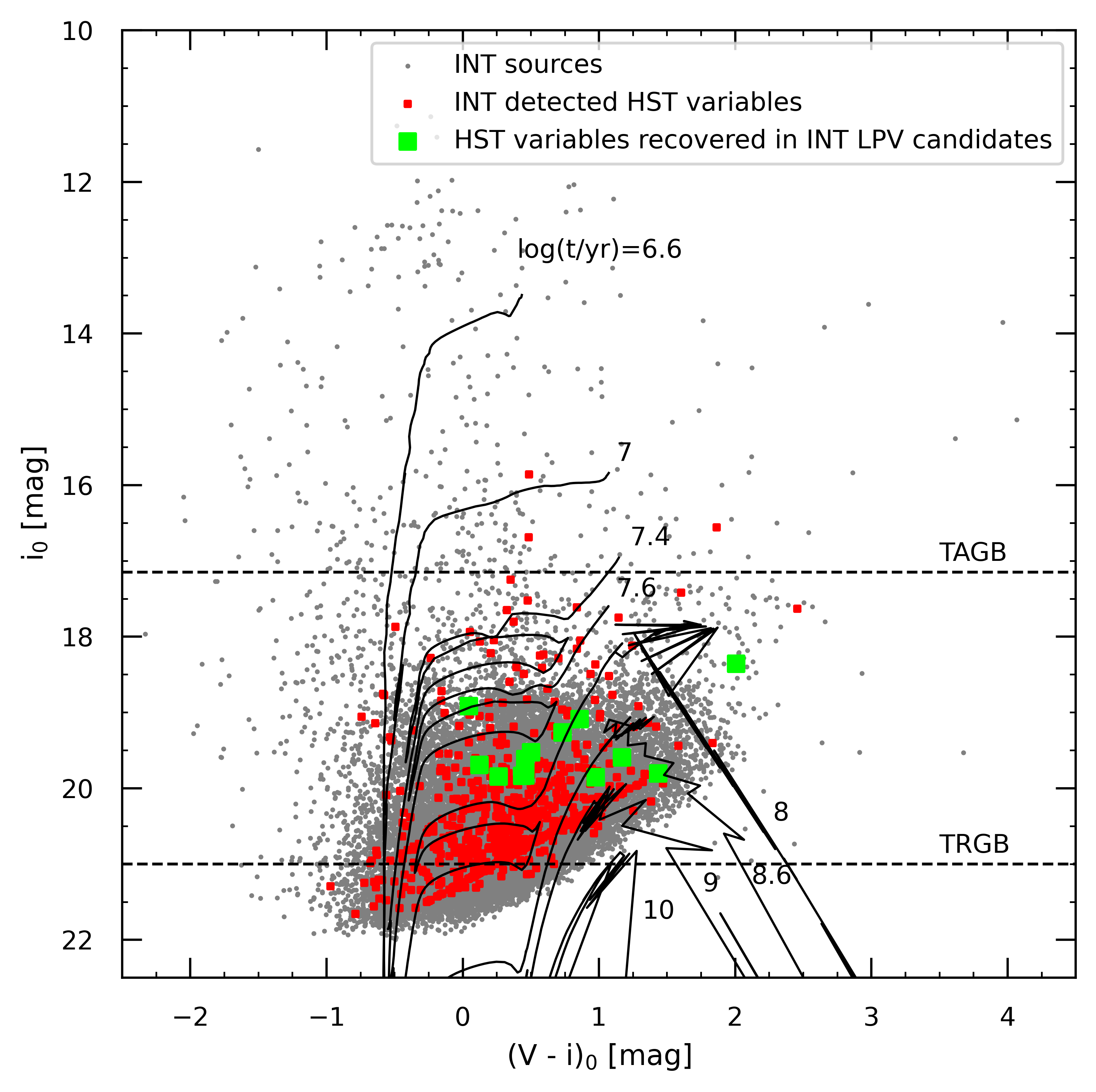}
\caption{CMD of the i$_0$--band versus color (V--i)$_0$ for IC\,10 sources. The HST variables recovered in our catalog are shown as red squares. In 
total, 34 of our variables were recovered from the HCV, of which 17 are considered LPV candidates (green squares) within the CCD4 area. Isochrones from \cite{Marigo17} are also overplotted.}
\label{fig:hst}
\end{figure}

We performed a cross-correlation between the INT survey and HST variables, as presented in Fig.~\ref{fig:hst}. In our optical catalog, we recovered 669 out of 967 ($\sim$ 70\,\%) variables classified as SFVCs within the CCD4 area. These SFVCs were identified based on measurements in one filter (F814W) from the HCV catalog. Additionally, we identified 22 of our variables among the HCV variables, of which 13 are LPV candidates, exhibiting an amplitude greater than 0.20 mag.

\subsection{RSGs}

\subsubsection{RSGs from \cite{Ren21}}

\cite{Ren21} conducted a study on the RSGs in 12 low-mass galaxies within the LG, utilizing UKIRT/WFCAM and 2MASS photometry. For ten dwarf galaxies, they acquired 
JHK photometry covering the region within the radius where the number of stars declines to a level of 5$\sigma$. They applied extinction correction using the 2D 
dust map of SFD98 \citep{Schlegel98}. They identified 1,340 RSGs in IC\,10, predominantly in the central region where star-forming activity is concentrated.

\begin{figure}[]
\centering
\includegraphics[width=1\linewidth,clip]{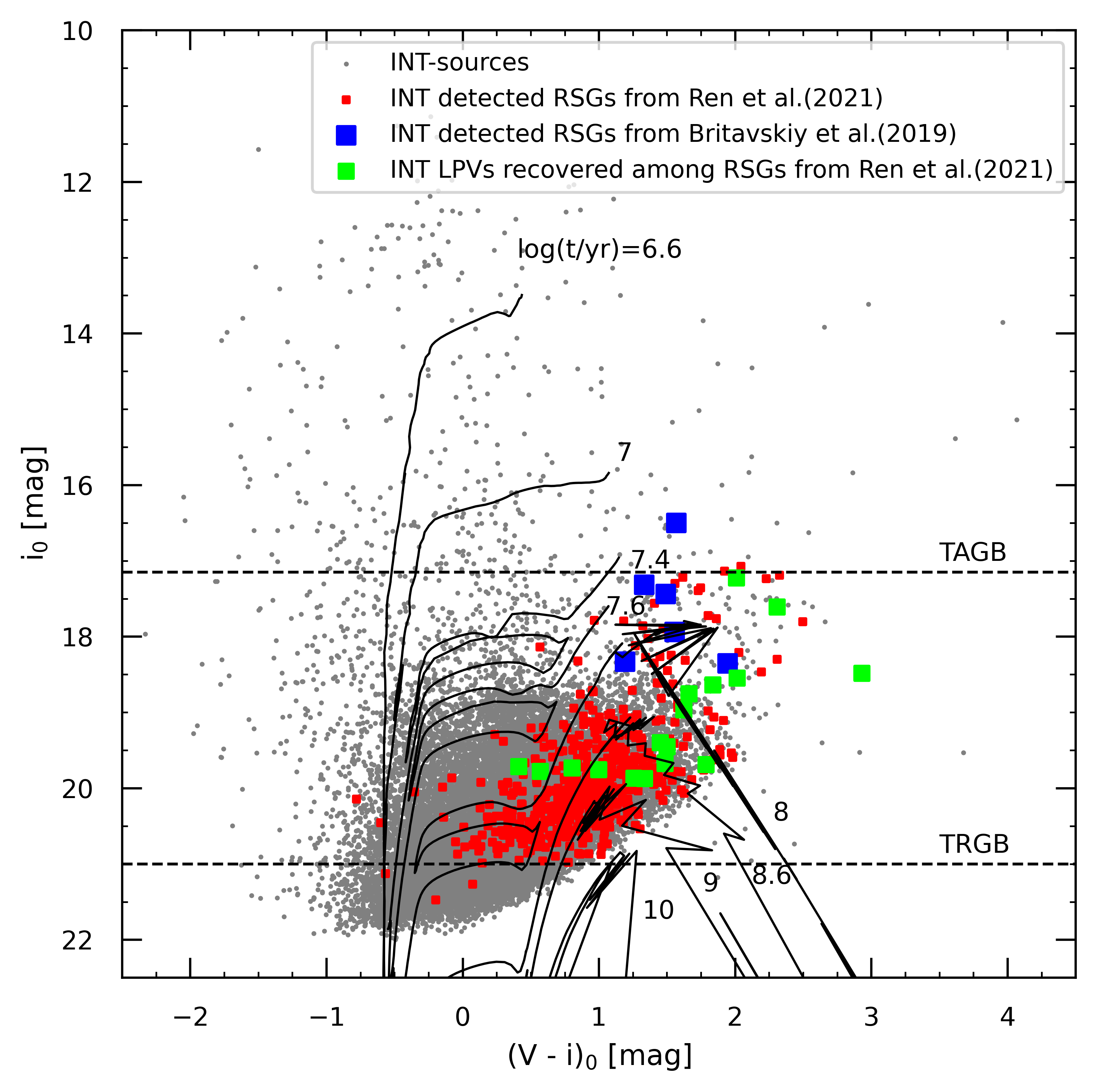}
\caption{CMD of the i$_0$--band versus color (V--i)$_0$ for IC\,10 sources within the area of CCD4. The red squares represent the RSGs from Ren's catalog detected by us, with LPV candidates labeled with green squares. Blue squares indicate the RSGs from \cite{Britavskiy19}. Isochrones from \cite{Marigo17} are also overplotted.}
\label{fig:ren-ebv}
\end{figure}

\begin{figure*}[]
\centering
\includegraphics[width=1\linewidth,clip]{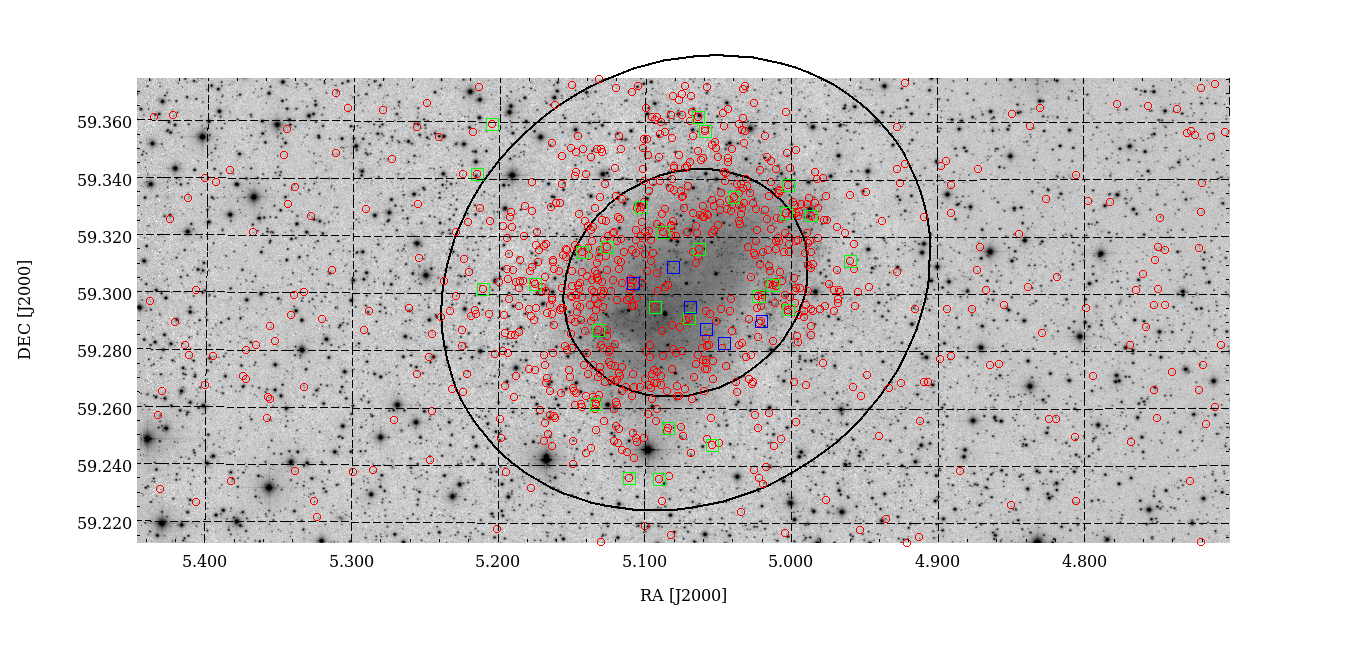}
\caption{RSGs from \cite{Ren21} recovered in our catalog are shown with red circles, across the entire area of CCD4 and the median WFC image of IC\,10. The RSGs recovered among our LPV candidates are labeled with green open boxes, while blue open boxes represent RSGs from \cite{Britavskiy19} recovered in our catalog. Black ellipses mark the r$_{\rm h}$ and 2r$_{\rm h}$ areas.}
\label{fig:median-ren}
\end{figure*}

Through cross-correlation between our catalog and \cite{Ren21}'s data, we identified 826 RSGs in common within the CCD4 area (Fig.~\ref{fig:ren-ebv}). The 513 RSGs from \cite{Ren21} that we did not recover are located outside the CCD4 area. Among the RSGs, 659 are within the 2r$_{\rm h}$ area, and 168 are outside of it. We identified 27 LPV candidates among these RSGs (see green squares in Fig.~\ref{fig:ren-ebv} and green open squares in Fig.~\ref{fig:median-ren}). The spatial distribution of \cite{Ren21}'s RSG candidates (Fig.~\ref{fig:median-ren}) suggests that most of them are likely RGB or foreground stars; thus, we do not confirm them as RSG candidates in IC\,10.

\subsubsection{RSGs from \cite{Britavskiy19}}

\cite{Britavskiy19} studied the fundamental physical parameters of seven star-forming dwarf irregular galaxies in the LG using photometry from images taken by the 
{\it Spitzer} Space Telescope in the 3.6-$\mu$m and 4.5-$\mu$m bands. They corrected each star for extinction, which ranged from 0.1 to 4 mag, using 
the extinction law from \cite{Maiz14} with R$_V$ = 3.1. They identified six RSGs based on CMD analysis with PARSEC evolutionary tracks \citep{Bressen12, Chen15}, 
estimated from H$_\alpha$ and Ultraviolet data covering the past 50 Myr, four of which are confirmed as RSGs by spectroscopy. We cross-matched the RSGs from 
\cite{Britavskiy19} with our sources within the CCD4 area. All six RSGs detected by \cite{Britavskiy19} are found in our catalog (see blue squares in Fig.~\ref{fig:ren-ebv} and Fig. \ref{fig:median-ren}).

\section{CONCLUSION}\label{sec:CON}

We utilized INT optical survey to monitor the LG's IC\,10 dwarf irregular galaxy with the WFC in the i--band and V--band over nine epochs. We created a 
PSF-fitting photometric catalog, resulting in the most comprehensive optical catalog of LPV candidates in IC\,10 to date. This catalog contains 53,579 stars within the entire area of CCD4 (0.07 deg$^2$, corresponding to 13.5 kpc$^2$ at the distance of IC\,10), of which 536 stars are identified as LPV--candidates.

We estimated the reddening for each star in our catalog using the SFD98 map, which indicated that most detected sources have E(B--V) $\sim$ 1.5 mag. We also examined the effect of applying a constant reddening to all sources. Based on this comparison, we found that the SFD98 dust map does not perform well for IC\,10. Therefore, we corrected our catalog's sources for a constant extinction.

Furthermore, we cross-correlated our catalog with those from Pan-STARRS, {\it Spitzer}, and HST optical monitoring surveys, as well as with carbon stars from the CFHT survey. We recovered more than 70\,\% of these sources within 2r$_{\rm h}$ centered on IC\,10 in our INT catalog. Our analysis of IC\,10 reveals several dusty AGB stars with large amplitudes in the i--band and red colors, including 10 x-AGB stars and a population of RSGs. Additionally, we recovered all LPV candidates confirmed by {\it Gaia} DR3 in our LPV candidates list. Although the {\it Gaia} LPV catalog introduced the largest all-sky LPV catalog to date, our optical LPV catalog is the most comprehensive, covering a wide range of magnitudes. Due to the large distance to this galaxy, the {\it Gaia} DR3 catalog contains bright LPVs for IC\,10. Our catalog will be made publicly available at CDS. In subsequent papers in this series, we will use our photometric catalog to determine the SFH and dust production of IC\,10.

\section{ACKNOWLEDGMENTS}

The observing time for this survey was primarily provided by the Iranian National Observatory (INO), complemented by UK-PATT allocation of time to programs I/2016B/
09 and I/2017B/04 (PI: J.\ van Loon). The INO and the School of Astronomy at the Institute for Research in Fundamental Sciences (IPM) provided financial support for 
this project. MGh is grateful to Prof.\ Peter Stetson for sharing his photometry software package. We are grateful to the anonymous referee for carefully reading the manuscript and for helpful comments and suggestions, which helped us to improve the quality of the paper.

\def\apj{{ApJ}}    
\def\nat{{Nature}}    
\def\jgr{{JGR}}    
\def\apjl{{ApJ Letters}}    
\def\aap{{A\&A}}   
\def\mnras{{MNRAS}}
\def\aj{{AJ}}
\let\mnrasl=\mnras

\appendix
\section{Supplementary Material}
\label{sec:apndix}

As mentioned in Sec.~\ref{sec:amp}, we identified 536 LPV candidates within the area of CCD4 of the WFC on the INT 
telescope. Here, we present the light--curves of these LPV candidates. The complete set of figures (20 images) is 
available in the online journal.

\renewcommand{\thefigure}{A\arabic{figure}}
\renewcommand{\thetable}{A\arabic{table}}
\setcounter{figure}{0} 
\setcounter{table}{0}  

\clearpage 

\begin{figure}
    \centering
    \includegraphics[width=0.9\textwidth]{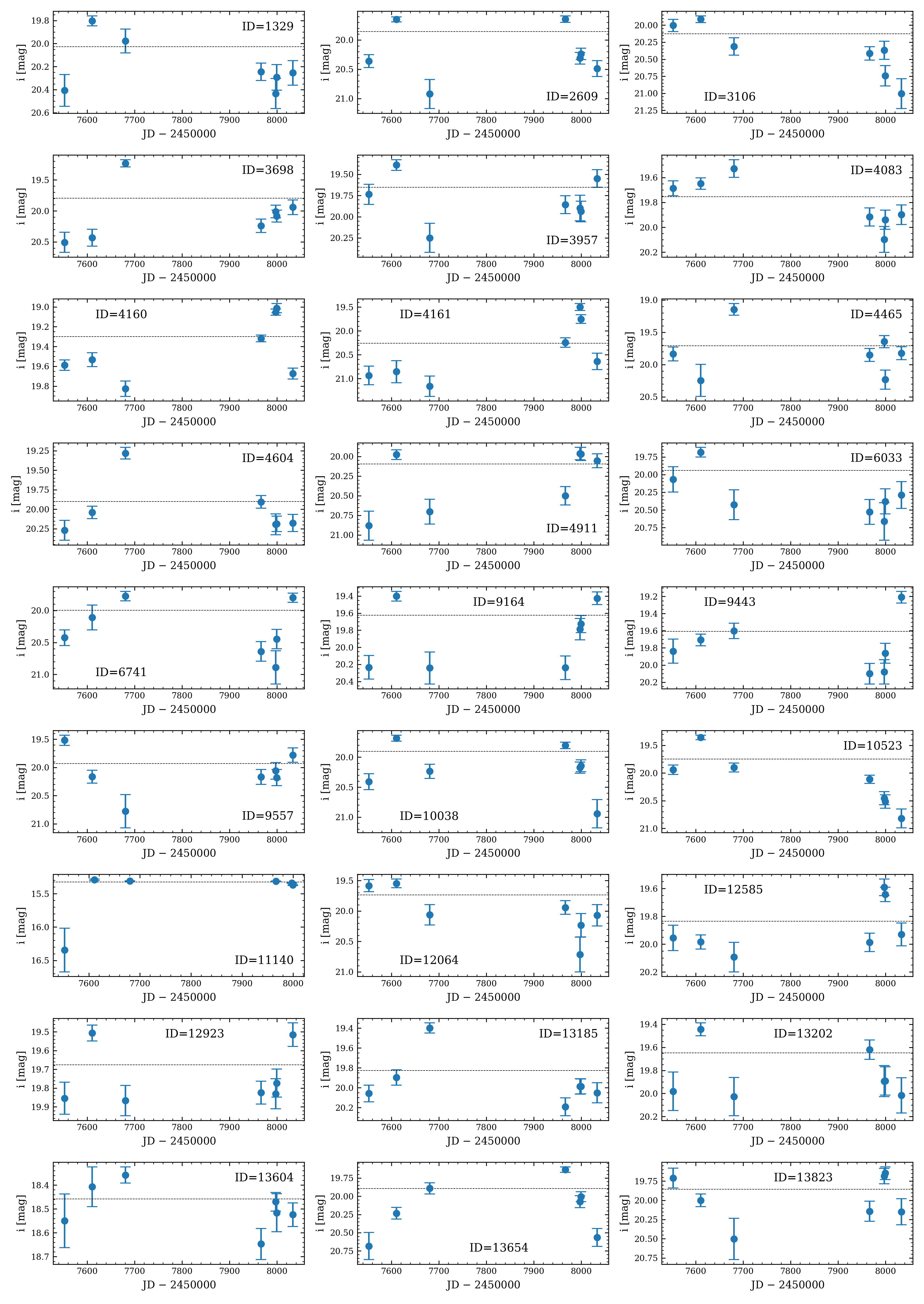}
    \begin{minipage}{0.85\textwidth} 
        \caption{Light curves of LPV candidates. The complete figure set (20 images) is available in the online journal.}
        \label{fig:SED_fits_147}
    \end{minipage}
    \figsetstart
    \figsetnum{A1}
    \figsettitle{Light Curves of LPV Candidates}
    \figsetgrpstart
    \figsetgrpnum{A1.1}
    \figsetgrptitle{Light Curve (a)}
    \figsetplot{page_1.png}
    \figsetgrpnote{Light curve of LPV candidate (a).}
    \figsetgrpend
    \figsetgrpstart
    \figsetgrpnum{A1.2}
    \figsetgrptitle{Light Curve (b)}
    \figsetplot{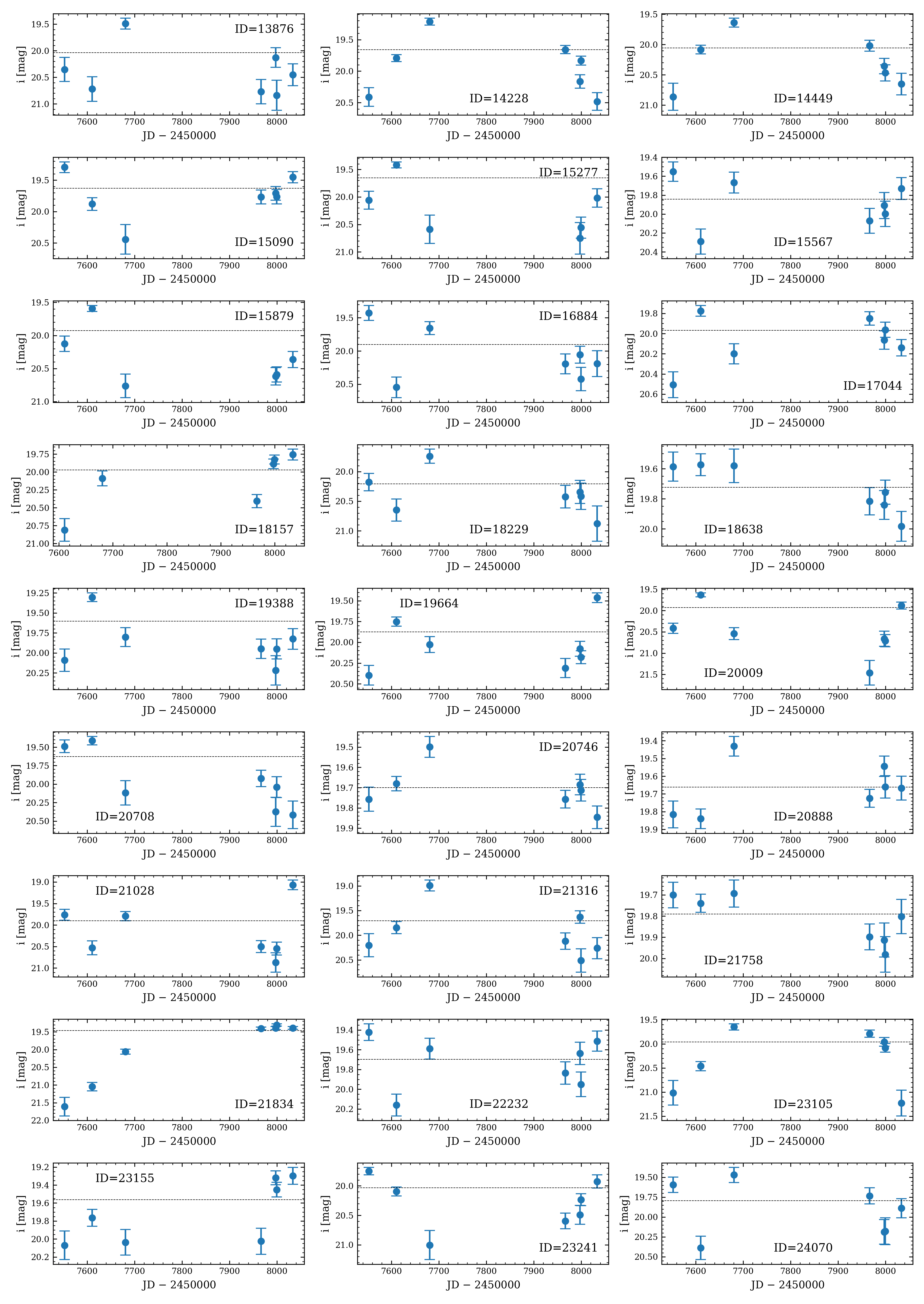}
    \figsetgrpnote{Light curve of LPV candidate (b).}
    \figsetgrpend
    \figsetgrpstart
    \figsetgrpnum{A1.3}
    \figsetgrptitle{Light Curve (c)}
    \figsetplot{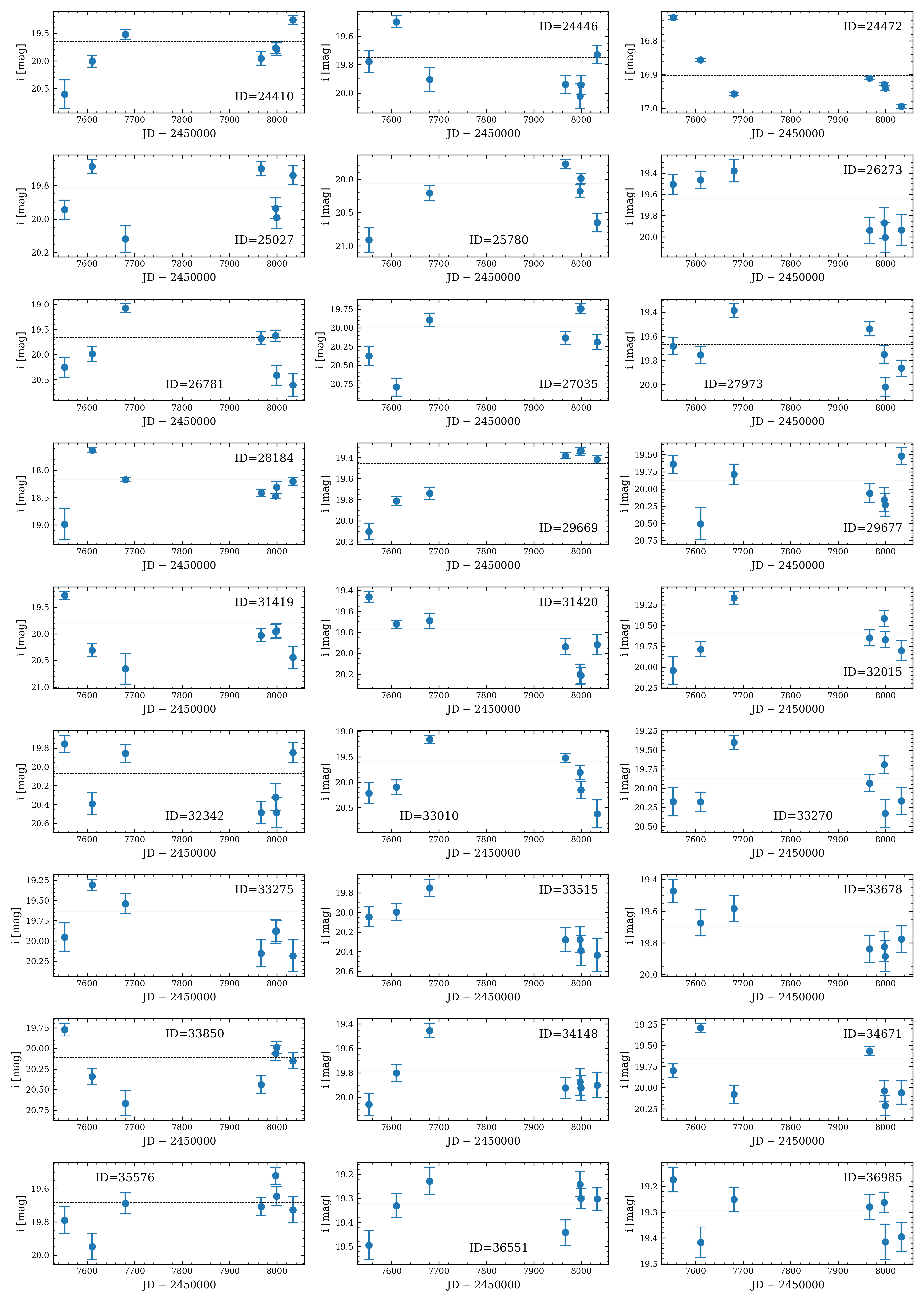}
    \figsetgrpnote{Light curve of LPV candidate (c).}
    \figsetgrpend
    \figsetgrpstart
    \figsetgrpnum{A1.4}
    \figsetgrptitle{Light Curve (d)}
    \figsetplot{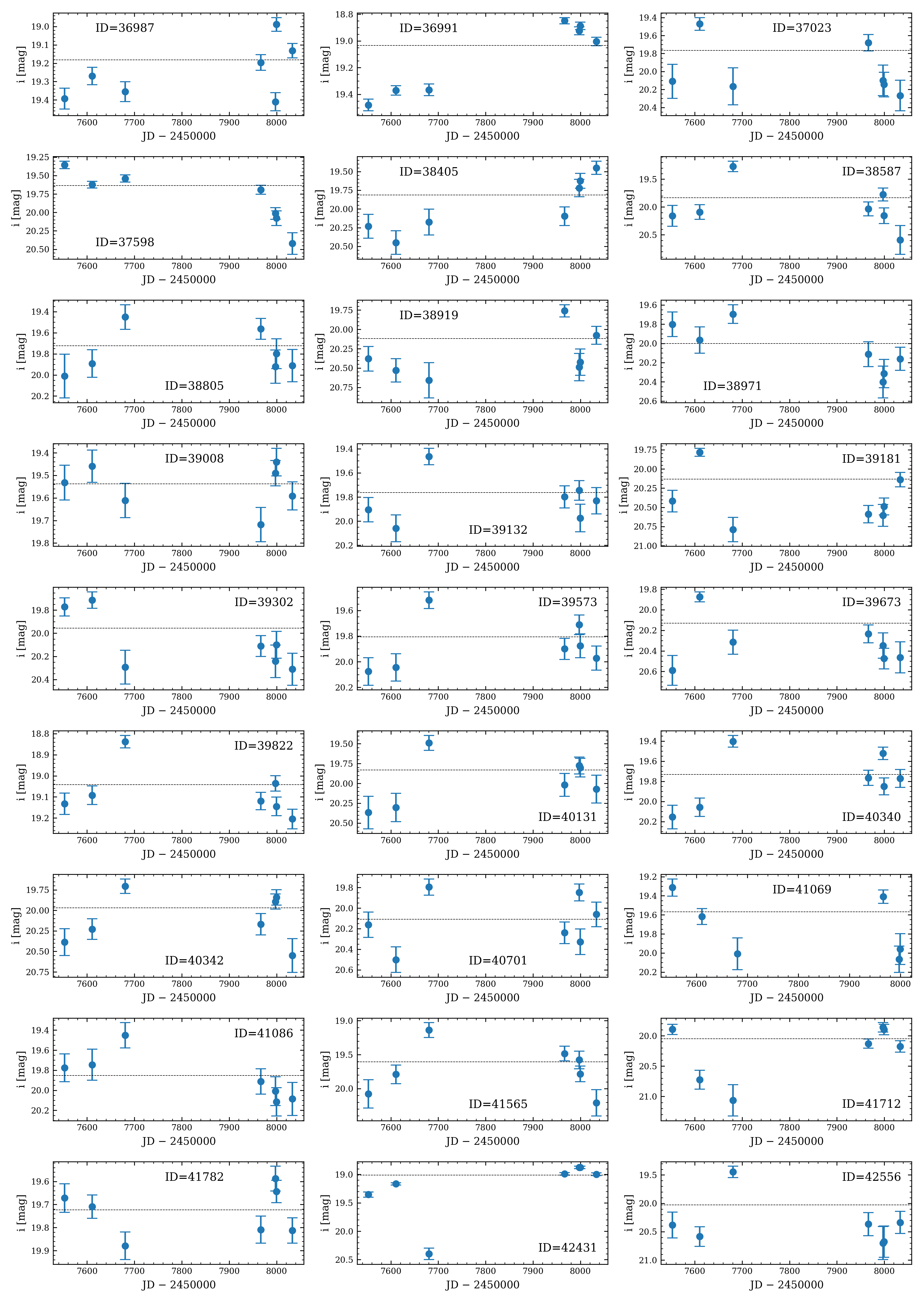}
    \figsetgrpnote{Light curve of LPV candidate (d).}
    \figsetgrpend
    \figsetgrpstart
    \figsetgrpnum{A1.5}
    \figsetgrptitle{Light Curve (e)}
    \figsetplot{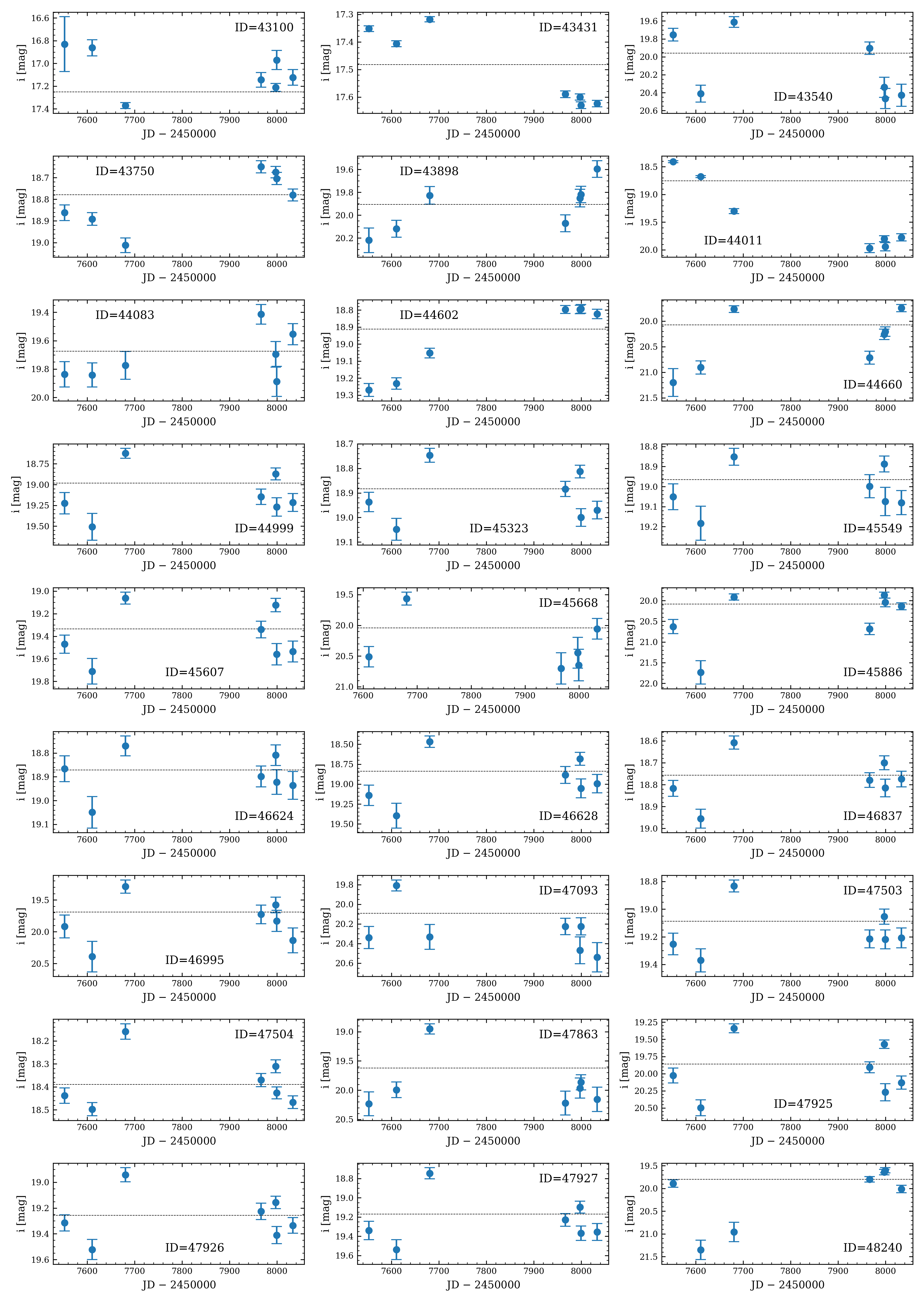}
    \figsetgrpnote{Light curve of LPV candidate (e).}
    \figsetgrpend
    \figsetgrpstart
    \figsetgrpnum{A1.6}
    \figsetgrptitle{Light Curve (f)}
    \figsetplot{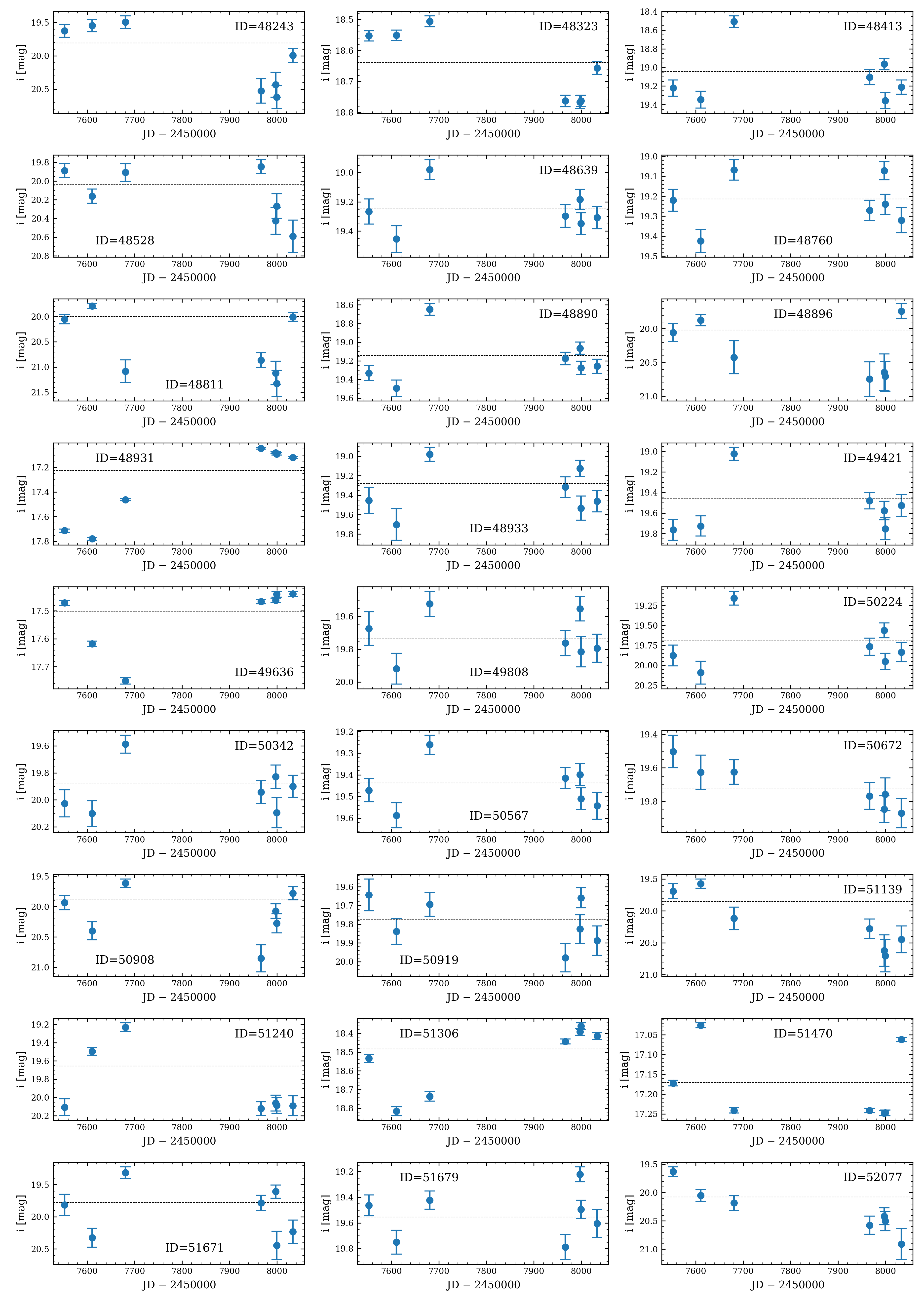}
    \figsetgrpnote{Light curve of LPV candidate (f).}
    \figsetgrpend
    \figsetgrpstart
    \figsetgrpnum{A1.7}
    \figsetgrptitle{Light Curve (g)}
    \figsetplot{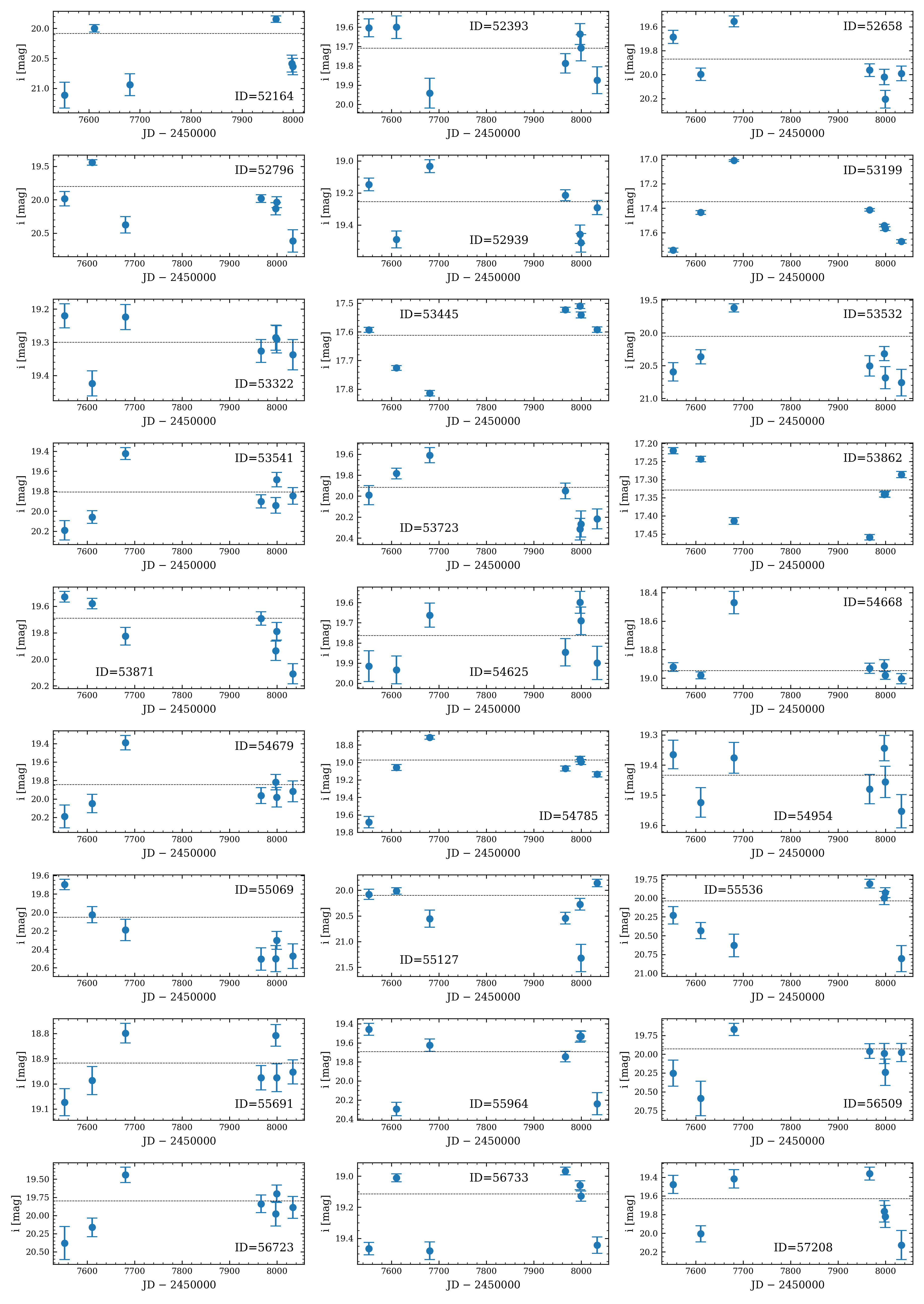}
    \figsetgrpnote{Light curve of LPV candidate (g).}
    \figsetgrpend
    \figsetgrpstart
    \figsetgrpnum{A1.8}
    \figsetgrptitle{Light Curve (h)}
    \figsetplot{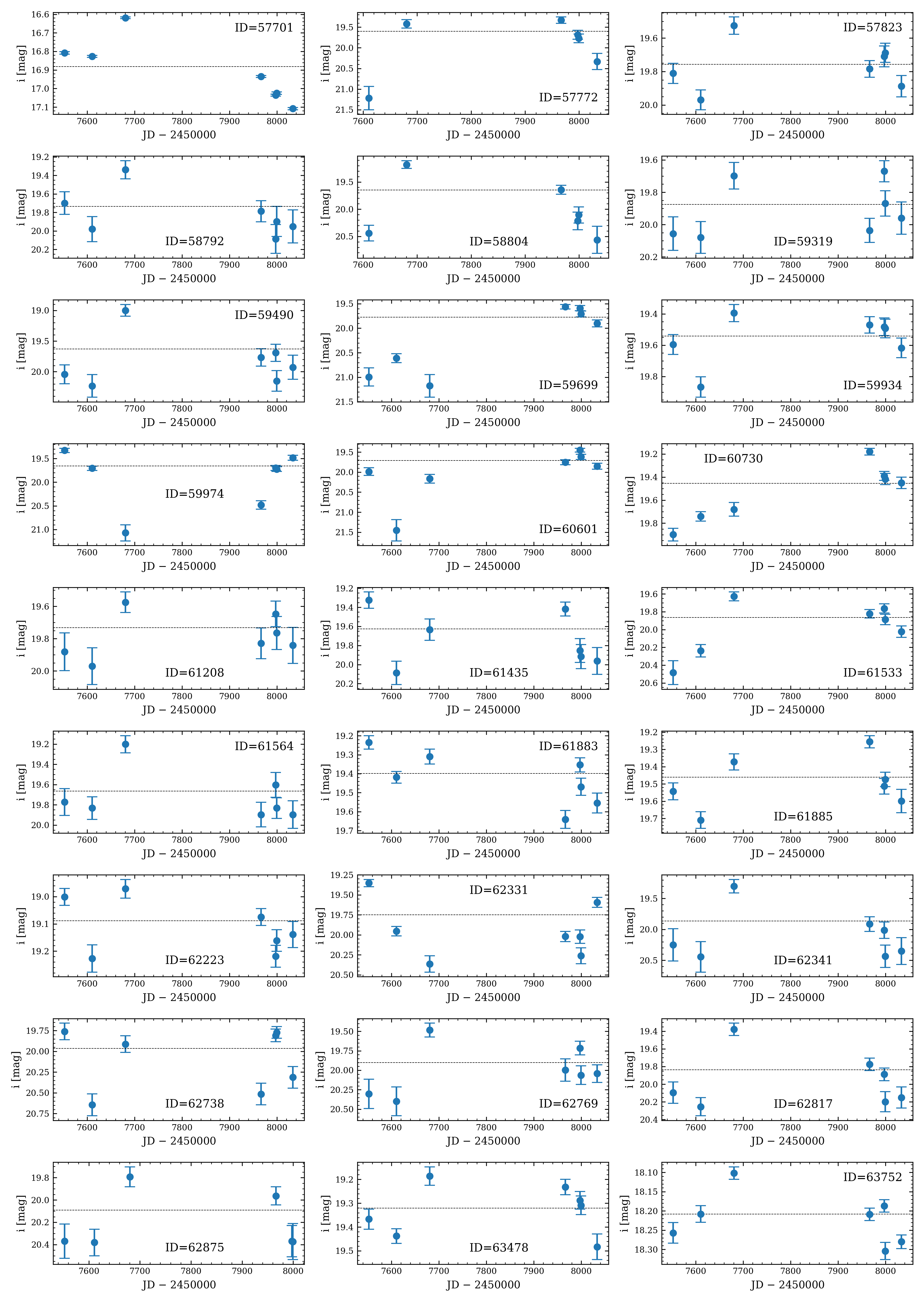}
    \figsetgrpnote{Light curve of LPV candidate (h).}
    \figsetgrpend
    \figsetgrpstart
    \figsetgrpnum{A1.9}
    \figsetgrptitle{Light Curve (i)}
    \figsetplot{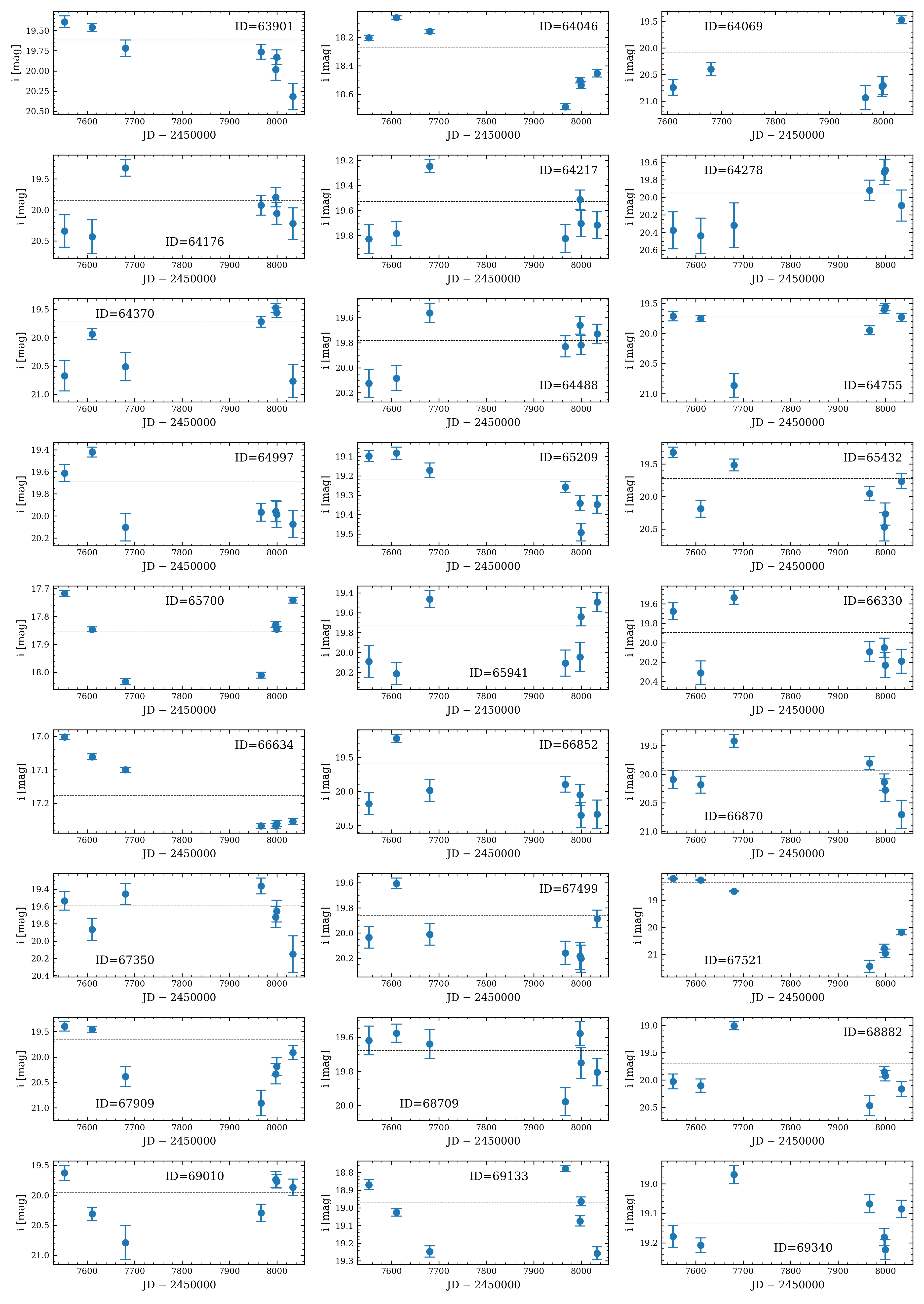}
    \figsetgrpnote{Light curve of LPV candidate (i).}
    \figsetgrpend
    \figsetgrpstart
    \figsetgrpnum{A1.10}
    \figsetgrptitle{Light Curve (j)}
    \figsetplot{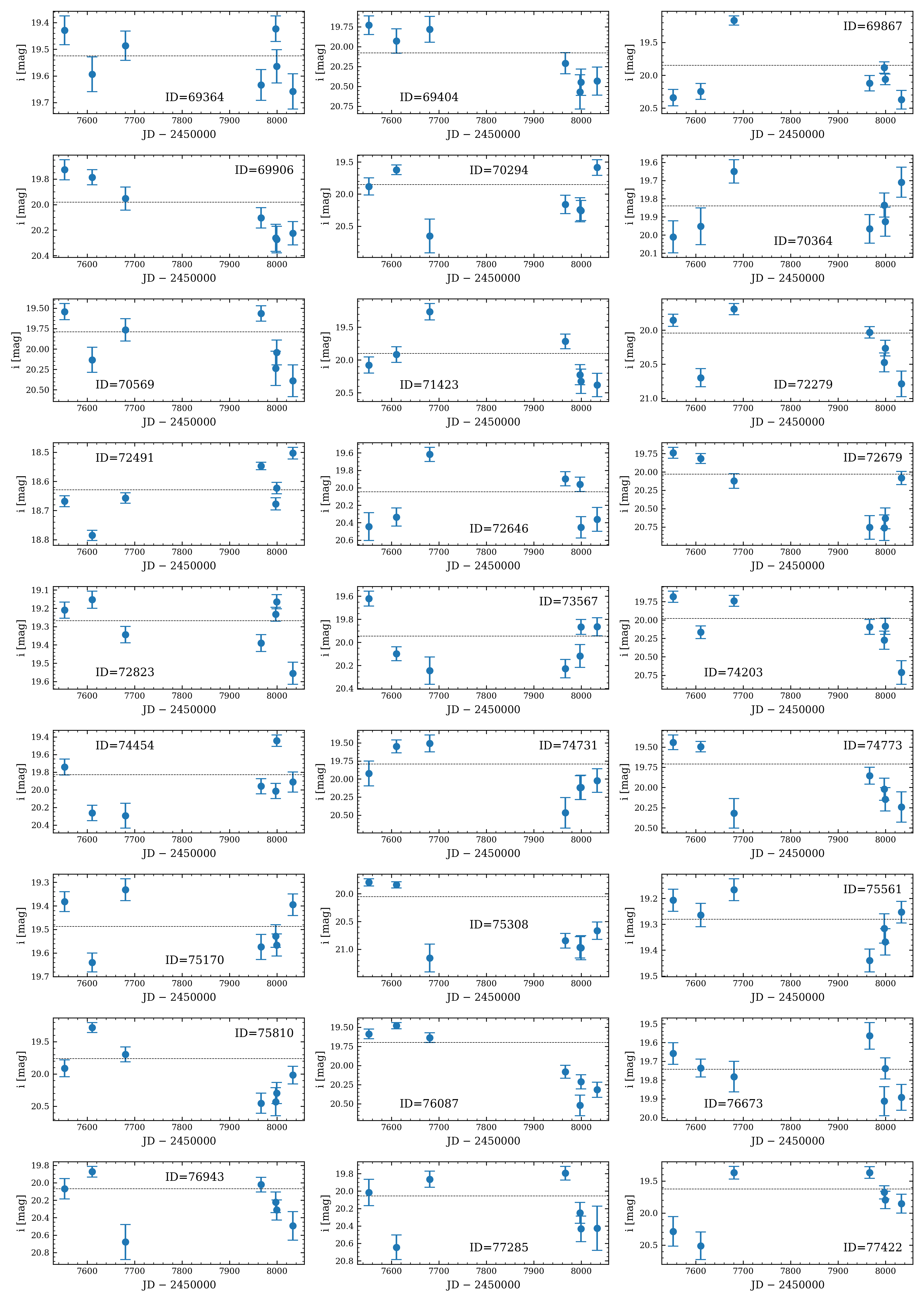}
    \figsetgrpnote{Light curve of LPV candidate (j).}
    \figsetgrpend
    \figsetgrpstart
    \figsetgrpnum{A1.11}
    \figsetgrptitle{Light Curve (k)}
    \figsetplot{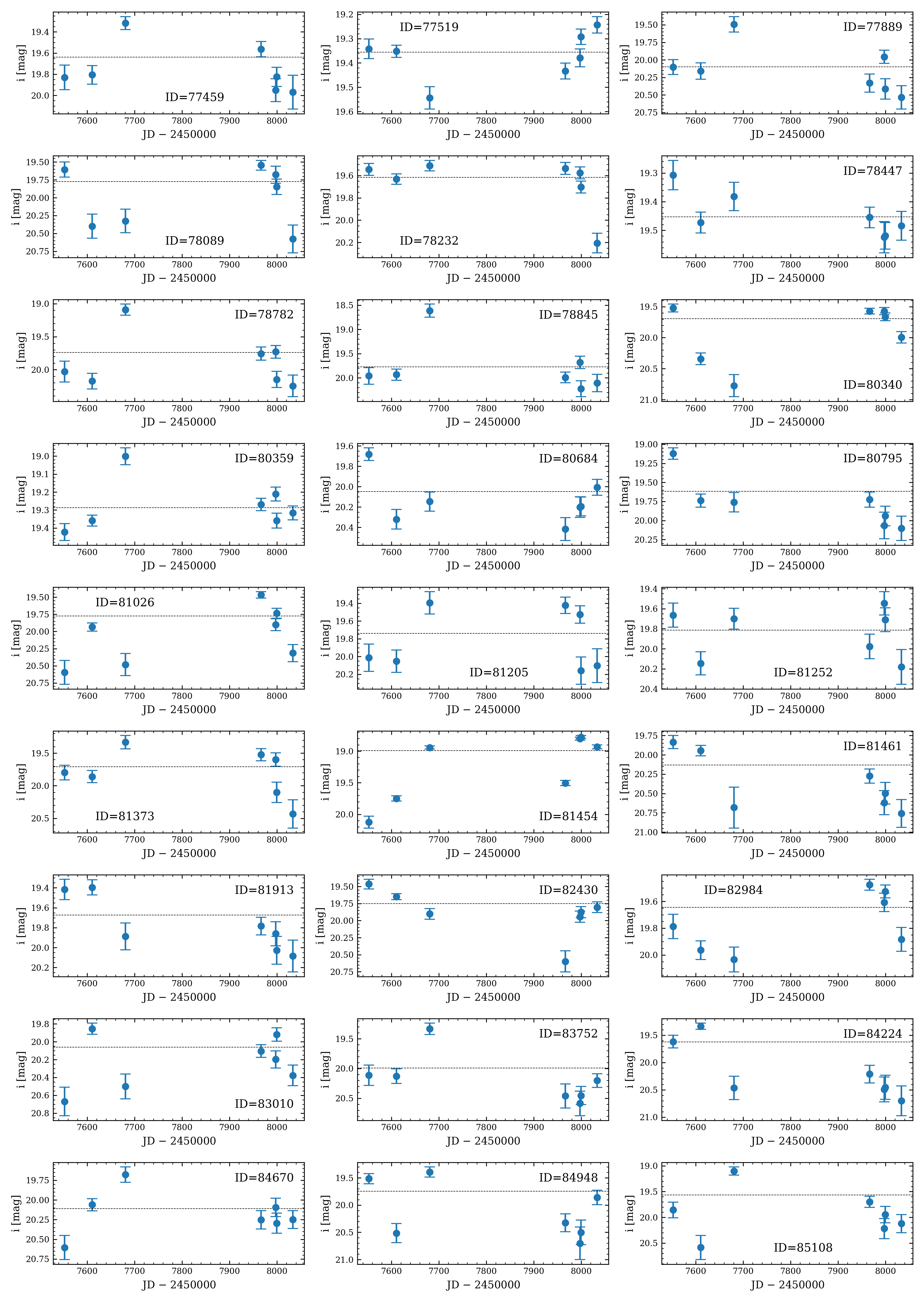}
    \figsetgrpnote{Light curve of LPV candidate (k).}
    \figsetgrpend
    \figsetgrpstart
    \figsetgrpnum{A1.12}
    \figsetgrptitle{Light Curve (l)}
    \figsetplot{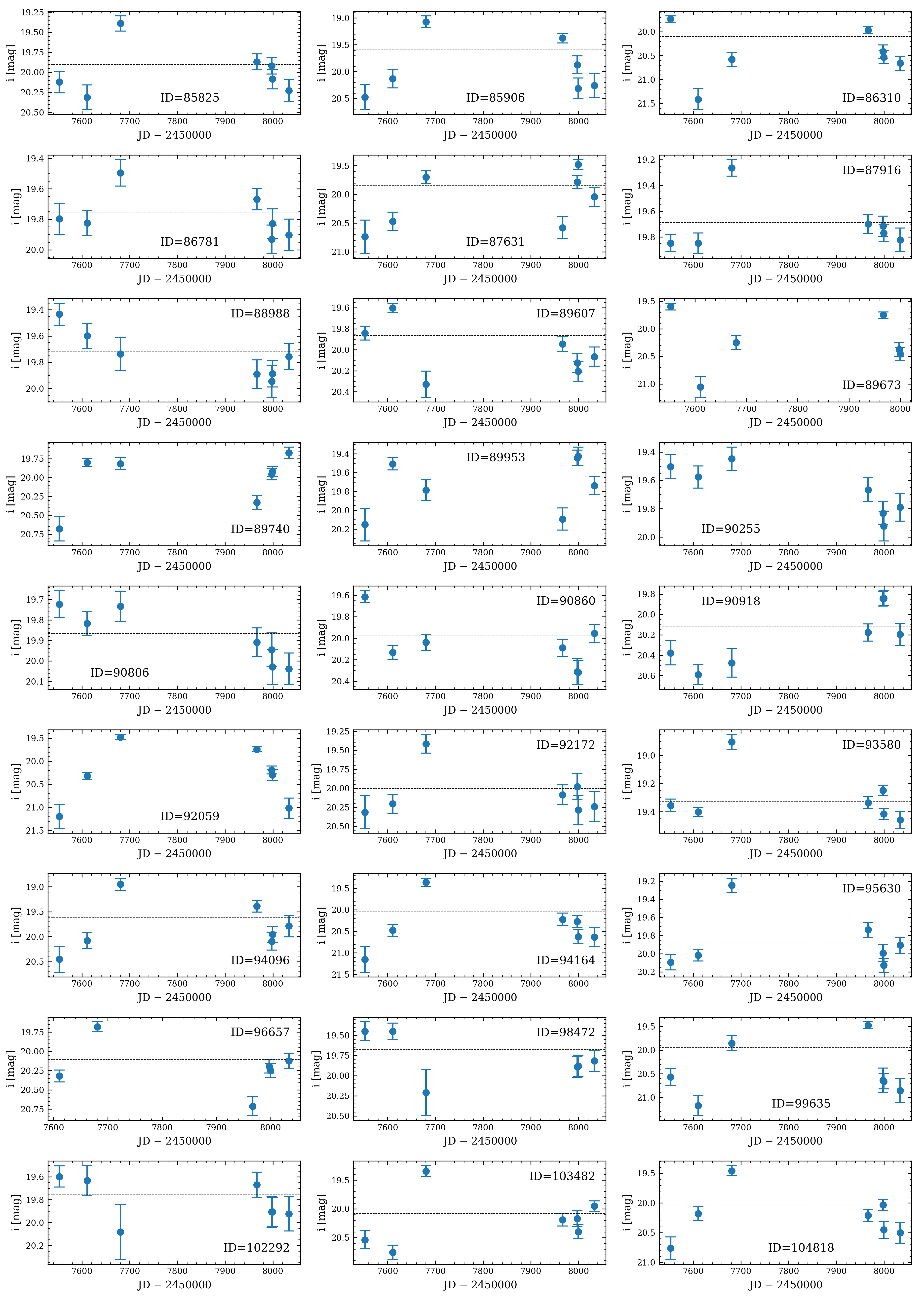}
    \figsetgrpnote{Light curve of LPV candidate (l).}
    \figsetgrpend
    \figsetgrpstart
    \figsetgrpnum{A1.13}
    \figsetgrptitle{Light Curve (m)}
    \figsetplot{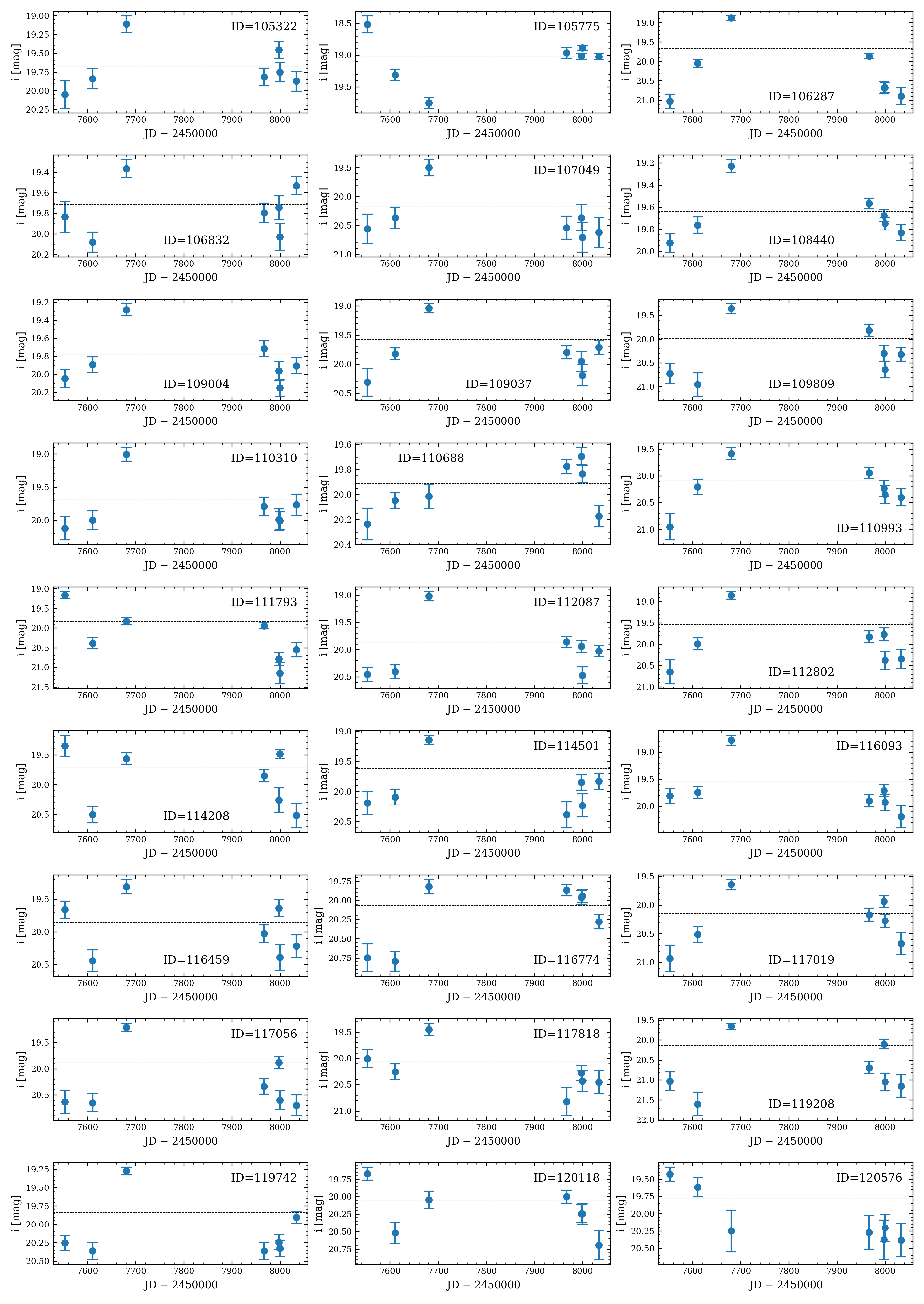}
    \figsetgrpnote{Light curve of LPV candidate (m).}
    \figsetgrpend
    \figsetgrpstart
    \figsetgrpnum{A1.14}
    \figsetgrptitle{Light Curve (n)}
    \figsetplot{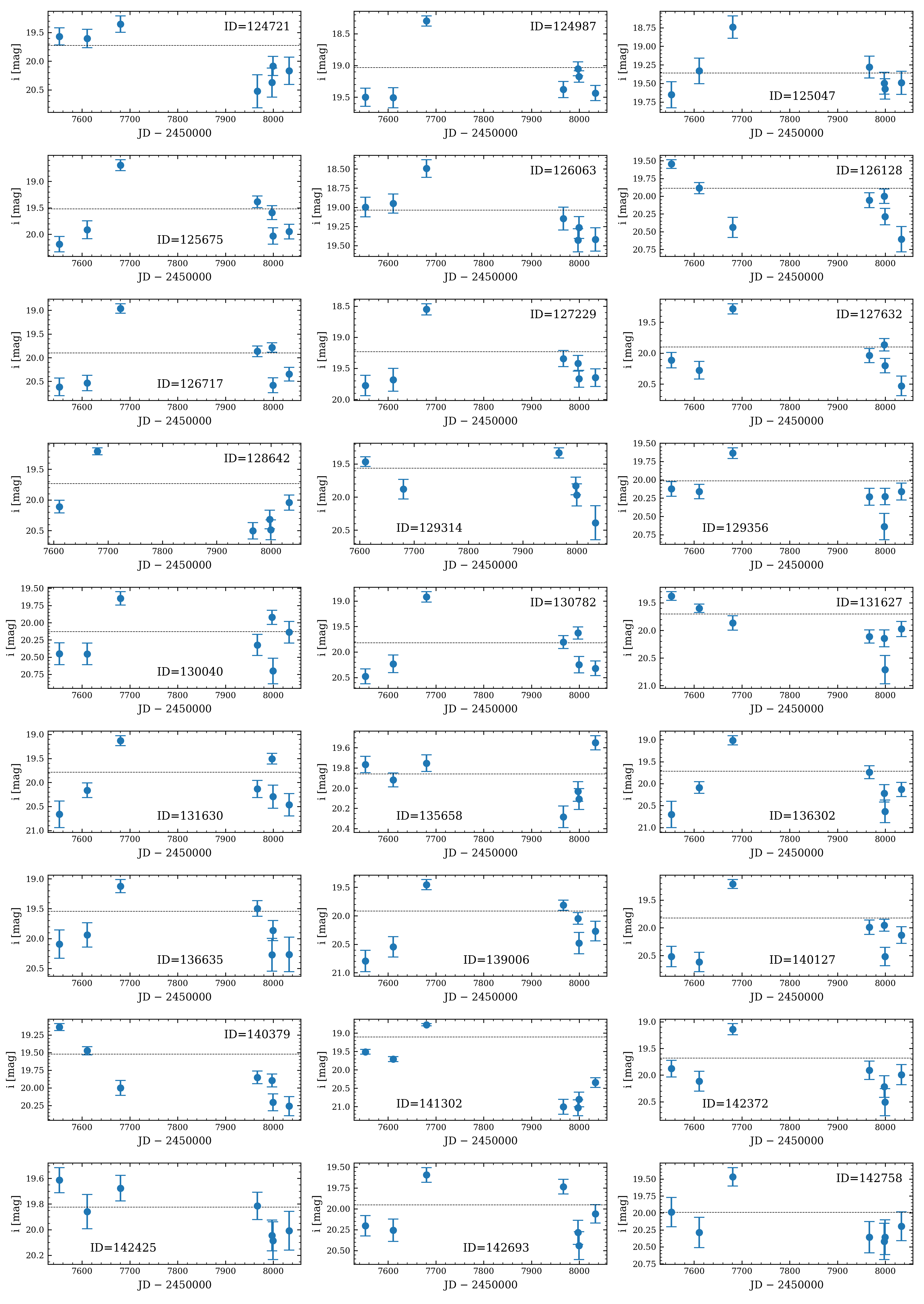}
    \figsetgrpnote{Light curve of LPV candidate (n).}
    \figsetgrpend
    \figsetgrpstart
    \figsetgrpnum{A1.15}
    \figsetgrptitle{Light Curve (o)}
    \figsetplot{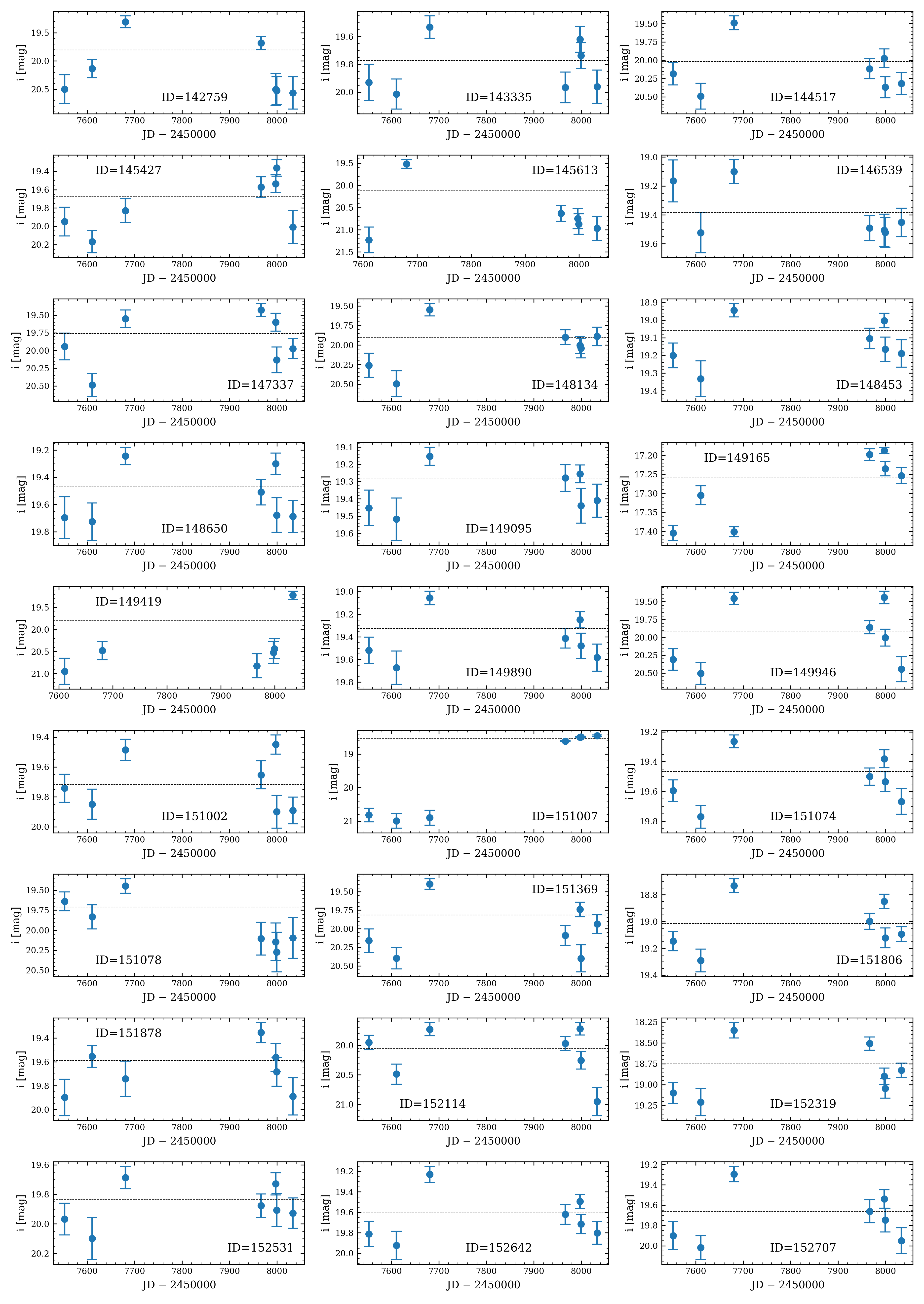}
    \figsetgrpnote{Light curve of LPV candidate (o).}
    \figsetgrpend
    \figsetgrpstart
    \figsetgrpnum{A1.16}
    \figsetgrptitle{Light Curve (p)}
    \figsetplot{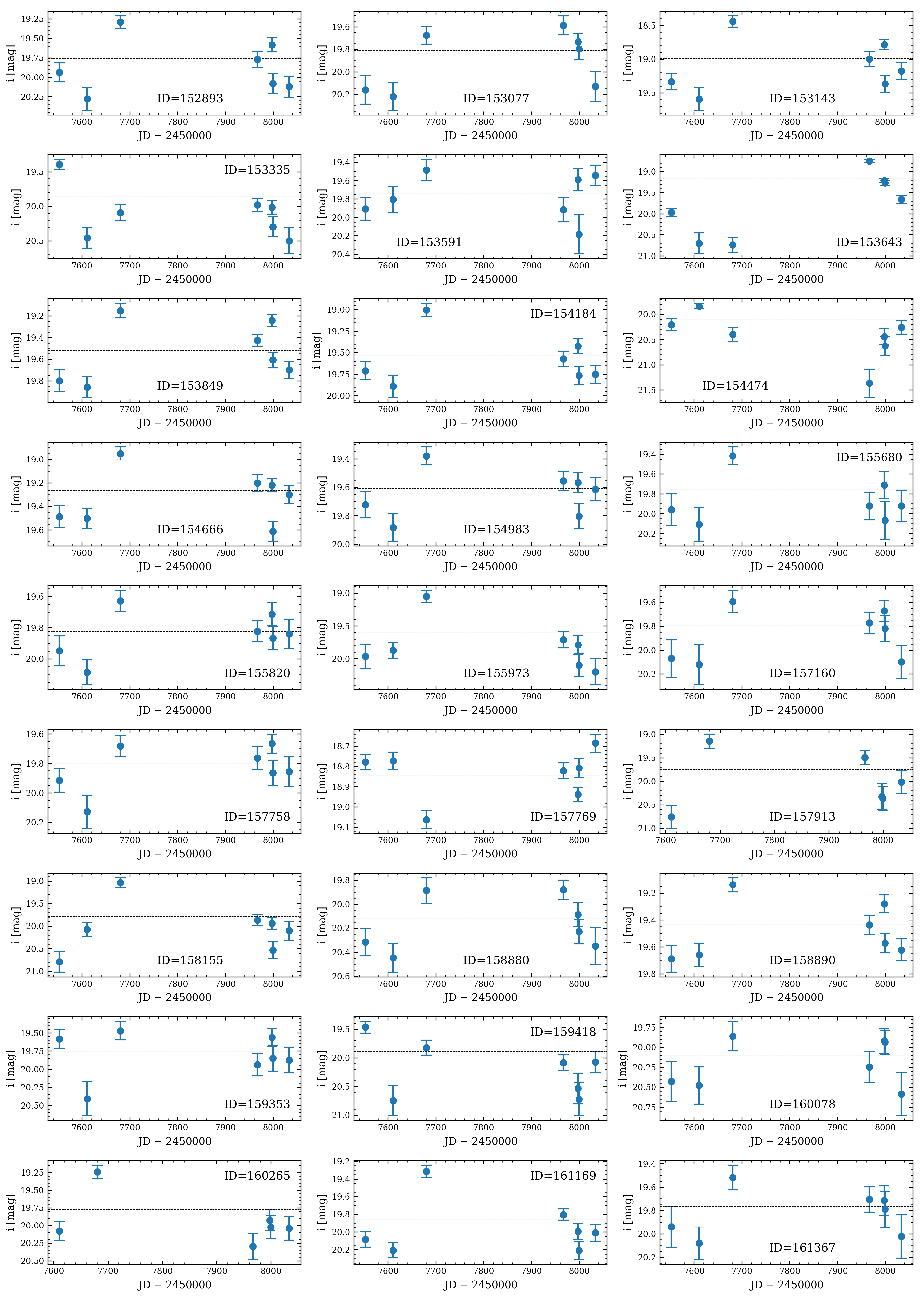}
    \figsetgrpnote{Light curve of LPV candidate (p).}
    \figsetgrpend
    \figsetgrpstart
    \figsetgrpnum{A1.17}
    \figsetgrptitle{Light Curve (q)}
    \figsetplot{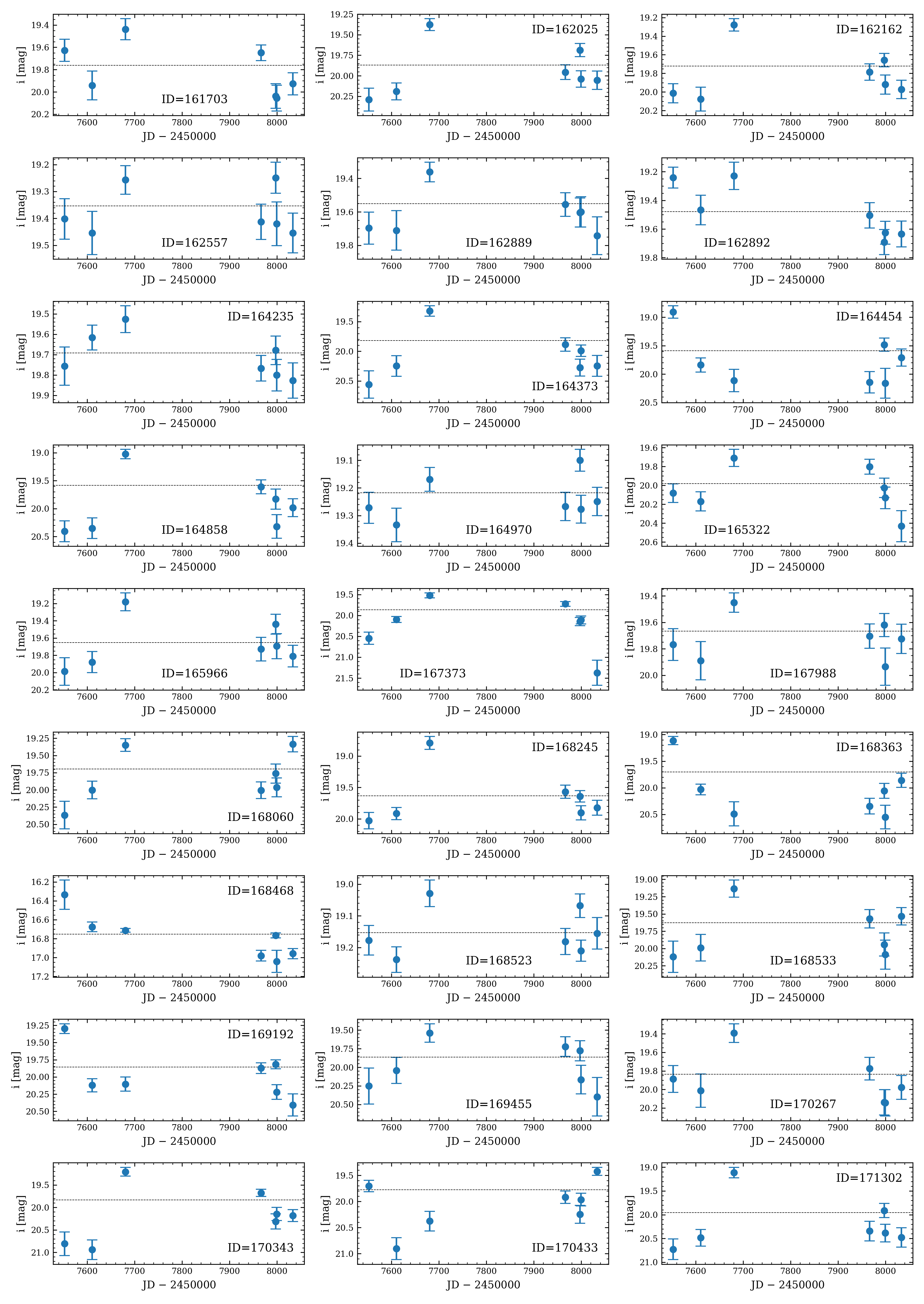}
    \figsetgrpnote{Light curve of LPV candidate (q).}
    \figsetgrpend
    \figsetgrpstart
    \figsetgrpnum{A1.18}
    \figsetgrptitle{Light Curve (r)}
    \figsetplot{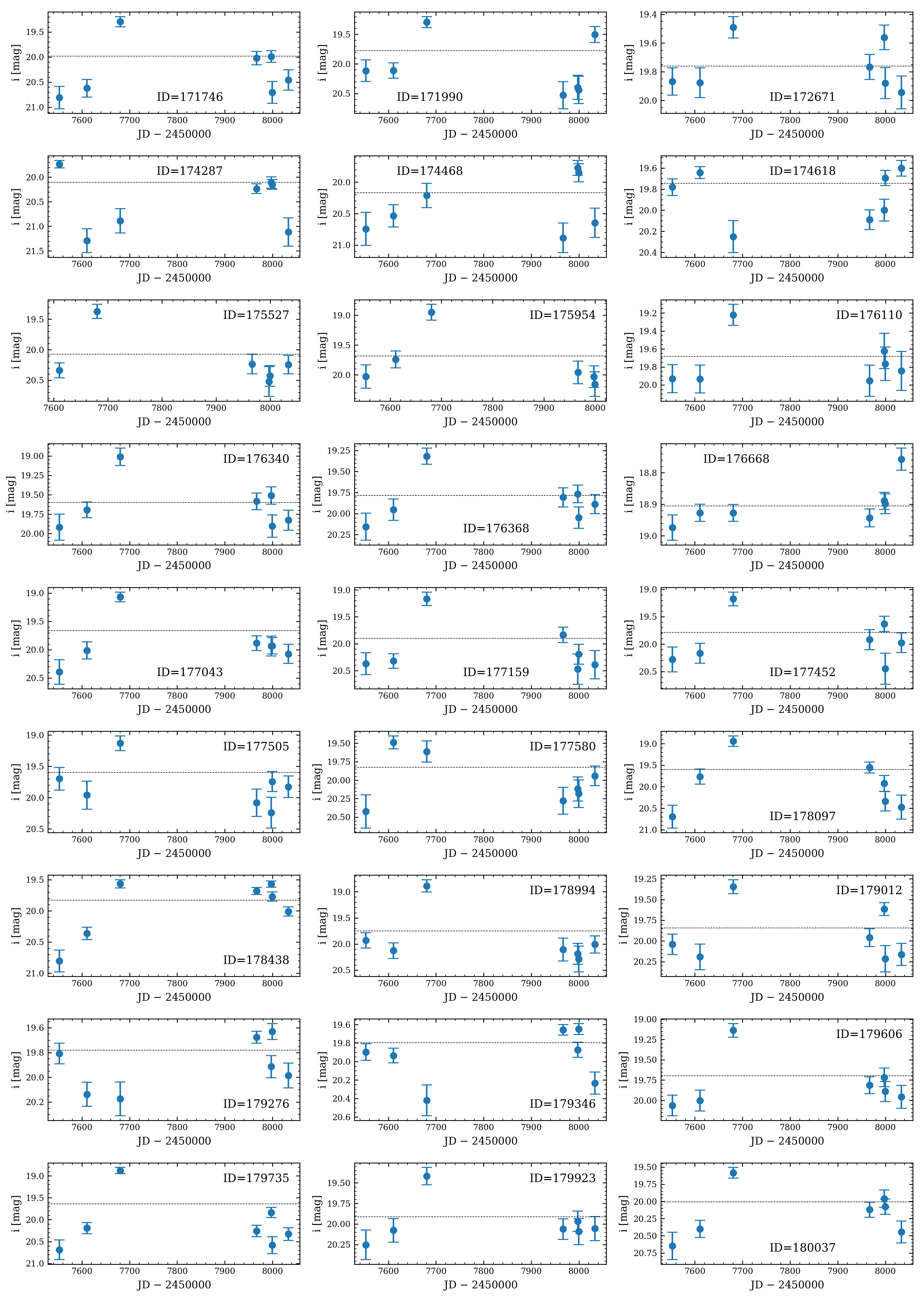}
    \figsetgrpnote{Light curve of LPV candidate (r).}
    \figsetgrpend
    \figsetgrpstart
    \figsetgrpnum{A1.19}
    \figsetgrptitle{Light Curve (s)}
    \figsetplot{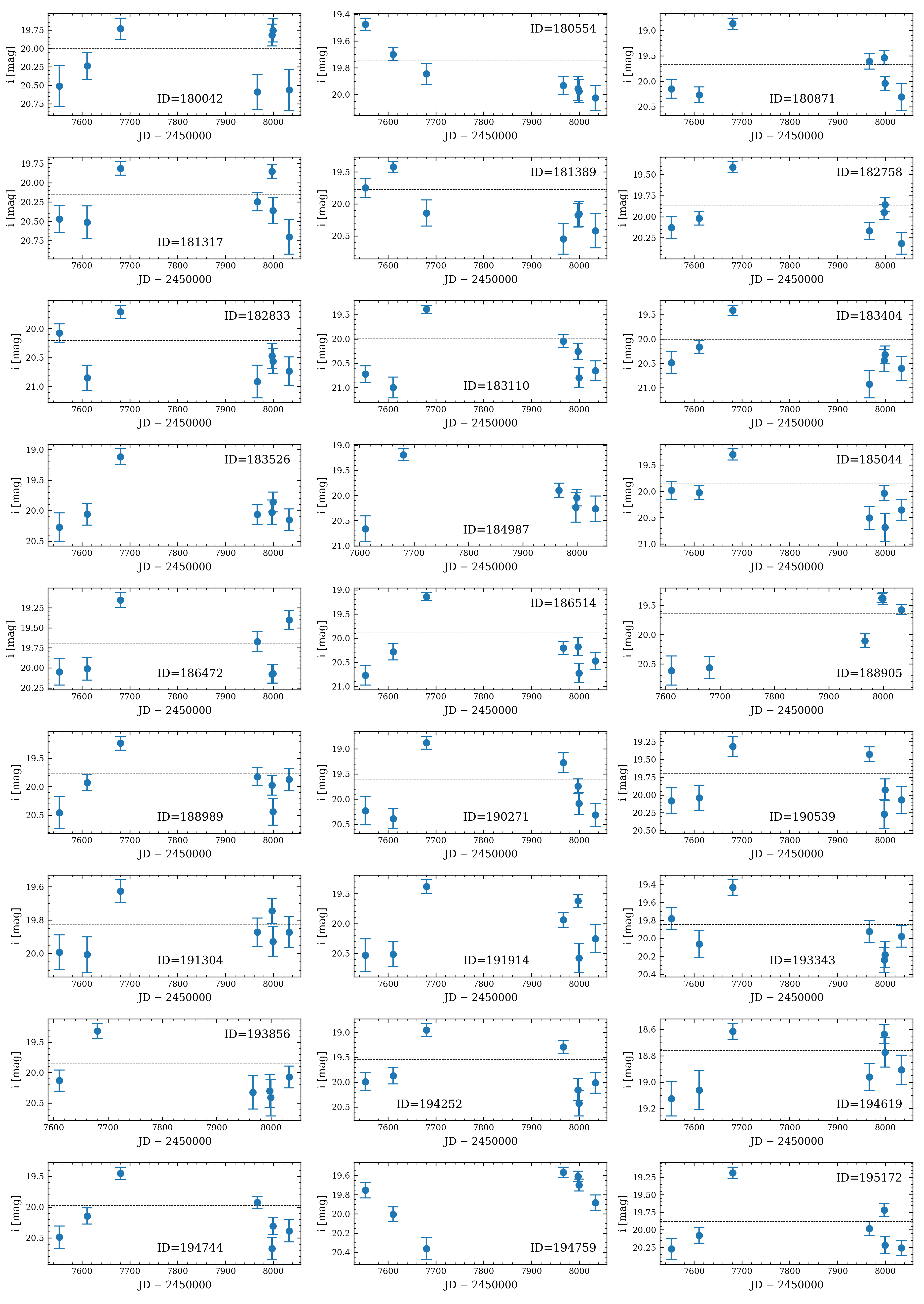}
    \figsetgrpnote{Light curve of LPV candidate (s).}
    \figsetgrpend
    \figsetgrpstart
    \figsetgrpnum{A1.20}
    \figsetgrptitle{Light Curve (t)}
    \figsetplot{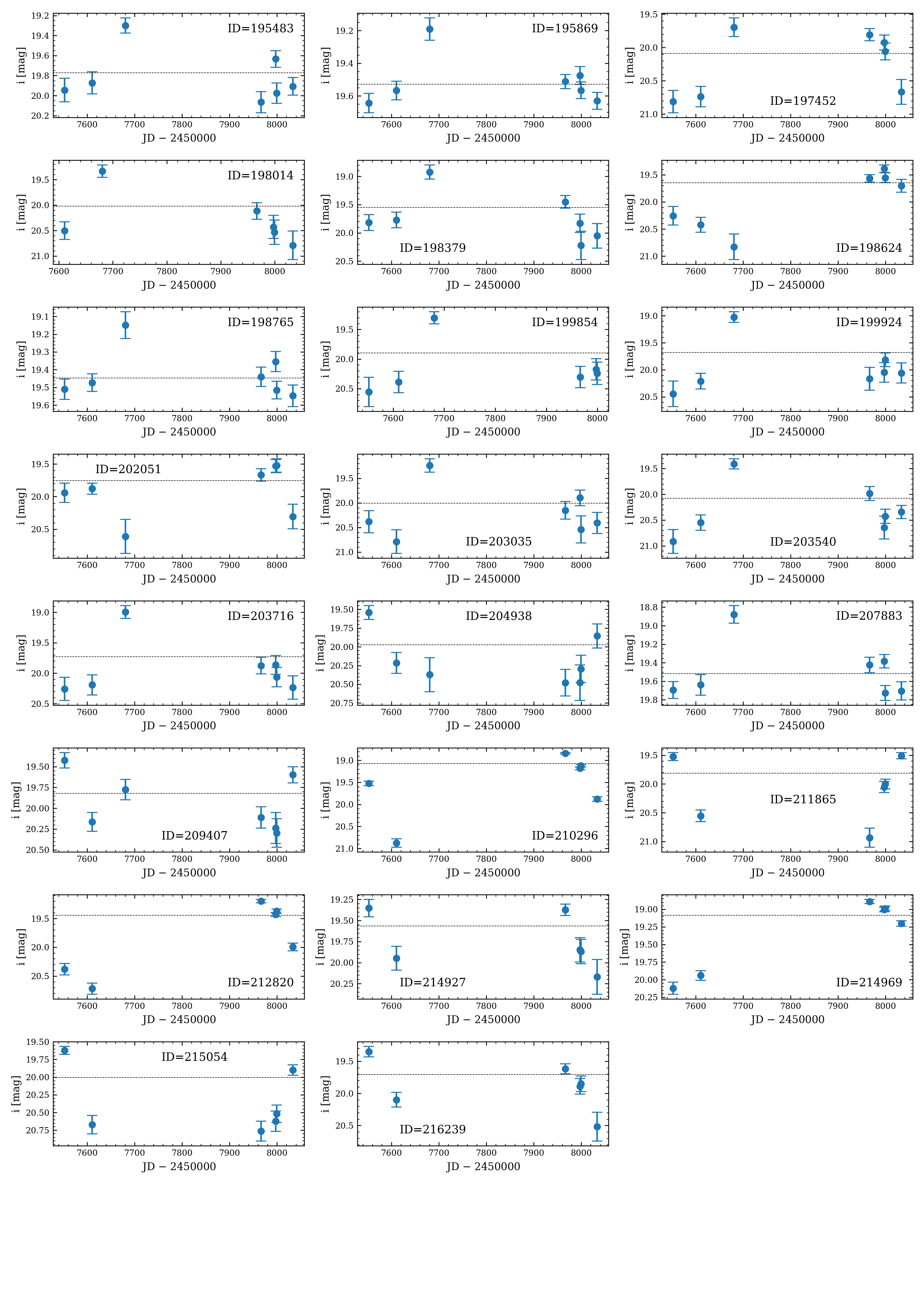}
    \figsetgrpnote{Light curve of LPV candidate (t).}
    \figsetgrpend
    \figsetend
\end{figure}
\clearpage 

\end{document}